\documentclass[a4paper,11pt]{article}
\pdfoutput=1
\usepackage{jcappub} 

\usepackage[T1]{fontenc} 
\usepackage{graphicx}	
\usepackage{amsmath}	
\usepackage{amssymb}	
\usepackage{nicefrac}
\usepackage[normalem]{ulem}
\usepackage{newtxtext,newtxmath}
\usepackage{tikz}
\usepackage{xfrac}
\usepackage{bbold}
\usepackage{slashed}
\usepackage{textcomp}  
\usepackage{scalerel}  
\usepackage{orcidlink}
\usepackage{booktabs}
\usepackage{pifont}
\RequirePackage{silence}
\usepackage{hyperref}

\definecolor{darkgreen}{rgb}{0,0.7,0}
\definecolor{mblue}{rgb}{0,0.4,0.9}

\usetikzlibrary{shadows.blur}
\usetikzlibrary{shapes.symbols}
\usetikzlibrary{positioning, shapes.geometric}

\newcommand*\circled[1]{\tikz[baseline=(char.base)]{
    \node[shape=circle,draw,inner sep=2pt] (char) {#1};}}

\newcommand*\circledred[1]{\tikz[baseline=(char.base)]{
    \node[shape=circle, draw, fill=red, color=red, text=white, inner sep=2pt,
    drop shadow={shadow xshift=0.3mm, shadow yshift=-0.3mm, opacity=0.5}] (char) {#1};}}

\newcommand*\circledgreen[1]{\tikz[baseline=(char.base)]{
    \node[shape=circle, draw, fill=darkgreen, color=darkgreen, text=white, inner sep=2pt,
    drop shadow={shadow xshift=0.3mm, shadow yshift=-0.3mm, opacity=0.5}] (char) {#1};}}

\newcommand*\circledblue[1]{\tikz[baseline=(char.base)]{
   \node[shape=circle, draw, fill=blue, color=blue, text=white, inner sep=2pt,
    drop shadow={shadow xshift=0.3mm, shadow yshift=-0.3mm, opacity=0.5}] (char) {#1};}}

\newcommand*\colourcheck[1]{%
  \expandafter\newcommand\csname #1check\endcsname{\textcolor{#1}{\ding{52}}}%
}
\newcommand*\colourcross[1]{%
  \expandafter\newcommand\csname #1cross\endcsname{\textcolor{#1}{\ding{55}}}%
}
\newcommand*\colourtilde[1]{%
  \expandafter\newcommand\csname #1tilde\endcsname{\textcolor{#1}{{\LARGE \bf \textasciitilde}}}%
}
\colourtilde{orange}
\colourcheck{green}
\colourcross{red}

\newcommand{\vecb}[1]{\ensuremath{\boldsymbol{#1}}}
\providecommand{\dd}{\ensuremath{{\rm d}}}

\newcommand{\fett}[1]{\boldsymbol{#1}}
\newcommand{\be}{\begin{equation}}
\newcommand{\ee}{\end{equation}}
\newcommand{\nabQ}{\fett \nabla_{\text{\fontsize{6}{6} \!\!$\fett Q$}}}

\newcommand{\nabX}{\fett \nabla_{\text{\fontsize{6}{6} \!\!$\fett X$}}}
\newcommand{\nabx}{\fett \nabla_{\text{\fontsize{6}{6} \!\!$\fett x$}}}
\newcommand{\nabq}{\fett \nabla_{\fett q}}
\newcommand{\nabqinverse}{\fett \nabla_{\fett q}^{-1}}
\newcommand{\nabQinverse}{\fett \nabla_{\fett Q}^{-1}}

\newcommand{\pD}{\partial_{\text{\fontsize{7}{7}\selectfont\it D}}}

\newcommand{\BF}{\textsc{BullFrog}}
\newcommand{\ini}{{\rm ini}}
\newcommand{\mPi}{\mathit{\Pi}}
\newcommand{\PF}{{{\text{\fontsize{6}{6}\selectfont ${\rm PF}$}}}}
\newcommand{\BFsmall}{{{\text{\fontsize{6}{6}\selectfont ${\rm BF}$}}}}
\newcommand{\ZA}{{{\text{\fontsize{6}{6}\selectfont ${\rm ZA}$}}}}
\newcommand{\mis}{{{\text{\fontsize{6}{6}\selectfont ${\rm mis}$}}}}
\newcommand{\twoLPT}{{{\text{\fontsize{6}{6}\selectfont ${\rm 2LPT}$}}}}
\newcommand{\FastPMsmall}{{{\text{\fontsize{6}{6}\selectfont ${\rm FPM}$}}}}

\newcommand{\EdS}{{{\text{\fontsize{6}{6}\selectfont ${\rm app}$}}}}
\newcommand{\ma}{\mathfrak{a}}
\newcommand{\mb}{\mathfrak{b}}
\newcommand{\mabar}{\bar{\mathfrak{a}}}
\newcommand{\mbbar}{\bar{\mathfrak{b}}}
\newcommand{\COLA}{\textsc{COLA}}
\newcommand{\fastpm}{\textsc{FastPM}}

\newcommand{\subcurl}[2]{#1_{\text{\fontsize{7}{7}\selectfont \medmuskip=2mu plus 2mu minus 2mu $#2$}}}

\newcommand{\upcurl}[2]{#1_{\text{\fontsize{7}{7}\selectfont \medmuskip=2mu plus 2mu minus 2mu $#2$}}}

\def\({\left(}
\def\){\right)}
\def\[{\left[}
\def\]{\right]}
\def\<{\left\langle}

\newsavebox\MBox

\makeatletter\def\Hy@Warning#1{}\makeatother

\def\FROG{\scalerel*{\includegraphics{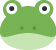}}{\textrm{\textbigcircle}}}

\title{\FROG \, \BF: Multi-step perturbation theory as a time integrator for cosmological simulations}

\author[\orcidlink{0000-0001-5947-9376}\,a,b,c,\star]{Cornelius Rampf,}
\author[\orcidlink{0000-0002-3741-179X}\,b,\star]{Florian List}
\author[\orcidlink{0000-0001-9440-1152}\,b,c]{and Oliver Hahn}

\note[$\star$]{These authors contributed equally.}

\affiliation[a]{Division of Theoretical Physics, Ru\dj er Bo\v{s}kovi\'c Institute, Bijeni\v{c}ka cesta 54, 10000 Zagreb, Croatia}
\affiliation[b]{Department of Astrophysics, University of Vienna, T\"{u}rkenschanzstra{\ss}e 17, 1180 Vienna, Austria}
\affiliation[c]{Faculty of Mathematics, University of Vienna, Oskar-Morgenstern-Platz 1, 1090 Vienna, Austria}

\emailAdd{cornelius.stefan.rampf@irb.hr} 
\emailAdd{florian.list@univie.ac.at}
\emailAdd{oliver.hahn@univie.ac.at}

\abstract{
Modelling the cosmic large-scale structure can be done through numerical $N$-body simulations or by using perturbation theory. Here, we present an $N$-body approach that effectively implements a multi-step forward model based on Lagrangian Perturbation Theory (LPT) in a $\Lambda$CDM Universe. This is achieved by introducing the second-order accurate \BF\ integrator, which automatically performs 2LPT time steps to second order without requiring the explicit computation of 2LPT displacements. Importantly, we show that \BF\ trajectories rapidly converge to the exact solution as the number of time steps increases, at any moment in time, even though 2LPT becomes invalid after shell-crossing. As a validation test, we compare \BF\ against other $N$-body integrators and high-order LPT, both for a realistic $\Lambda$CDM cosmology and for simulations with a sharp UV cutoff in the initial conditions. The latter scenario enables controlled experiments against LPT and, in practice, is particularly relevant for modelling coarse-grained fluids arising in the context of effective field theory. We demonstrate that \textsc{BullFrog} significantly improves upon other LPT-inspired integrators, such as \textsc{FastPM} and \textsc{COLA}, without incurring any computational overhead compared to standard $N$-body integrators. Implementing \textsc{BullFrog} in any existing $N$-body code is straightforward, particularly if \textsc{FastPM} is already integrated.
}


\subheader{
\footnotesize
\vspace*{-3.7em}
\begin{flushright}
RBI-ThPhys-2024-19
\end{flushright}
}

\begin{document}
\maketitle
\flushbottom

\section{Introduction}\label{sec:intro}

Perturbation theory (PT) is an indispensable tool for interpreting cosmological observables on a wide range of spatio-temporal scales \cite{Bernardeau:2001qr,2020moco.book.....D}. In general, PT is useful whenever the underlying physics is approximately linear; prominent examples are the computation of the cosmic microwave anisotropies imprinted in the surface of last scattering \cite{Hu:2001bc,WMAP:2012nax,Planck:2018vyg}, the analysis of the baryonic acoustic oscillation feature in the cosmic large-scale structure (LSS) \cite{Beutler:2011hx,Blake:2011en,Slepian:2016nfb}, as well as for the general purpose of data inference of the cosmological parameters from large-scale observations \cite{Croft:1997jf,2012MNRAS.420...61K,2012JCAP...11..029G,2014JCAP...01..042D,2018JCAP...04..030S,2020JCAP...11..035C}.

On large cosmological scales close to the horizon, the patterns of cosmic structures are well described by linear theory as they mostly follow the overall expansion of the Universe; therefore, PT predictions are very reliable on such scales. Even on smaller scales, perturbative predictions remain fairly accurate as long as the nonlinearities are sufficiently suppressed w.r.t.\ the bulk part; see e.g.\ refs.\,\cite{Rampf:2021rqu,2022LRCA....8....1A} for reviews and refs.\,\cite{2006MNRAS.373..369C,2008JCAP...10..036P,2008PhRvD..77f3530M,2013MNRAS.429.1674C,2013JCAP...09..024B,Vlah:2014nta,Rampf:2022apj,Garny:2022kbk} for rather recent findings. However, standard (Eulerian) PT becomes severely hampered at smaller spatial scales, essentially due to the nonlinear accumulation of collisionless matter at focused locations, which gradually generates local density fluctuations of order unity and well beyond.

A strategy for prolonging the validity of PT -- at least for some time -- is to solve the underlying equations perturbatively by employing Lagrangian coordinates \cite{1989A&A...223....9B,1991ApJ...382..377M,1992ApJ...394L...5B,1992MNRAS.254..729B,1995A&A...296..575B,1997GReGr..29..733E,Rampf:2012xa,Zheligovsky:2013eca}. Indeed, the so-called Lagrangian Perturbation Theory (LPT) converges until trajectories of collisionless matter intersect for the first time \cite{Rampf:2017jan,Saga:2018nud,2021MNRAS.501L..71R,2023PhRvD.108j3513R}. This instance is called shell-crossing and is accompanied by extreme matter densities; furthermore, it implies the breakdown of conventional/standard PT \cite{2018JCAP...06..028P,2021MNRAS.505L..90R}. 
Beyond its validity regime, LPT diverges in the sense that higher-order approximations perform worse than lower-order ones (see e.g.\ \cite{Sahni:1995rr,Rampf:2022eiu}).
Recently, beyond-standard perturbative techniques have been developed, applied at the fine-grained fluid level, that allow one to go beyond shell-crossing \cite{Colombi:2014lda,Taruya:2017ohk,2021MNRAS.505L..90R,Saga:2023vno}; however, these approaches are not mature enough to follow the halo formation process, which comes with hundreds of overlapping fluid streams.

Another way for prolonging the validity of conventional PT is to employ effective or coarse-grained fluid descriptions; see e.g.\ refs.\ \cite{Baumann:2010tm,Pietroni:2011iz,Carrasco:2012cv,Cabass:2022avo,Ivanov:2022mrd}. There, one filters out small-scale density fluctuations in the initial conditions by applying suitable smoothing operations, so that the time of shell-crossing for the coarse-grained fluid is significantly extended. As a consequence, the coarse-grained fluid is still within the realm of standard PT for sufficiently strong smoothing, which, for example, enables systematic bias expansions that relate the matter dynamics to the distribution of galaxies \cite{Desjacques:2016bnm}. Relating the distinct evolution of the coarse- to the fine-grained fluid is still possible, but typically requires input from numerical simulations; specifically, this input is encoded in an effective sound speed for the leading-order treatment within the effective field theory of LSS. However, from the perspective of data inference (and taking observational systematics aside), the effective description remains fully predictive on mildly nonlinear scales, even in the absence of such external inputs \cite{Schmidt:2018bkr,Schmidt:2020tao,Andrews:2022nvv,Kostic:2022vok,Tucci:2023bag,Nguyen:2024yth}, provided that 
(1) appropriate smoothing operations are applied in the analysis of the observable at hand, and 
(2) that the modelling of the coarse-grained fluid is sufficiently accurate. 
Within a perturbative context, the last requirement implies that the smoothing needs to be sufficiently strong, although this can depend on specifics of the employed forward model (see below). At the same time, it is clear that including smaller spatial scales in the analysis would unlock more information.

In principle, drawing from the information-rich nonlinear regime of the LSS is feasible by numerical simulations. These simulations are usually implemented as $N$-body methods, where the continuous distribution of collisionless matter is solved by following a discrete set of $N$ tracer particles; see ref.\,\cite{2022LRCA....8....1A} for a recent review and an overview of related simulation techniques. The general strategy of such methods is to time-integrate the Hamiltonian equations of motion using a leapfrog/Verlet scheme, which provides asynchronous updates for the particle positions (`drift') and momenta (`kick') within a time step, consisting of three consecutive operations; these three operations are either drift-kick-drift or kick-drift-kick. Such schemes are second-order accurate in time, provided that the gravitational force field is sufficiently regular. Here, `second-order accurate' means that at a given time, the deviation between the numerical and exact solution for the phase-space positions scales as the square of the time step.

While the drift calculation is trivial for $N$-body methods, the computational bottleneck relates to the kick update, as it requires evaluating the current gravitational force field, which is subject to solving the cosmological Poisson equation. The computational complexity for the kick update is typically $O(N \log N)$ \cite{2022LRCA....8....1A}, and thus is sensitive to the number of tracer particles. Unfortunately, to cover the vast range of scales of current and upcoming galaxy and radio surveys \cite{LSST:2008ijt,Racca:2016qpi,Weltman:2018zrl,2020arXiv200304962X,2021MNRAS.507.1746E}, very expensive simulations are required \cite{Habib:2014uxa,Potter:2016ttn,Maksimova:2021ynf,Ishiyama:2020vao}, involving $10^{12} - 10^{13}$ and more tracer particles. Even worse, a statistical analysis for retrieving the unknown cosmological parameters requires not just one, but a large set of expensive simulations (although machine-learning techniques, or the use of differentiable simulations could greatly reduce this complexity \cite{Villaescusa-Navarro:2019bje,CAMELS:2020cof,2021A&C....3700505M,2022arXiv221109815L,2022mla..confE..52Z}). Therefore, when leveraging cosmological simulations for the purpose of parameter inference, it is crucial to minimise their computational costs.

One way to improve the efficiency of an $N$-body simulation is the development of fast time-integration methods \cite{2013JCAP...06..036T,Howlett:2015hfa,2016MNRAS.463.2273F,2021JCAP...01..016B,List:2023jxz,List:2023kbb} (see e.g.\ refs.\,\cite{Scoccimarro:2001cj,Chartier:2020pmu,Partmann:2020qzb,2021arXiv210414568A,Kokron:2021xgh,Chartier:2021frd} for other fast simulation/hybrid techniques). There, the time integrator of the simulation is fine-tuned to incorporate insights for modelling cosmic structure formation, usually provided by PT (the sky is the limit). As a result, only a few time steps (i.e.\ $10 - 100$) are required to accurately reproduce the statistics on mildly nonlinear scales. Obviously, `less time steps' imply `less force calculations', and thus, fast methods have the potential to alleviate the core bottleneck of numerical simulations.

\begin{figure}
    \centering
    \resizebox{0.66\textwidth}{!}{
    \includegraphics{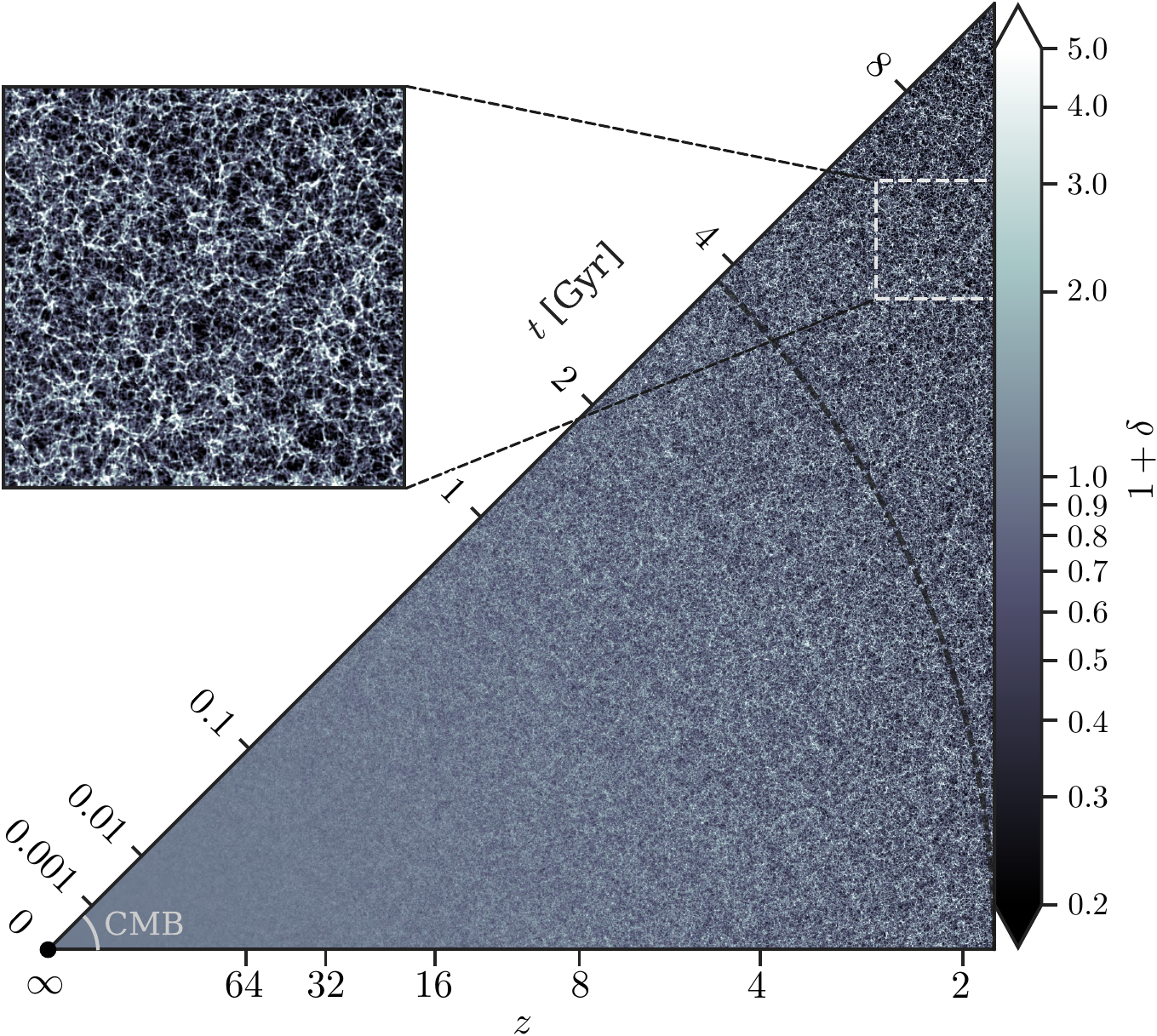}
    }
    \caption{Light cone of cosmic structures as seen by photons propagating from the Big Bang until today, obtained by exploiting our unified framework. Here, $1+\delta$ denotes the matter overdensity.
   }
   \label{fig:lightcone}
\end{figure}

An important example of a fast time-integrating method is \fastpm~\cite{2016MNRAS.463.2273F}, which, to the leading order, comes with built-in Zel'dovich motion a.k.a.\ first-order LPT, where the latter comprises the bulk part of the evolution of collisionless matter on very large scales. This integrator is second-order accurate and symplectic, as recently analysed in ref.\,\cite{List:2023jxz}. In fact, \fastpm~belongs to a class of cosmological integrators \cite{List:2023jxz}, all of which can be made Zel'dovich consistent before shell-crossing and up to higher-order terms (with the term `Zel’dovich consistency' elucidated at the end of this section). The essence of such {\it $D$-time integrators} is that the momentum is not canonically related to the position of particles, but instead defined by the rate of change of particle positions with respect to the growth time~$D=D_+$ of linear density fluctuations in a $\Lambda$CDM Universe. Amongst others, ref.\,\cite{List:2023jxz} introduced \textsc{PowerFrog}, which is their most powerful $D$-time integrator in terms of computational performance and efficiency. Unlike \fastpm, this integrator accurately matches second-order LPT for sufficiently early times. This capability allows us to start $N$-body simulations at time zero and analyse the emergence of cosmic structures within a unified framework (i.e., without switching from LPT to $N$-body at some intermediate time as commonly done); see figure~\ref{fig:lightcone}.

\begin{figure}
    \centering
    \resizebox{1\textwidth}{!}{\includegraphics{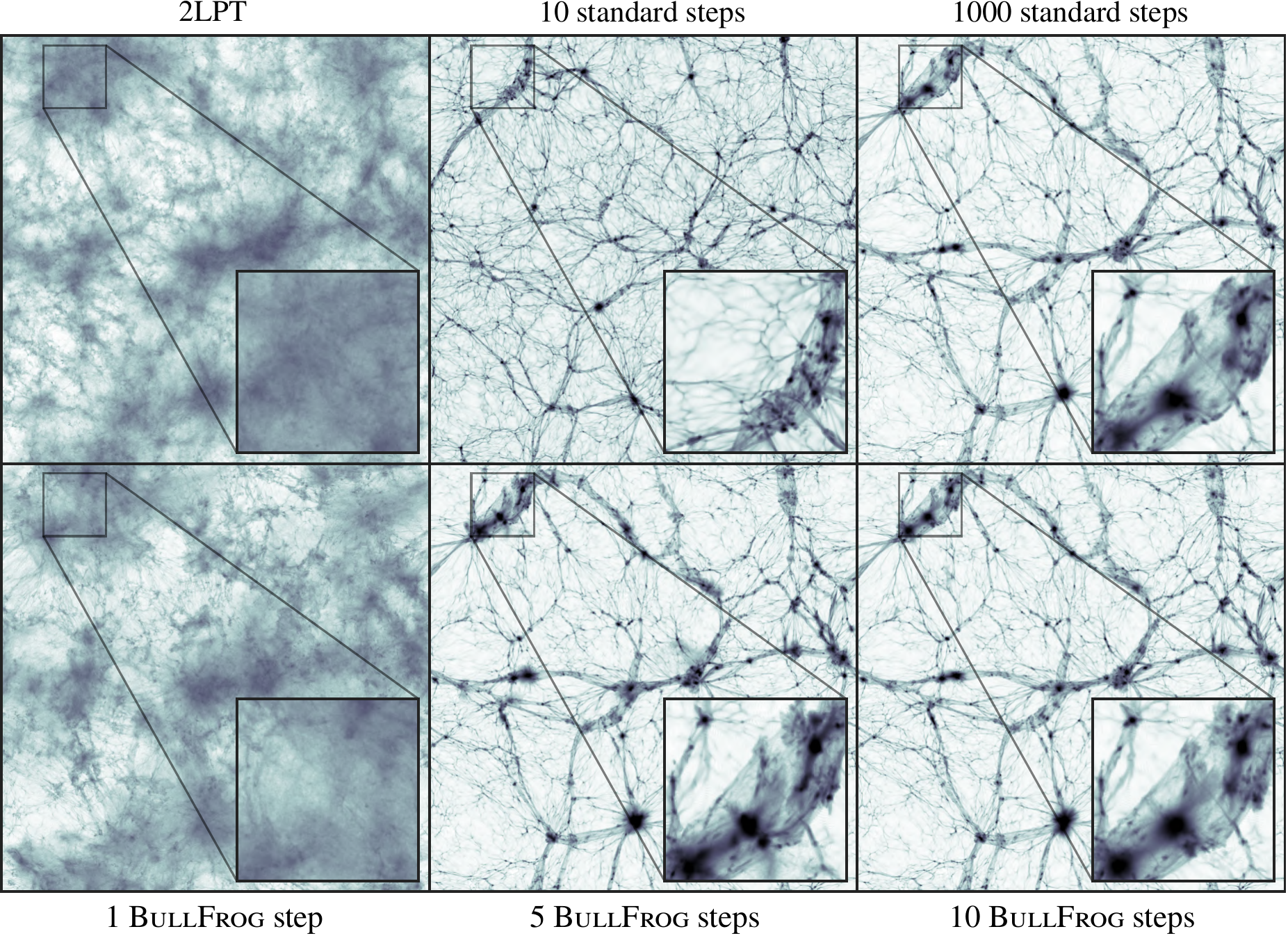}
    }
    \caption{Illustration of razor-thin slices through today's overdensity field obtained using various forward models for a $(25\,h^{-1}\,{\rm Mpc})^3$ box containing $512^3$ particles  (see section~\ref{sec:res} for implementation details and quantitative results). Simulations with `standard steps' were conducted using a standard symplectic leapfrog integrator, initialised at $z = 127$ with 2LPT. While 10 standard steps lead to the formation of small-scale structure, the growth of large-scale structure deviates significantly from the reference simulation, which was obtained with 1000 standard steps. By contrast, \BF\ already performs slightly better than 2LPT for a single step, and converges rapidly across all scales as the number of steps increases.
    }  \label{fig:field_level_intro}
\end{figure}

In this paper, we introduce \BF, short for \uline{B}lazingly fast \uline{U}V-complete \uline{L}PT-informed \uline{L}eap\uline{\textsc{Frog}}, which is a $D$-time integrator as well. \BF\ is the first integrator that, to second order, is in accordance with second-order LPT in $\Lambda$CDM after each completed time step (which \textsc{PowerFrog} achieved only for the initial step starting at $D=0$).  In essence, \BF\ implements a multi-time-stepping variant of 2LPT, where the latter is, in fact, a single time-step approach (since the LPT displacement is the result of integrating the fluid equations from initial to final time). Even more, building 2LPT into \BF\ enables high accuracy, far beyond the 2LPT level, even with a limited number of time steps; see figure~\ref{fig:field_level_intro} for a demonstration.

From the perspective of LPT, one can explain the performance gain of \BF\ as follows:
Standard LPT takes as input the initial gravitational field, and seeks power-series solutions around a unique temporal expansion point (usually about $D=0$). By contrast, \BF\ receives consecutive 2LPT updates on the evolved gravitational field during each kick calculation and thus has a significant advantage over single-step LPT, which basically needs to forecast the evolution of the gravitational field based on the initial conditions. In many respects, \BF\ could be viewed as a near-optimal (and easy to access!) implementation of LPT, as arbitrarily accurate-in-time solutions are achieved, simply by executing as many time steps as needed to reach a certain target accuracy. Alternatively, one could use `single-time-step' LPT to (much) higher orders to achieve comparable accuracy, at least until shell-crossing; however, as it turns out, generating $n$LPT solutions with $n \gtrsim 4$ is in general not recommended, mainly due to a sweeping memory footprint at high orders, needed to store and de-alias the (longitudinal and transverse) perturbation fields \cite{2021MNRAS.501L..71R,Schmidt:2020ovm}. By contrast, \BF\ is UV-complete in the temporal sense -- even after shell-crossing -- and is basically only limited by the number of force computations. For future applications, we envision \BF\ not only to be used as a standalone technique for rapid LSS modelling, but also as a valuable tool in the development of hybrid methods, such as within the framework of effective field theory where \BF\ could replace the commonly used perturbative forward model.

This paper is organised as follows. In section~\ref{sec:nut}, we outline the essence of \BF. Sections~\ref{sec:basics}  and~\ref{sec:main} respectively introduce the basic framework and provide more details on \BF. Experienced readers interested only in the results may skip directly to section~\ref{sec:res}. We conclude in section~\ref{sec:conclusions}.\\[-0.3cm]

\noindent {\it Notation:} We employ comoving Eulerian coordinates $\fett x = \fett r/a$, where $\fett r$ are physical coordinates, and~$a$ is the cosmic scale factor, normalised to unity at the present time. Lagrangian coordinates are dubbed $\fett q$. We use $\nabq$ and $\fett \nabla_{\fett q}^{-2}$ to denote respectively the partial derivative and inverse Laplacian w.r.t.\  $\fett q$, and we define $\nabqinverse := \fett \nabla_{\fett q}^{-2} \nabq$. Partial derivatives w.r.t.\ the spatial component~$q_i$ are abbreviated with $g_{,i} =: \nabla_{q_i} g$. We employ the unnormalised $\Lambda$CDM linear growth function $D(a) \!=\! a \,{}_2F_1\left( 1/3,  1, 11/6; - \Lambda a^3 \right) = a - \!(2\Lambda/11) a^4 \! +\! O(a^7)$ as a temporal variable (represented by the Gauss hypergeometric function), where $\Lambda = \Omega_{\Lambda0}/\Omega_{\rm m0}$ is the ratio of the present-day values of the dark-energy and matter densities. $D$-time derivatives are denoted with $T' =:\pD T$ for any given time-dependent function $T =T(D)$.  Temporal advancements from $D: D_n \to D_{n+1}$ are linked to the (potentially nonuniform) increment $\Delta D := D_{n+1}- D_n >0$ from step~$n$ to~$n+1$, where we omit the index `$n$' on $\Delta D$ to avoid cluttering. Temporal evaluations w.r.t.\ step $n$ are abbreviated with~$T_{n} :=  T|_{D= D_n}$. \\[0.1cm]
{\it Nomenclature:} 
We call an integrator `ZA consistent' or `2LPT consistent', if its particle trajectories and momenta agree respectively with the predictions of first- or second-order Lagrangian perturbation theory after each completed time step, up to higher-order terms, and before shell-crossing. The integrators \BF, \textsc{PowerFrog} and \fastpm\ are frequently abbreviated with BF, PF, and FPM, respectively. Performing $n$ linear-in-$D$ steps with \BF\ is denoted with $n$BF, and similarly for the other $D$-time integrators. For COLA, $n$ linear-in-$a$ steps are dubbed $n$COLA.

\section{\BF\ boiled down: a summary}\label{sec:nut}

\BF\ incorporates analytical insights from LPT up to second order, achieving a significant leap in performance for predicting the LSS compared to traditional methods. For this, no perturbative input is required; instead, the \BF\ integrator computes the gravitational force at each time step by solving the Poisson equation, just like other integrators. The key idea of \BF\ is then to select temporal weights in the kick step such that, as long as LPT remains valid and the force (and hence the $N$-body motion) could be computed perturbatively, the resulting trajectories align with the LPT prediction to second order. In the non-perturbative regime, by contrast, the gravitational force computed with \BF's Poisson solver \textit{automatically} deviates from the LPT prediction. In fact, as we demonstrate in this paper, this departure from 2LPT is crucial for \BF\ to converge toward the correct solution, regardless of shell-crossing.

Let us now outline how the above is technically implemented in \BF. We employ the linear $\Lambda$CDM growth function~$D= a + O(\Lambda a^4)= D_n$ as a time variable, where $n$ is a step~counter linked to the temporal advancement of the system with (potentially non-uniform) increment $\Delta D = D_{n+1} - D_n >0$. Considering a drift-kick-drift (DKD) scheme from step $n \to n+1$ with intermediate updates provided at $D_{n+\nicefrac{1}{2}} = D_n + \Delta D / 2$, \BF\ determines
(dropping $N$-particle indices for simplicity)
\begin{subequations} \label{eqs:DKDintro}
\begin{align}
   \upcurl{\fett x}{n + \nicefrac{1}{2}} &= \upcurl{\fett x}{n} + \frac{\Delta D}{2} \upcurl{\fett v}{n}\,,   
        \qquad  &\circledgreen{D}   \label{eq:firsthalfdriftintro} \\
    \upcurl{\fett v}{ n+1 } &= \alpha \, \upcurl{\fett v}{ n } +  \beta \, \subcurl{D}{n+\nicefrac{1}{2}}^{-1} \fett A (\upcurl{\fett {x}}{ n+\nicefrac{1}{2}}) \,, \qquad & \circledred{K}    \label{eq:vupdateintro} \\
    \upcurl{\fett x}{ n+1 } &= \upcurl{\fett x}{ n+\nicefrac{1}{2} } + \frac{\Delta D}{2} \upcurl{\fett v}{n+1} \,.
       \qquad \qquad  & \circledgreen{D}   \label{eq:secondhalfdriftlaterstepsintro}
\end{align}
\end{subequations}
Here, $\fett x_n$  and $\fett v_n\! =\! \pD \fett x_n$ denote respectively fluid-particle positions and velocity evaluated at step~$n$, while $\fett A$ is the gravitational field, associated to the Poisson equation $\nabx \cdot \fett A = - \delta$, where $\delta = \rho/\bar \rho -1$ is the density contrast. Furthermore, $(\alpha,\beta)(D_n, \Delta D)$ are temporal weights that, for \BF, are fixed by a 2LPT matching condition, a procedure that is briefly outlined below; see eq.\,\eqref{eq:BFcoeff} for the results of $(\alpha, \beta)$.

\begin{figure}
    \centering
    \resizebox{0.99\textwidth}{!}{
    \includegraphics{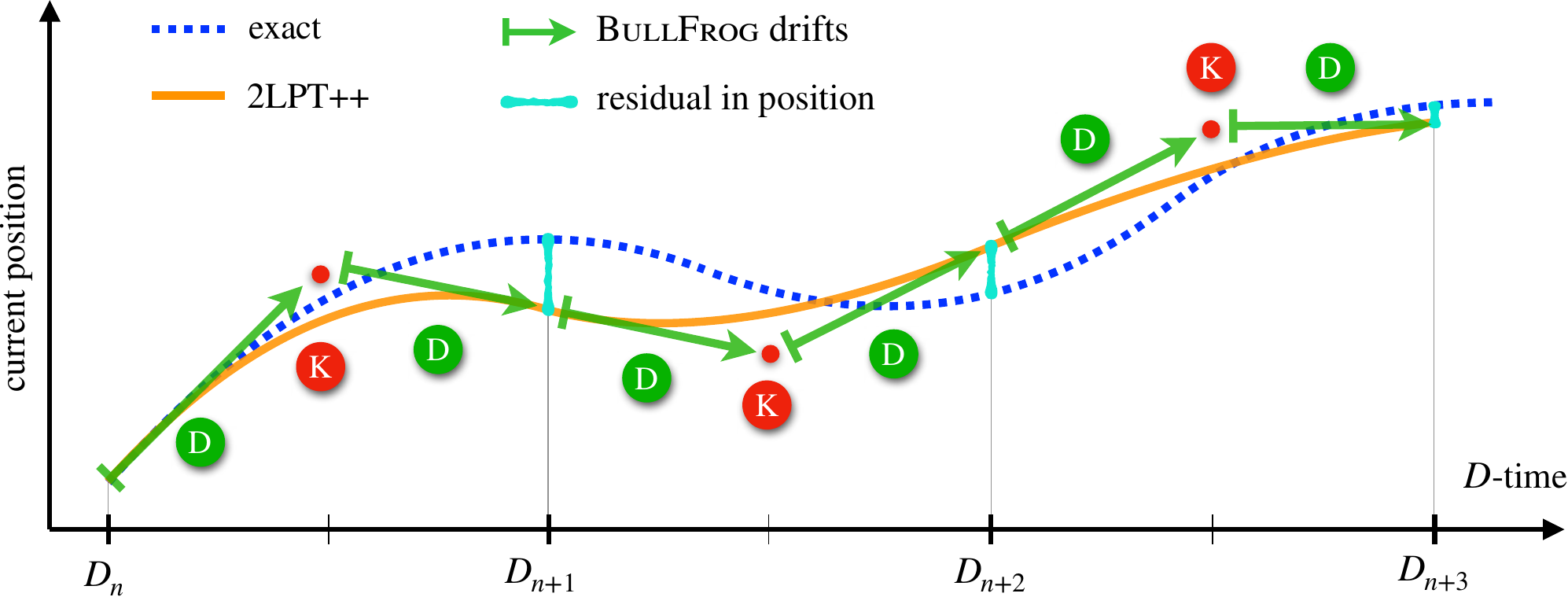}
    }
    \caption{%
    Sketch of temporal integrator operations built into \BF\ (shown as green arrows and red dots, labelled with `D' for drift and `K' for kick), compared against the exact solution which is achieved by $\infty$LPT before shell-crossing (blue dotted curve). Due to temporal discretisation, the $N$-body integrator evaluates the force field at a slightly misplaced position after each first half drift (red dots) relative to the exact position. \BF\ takes the force field from this position as input and adjusts the temporal weights $(\alpha, \beta)$ during the kick operation, such that the truncated 2LPT trajectory is exactly recovered after each completed time step (notably without explicitly computing the spatial 2LPT kernel). On top of that, an infinite tower of higher-order terms is excited during the force calculation; therefore, we call the resulting trajectory 2LPT++ (orange curve). \BF\ receives a refined update on the evolving gravitational field at each step, and its trajectory converges to the exact solution as the step size decreases -- even after shell-crossing. \BF\ can be initialised directly at time zero, i.e., without the use of (external) initial-condition generators.
    }
    \label{fig:DKD-intro}
\end{figure}

Since the leapfrog integrator approximates the continuous particle trajectories in a piecewise linear manner, the positions of the $N$-body particles after the first half drift~\eqref{eq:firsthalfdriftintro} will, in general, not agree with the exact solution (blue dotted curve in figure~\ref{fig:DKD-intro}). As a consequence, for finite time step size, the force field during the kick operation is determined at a misplaced position (red dots in figure~\ref{fig:DKD-intro}), in the following dubbed~$\fett x_{\rm mis}$. To correct for this misplacement, we exploit Einstein's weak equivalence principle, and compute the gravitational field along the particle accelerations of the misplaced trajectories, 
\be
   \( \nabx \cdot \fett A \) |_{\fett x = \fett x_{\rm mis}} = - \delta(\fett x_\mis)  \,. \label{eq:accintro}
\ee
Here, the nonlinear density contrast $\delta(\fett x_\mis)$ is evaluated at the time of the half step. Straightforward calculations (in Lagrangian coordinates) then allow us to determine $\fett A$ by perturbative means. Crucially, the computation of $\fett A$ generates an infinite tower of nonlinear terms, regardless of the level of (perturbative) approximation inherent in the misplaced trajectories. Considering then the {\it leading nonlinear correction} in $\fett  A$ allows us to determine the weights $(\alpha, \beta)$, so that the drift and kick updates are 2LPT consistent after each completed time step at $D_{n+1}$ for any $n\geq 0$, up to higher-order effects and before shell-crossing. In summary, we find for the two weights of the \BF\ integrator the following expressions 
\be
 \boxed{ 
   \alpha = \frac{\subcurl{E'}{n+1} - \subcurl F {n+\nicefrac 1 2}}{ \subcurl{E'}{n} - \subcurl F {n+\nicefrac 1 2}} 
     \,\stackrel{\text{EdS}}{\asymp} \,\frac{4 \mathfrak{n} ( 4 \mathfrak{n} + 1)-5}{4 \mathfrak{n} ( 4 \mathfrak{n} + 7)+7} \,, 
    \qquad\quad \beta =  1 -\alpha
   \,\stackrel{\text{EdS}}{\asymp}\,  \frac{ 24 \mathfrak{n} + 12}{4 \mathfrak{n} (4 \mathfrak{n}+7)+7} 
   }
   \,,  
  \label{eq:BFcoeff}
\ee
where $E_n' = \pD E|_{D=D_n}$, `EdS' refers to the Einstein--de Sitter cosmological model ($\Omega_{\mathrm{m}0} = 1$), and we have defined $\mathfrak{n} := a_n/\Delta a$ (which is equal to $n$ for uniform time stepping). Furthermore, $E$ is the second-order growth function of the LPT displacement in $\Lambda$CDM \cite{1995A&A...296..575B,1995PThPh..94.1151M}, and 
\be
  \subcurl F {n+\nicefrac 1 2} :=  D_{n+ \nicefrac 1 2}^{-1} \left( \subcurl{E}{n} + \subcurl{E'}{n} \tfrac{\Delta D}{2} \right) - D_{n+\nicefrac 1 2} \, .
\ee
Calculating the weights in $\Lambda$CDM involves the numerical integration of ordinary differential equations for $D$ and $E$ (see eqs.\,\ref{eqs:2LPT}); besides that, no further computational overhead is required for \BF\ compared to standard $N$-body integrators. See appendix~\ref{app:KDK-BF} for the KDK version of \BF.

Readers not interested in the basics and technical details can go directly to the results section~\ref{sec:res}.

\section{Basic setup}\label{sec:basics}

Before delving into the details of our approach, let us begin by building up the basic framework. A host of numerical and theoretical approaches for the LSS aim to solve the cosmological Vlasov--Poisson system, which encodes the evolution of the distribution $f(\fett x, \fett p,t)$ of collisionless matter, i.e., 
\be 
  \frac{\dd f}{\dd t} =
  \frac{\partial f}{\partial t} + \frac{\fett p}{a^2} \cdot \nabx f  - \nabx \phi \cdot  \fett {\nabla_p} f = 0 \,, \qquad  \nabx^2 \phi = 4\pi G \bar \rho  a^2 \delta \,.  \label{eq:VP}
\ee
Here, $\fett x = \fett r/a$ denote comoving coordinates composed of physical coordinates $\fett r$  and cosmic scale factor $a$ for a $\Lambda$CDM Universe, $\fett p$ is the momentum variable canonically conjugated to $\fett x$, and the density contrast $\delta = (\rho- \bar \rho)/\bar \rho$ is defined through
\be
   \rho(\fett x,t) = \bar \rho(t) \left[ 1+ \delta(\fett x, t) \right] = \int f(\fett x, \fett p,t) \, \dd^3 p \,,
\ee
where $\bar \rho(t) \propto a^{-3}$ is the background density. In this paper we focus on two distinct basic frameworks to solve eq.\,\eqref{eq:VP}, namely by perturbative means valid before shell-crossing (section~\ref{sec:LPT}), and $N$-body methods based on $D$-time integrators (section~\ref{sec:nbody}), both of which we briefly describe in the following.

\subsection{Perturbative treatment in Lagrangian coordinates} \label{sec:LPT}

\paragraph{Standard Lagrangian perturbation theory.}
For cold initial conditions, it is convenient to formulate eqs.\,\eqref{eq:VP} in Lagrangian space, where the initial/Lagrangian coordinates $\fett q$ serve as fluid labels for the matter elements that evolve to their current positions $\fett x(\fett q,t)$. Denoting with an overdot the total time derivative w.r.t.\ time $t$, the Lagrangian evolution equations in $\Lambda$CDM are
\be \label{eqs:EOM}
   \ddot {\fett x} +  2 H \dot {\fett x}  = - a^{-2}\nabx \phi \,, \qquad 
  \nabx^2 \phi = 4\pi G \bar \rho a^2 \delta \,,
\ee
where $(H/H_0)^2= \Omega_{\rm m0}a^{-3} + \Omega_{\Lambda 0}$ and, in Lagrangian coordinates and until shell-crossing, the density contrast is expressed by the Jacobian determinant $J(\fett x (\fett q,t)) = \det [\nabq \fett x]$,
\be
  \delta(\fett x(\fett q,t)) + 1 = J^{-1}(\fett x (\fett q,t)) \,.
\ee
In LPT, these equations are solved by a perturbative expansion of the displacement field $\fett \psi = \fett x- \fett q$; here we just state the well-known results up to second order and refer the reader to e.g.\ refs.\,\cite{1989A&A...223....9B,1991ApJ...382..377M,1997GReGr..29..733E,Zheligovsky:2013eca,Rampf:2021rqu,2022LRCA....8....1A, Rampf:2012xa} for details and reviews. Expanding the displacement up to first order in a small parameter $\epsilon>0$ and keeping only linear terms in $\epsilon$ in the equations of motion, one arrives at the well-known Zel'dovich approximation (ZA)
\begin{subequations} \label{eqs:2LPT}
\be
   \fett x^\ZA := \fett q + \fett \psi^\ZA\,, \qquad
   \fett \psi^\ZA := \fett \psi^{(1)} =  - D \nabq \varphi^\ini  \,, \qquad \text{with~~} \ddot D + 2H \dot D = 4\pi G \bar \rho D  \,. \label{eq:ZA}
\ee
Here, $D = a  - (2\Lambda/11)a^4 + O(\Lambda^2 a^7)$ is the $\Lambda$CDM growing-mode solution and $\Lambda =\Omega_{\Lambda0}/\Omega_{\rm m0}$, while 
\be
  \varphi^\ini(\fett q) := D^{-1}(a_{\rm today}) \, \nabq^{-2} \delta_{\rm lin}(\fett q, a_{\rm today}) 
\ee
is the gravitational potential backscaled to `initial time' $D=0$, expressed by today's linear matter density contrast,
retrieved from standard Einstein--Boltzmann solvers.%
\footnote{\label{foot:norm}%
    We normalise today's scale factor to unity, but work, unless stated otherwise, with the `unnormalised' growth $D=a+ O(\Lambda a^4)$, which is in particular relevant when implementing \BF. In this context, while the {\it normalised} growth function $D / D(a_{\mathrm{today}})$ could also be used as a time variable, normalising the second-order growth function $E$ involves $D(a_{\mathrm{today}})$ quadratically, requiring manual reversion of this normalisation when computing the kick coefficients.
}
Here and in the following, we initialise the fluid at time $D=0$ in the growing mode, which is well defined provided one imposes the initial conditions $\delta(D=0)=0$ and $(\pD \fett x) |_{D = 0} = -\nabq \varphi^\ini$; see e.g.\ ref.\,\cite{2003MNRAS.346..501B,Michaux:2020yis} for details.

Likewise, expanding the displacement to second order, one finds the well-known 2LPT solution 
\be 
   \fett x^\twoLPT := \fett q + \fett \psi^\twoLPT\,, \qquad \,\,
    \fett \psi^\twoLPT := \fett \psi^{(1)} + \fett \psi^{(2)}\,,
\ee
 where
\be 
  \fett \psi^{(2)} =  E\, \nabqinverse \mu_2 \,,  \quad\qquad\quad~ \text{with~~}  \ddot E + 2H \dot E =  4\pi G \bar \rho \big( E - D^2 \big) \,.
\label{eq:ODE_for_E}
\ee
\end{subequations}
Here, $\nabqinverse := \fett \nabla_{\fett q}^{-2} \nabq$, and
$\mu_2 = (1/2) [\varphi_{,ll}^\ini \varphi_{,mm}^\ini - \varphi_{,lm}^\ini \varphi_{,lm}^\ini]$ is a purely spatial kernel (summation over repeated indices is assumed), while 
$E= - (3/7) D^2 - (3\Lambda/1001) D^5 + O(\Lambda^2D^8)$
is the fastest growing-mode solution at second order in $\Lambda$CDM (which can also be expressed in terms of hypergeometric functions, see ref.\,\cite[eq.\,22]{1995PThPh..94.1151M}). Note that for \BF\ we do {\it not} employ the `$D^n$ approximation' $E \simeq - (3/7) D^2$ which would ignore explicit $O(\Lambda)$ corrections, nor do we employ the refined approximation of ref.\,\cite{1995A&A...296..575B}, as both would spoil the convergence of the proposed integrator; see section~\ref{sec:loss-conv} for details.

Since the ZA solution~\eqref{eq:ZA} is proportional to $D$, it turns out to be convenient -- in LPT but also for the proposed integrator -- to define a suitably rescaled velocity
\be
  \fett v = \pD \fett x  \label{eq:Lagvel}
\ee
such that, to the leading order, the rescaled velocity is constant in the $D$-time before shell-crossing. Here, $\pD$ is the Lagrangian $D$-time derivative, and the ZA and 2LPT velocities read respectively
\be
  \fett v^{\rm ZA} =  -\nabq \varphi^\ini \,, \qquad  \fett v^{\rm 2LPT} = \pD \fett \psi^{\rm 2LPT} = -\nabq \varphi^\ini + \pD E \nabqinverse \mu_2\,. \label{eq:vZAand2LPT}
\ee

\paragraph{Particle acceleration along approximate trajectories.} As a somewhat non-standard exercise, we will need an expression for the gravitational field $\fett A$, self-induced by the advected particles,
\be
  \boxed{\fett A := - \nabx \varphi ,  \qquad \nabx^2 \varphi := \delta }\,,
\ee
where we have rescaled the cosmological potential according to
\be
    \varphi = \frac{\phi}{4 \pi G \bar{\rho} a^2} \,.
\ee
The quantity $\fett A$ and its perturbative treatment play a central role in this paper. In particular, we need to determine~$\fett A$ along approximate trajectories, which could be based on ZA or 2LPT motion (or anything else). Crucially and distinct to standard LPT, however, the approximate trajectories are taken as input and are not perturbatively refined, while the resulting gravitational acceleration is truncated to arbitrary order (here: to second order). To demonstrate how~$\fett A$ can be solved along a given (approximate) trajectory, let us determine it for the case when~$\fett x =\fett x^{\rm ZA}$, for which we consider
\be
   \( \nabx \cdot \fett A \) |_{\fett x = \fett x^{\rm ZA}} = - \delta (\fett x^{\rm ZA}) \,.
\ee
To proceed, we convert the Eulerian derivative into Lagrangian ones, using
$\nabla_{x_i} = (\nabla_{x_i} q_j) \nabla_{q_j}$, where $\nabla_{x_i} q_j = \varepsilon_{ikl} \varepsilon_{jmn} x_{k,m} x_{l,n}/(2J)$, and $\varepsilon_{ikl}$ is the Levi-Civita symbol. Straightforward derivations then lead to 
\be
 \left[ 
     (1 - D \varphi_{,ll}^\ini) \delta_{ij} + D \varphi_{,ij}^\ini + h.o.t.
   \right] A_{i,j}^{\rm ZA}
   = 
   J_{\rm ZA} -1  \,,
\ee 
where $J_{\rm ZA} = 1 - D \varphi_{,ll}^\ini  + D^2 \mu_2 + h.o.t.$ 
Here and in the following, `$h.o.t.$' stands for higher-order terms. We solve this equation by a standard perturbative Ansatz $\fett A^{\rm ZA} := \fett A^{(1)} \epsilon + \fett A^{(2)}\epsilon^2 + \ldots$, yielding
\begin{align} \label{eq:accZA}
  \fett A^{\rm ZA} &=  - D \nabq \varphi^\ini 
    - D^2 \nabqinverse \mu_2 + h.o.t.\ 
     \stackrel{\text{EdS}}{\asymp} - a \nabq \varphi^\ini -  a^2 \nabqinverse \mu_2 + h.o.t.
\intertext{Likewise, we can determine the acceleration along 2LPT trajectories, for which we need $J_{\rm 2LPT} = 1 - D \varphi_{,ll}^\ini  + (E+D^2) \mu_2 + h.o.t.$ as updated input,  leading to} 
  \fett A^{\rm 2LPT} &= - D \nabq \varphi^\ini 
   + (E - D^2) \nabqinverse \mu_2 + h.o.t.\
   \stackrel{\text{EdS}}{\asymp} - a \nabq \varphi^\ini - \frac{10}{7} a^2 \nabqinverse \mu_2 + h.o.t., \label{eq:A2LPT}
\end{align}
where, notably, the second-order term has an altered temporal coefficient in comparison to the ZA acceleration~\eqref{eq:accZA}, due to the 2LPT contribution in the Jacobian. Here it is important to observe that, no matter whether the trajectories are approximated with ZA or with higher-order schemes, the associated gravitational acceleration always contains an infinite number of perturbative terms (except in the one-dimensional case). This point is in particular crucial for $N$-body integrators where a fairly similar hierarchy of nonlinear terms is generated when determining the $N$-body forces (see section~\ref{sec:main}).

Finally, for future reference, we remark that beyond second order, there will be transverse contributions to $\fett A$, virtually for the same reason as observed in standard LPT \cite{1993MNRAS.264..375B,1994MNRAS.267..811B}. To get the governing equation, we multiply the definition $A_l = - \nabla_{x_l} \varphi$ by the matrix with elements $\partial x_l/\partial q_j$, leading firstly to $x_{l,j} A_l = - \nabla_{q_j} \varphi$. Taking from this the Lagrangian curl $\varepsilon_{ijk} \partial_{q_k}$, we arrive at
\be 
  \varepsilon_{ijk} x_{l,j} A_{l,k}=0 \,, \label{eqs:Cauchy}
\ee
which is a vector equation that is related to the so-called Cauchy invariants (see e.g.\ \cite{Zheligovsky:2013eca} for details and historical context). Of course, eqs.\,\eqref{eqs:Cauchy} can also be evaluated along approximate trajectories.

\subsection[\texorpdfstring{$D$}{D}-time integrators for \texorpdfstring{$N$}{N}-body simulations]{\texorpdfstring{$\fett D$}{D}-time integrators for \texorpdfstring{$\fett N$}{N}-body simulations}\label{sec:nbody}

Standard $N$-body methods employ tracer particles that follow the evolution of the continuous phase-space distribution $f(\fett x, \fett p,t)$; see, e.g., refs.\,\cite{2008SSRv..134..229D,2020NatRP...2...42V,2022LRCA....8....1A} for reviews. This is effectively implemented by replacing the functional dependencies of the distribution function from $(\fett x, \fett p)$ to $(\fett x_i, \fett p_i)$, where $i=1,\ldots, N$ denote the $N$ particle labels; but note that, here and in the following, we will omit the particle labels to avoid unnecessary cluttering. The `discretised' formulation of Vlasov--Poisson relies on the associated Hamiltonian equations of motion $\partial_t \fett x = \fett p/a^2$ and $\partial_t \fett p = - \nabx \phi$ with $\nabx^2 \phi = 4\pi G \bar \rho  a^2 \delta$, where all expressions -- including the density -- are evaluated along $N$ fluid characteristics. Importantly, the continuous potential $\phi$ in the Poisson equation is approximated by the discrete potential sourced by the tracer particles. In this work, we will assume that the spatial resolution and the accuracy of the force computation are sufficient to adequately represent the underlying continuous dynamics, and focus on the \textit{temporal} discretisation of Vlasov--Poisson. In practice, this necessitates the use of `discreteness-reduction techniques' at early times, see e.g.\ refs.~\cite{2006PhRvD..73j3507M,2016MNRAS.461.4125G,Michaux:2020yis,List:2023kbb} for related avenues.

\setlength{\emergencystretch}{5pt}

Typically, the Hamiltonian equations are time integrated using symplectic leapfrog schemes, as for example executed with consecutive kick-drift-kick operations, implemented in widely employed codes such as \textsc{Gadget}, \textsc{Pkdgrav-3}, \textsc{Ramses}, and \textsc{Swift} \cite{2021MNRAS.506.2871S,2017ComAC...4....2P,2002A&A...385..337T,2024MNRAS.tmp..925S}. Symplectic leapfrog schemes employ canonically conjugate variables in their drift and kick updates, providing certain stability properties for energy-preserving systems. On cosmological scales, however, energy is continuously extracted from the Vlasov--Poisson system due to the presence of the Hubble drag term in comoving coordinates. As a result, the benefit of enforcing symplecticity on cosmological scales is questionable, although it is relevant for simulating virialised systems that have decoupled from the Hubble flow.

Here we follow a different strategy and work with `$D$-time integrators' (called $\mPi$ integrators in refs.~\cite{List:2023jxz,List:2023kbb}), which are closely related to the basic setup of LPT. Indeed, with $D$-time integrators, one works with the phase-space variables $(\fett x, \fett v)$ where, importantly, $\fett v := \pD \fett x$, which is {\it not} the canonical conjugate to $\fett x$. Employing  $D$-time integrators, a generic drift-kick-drift advancement from $D: \subcurl D n \to \subcurl D {n+1}$ is performed according to~\cite{List:2023jxz} 
\begin{subequations} \label{eq:pi_integrator}
\begin{align}
    \upcurl{\fett x}{n + \nicefrac{1}{2}} &= \upcurl{\fett x}{n} + \frac{\Delta D}{2} \upcurl{\fett v}{n}\,,  \label{eq:firsthalfdrift} \\
    \upcurl{\fett v}{ n+1 } &= \alpha \, \upcurl{\fett v}{ n } +  \beta \, \subcurl{D}{n +\nicefrac{1}{2}}^{-1}\,  \fett A (\upcurl{\fett x}{ n+\nicefrac{1}{2}}) \,,  \label{eq:vupdate} \\
    \upcurl{\fett x}{ n+1 } &= \upcurl{\fett x}{ n+\nicefrac{1}{2} } + \frac{\Delta D}{2} \upcurl{\fett v}{n+1}\,, \label{eq:secondhalfdriftlatersteps}
\end{align}
\end{subequations}
where $\alpha$ and $\beta$ are temporal weights that can be chosen such that the integrator has certain properties (most importantly convergence towards the correct solution as $\Delta D \to 0$). 

An important example of a $D$-time integrator is \fastpm~\cite{2016MNRAS.463.2273F,2021JCAP...01..016B,List:2023jxz}, which is ZA consistent after each completed time step (achieved by requiring $\alpha+\beta=1$), is symplectic and second-order accurate.
The weights for the DKD~version of \fastpm~(FPM) are \cite{List:2023jxz}
\begin{align}
  \alpha^\FastPMsmall =  \frac{\subcurl{\cal F}{n}}{\subcurl{\cal F}{n+1}} \,,
    \qquad \qquad 
  \beta^\FastPMsmall = 1- \alpha^\FastPMsmall  \,,  \label{eq:FastPMcoeff}
\end{align}
with
\be
    {\cal F}(a) :=  a^3 H \partial_a D \,, 
   \qquad \qquad {\cal F}_n := {\cal F}(a(D))|_{D=D_n} \,, \label{eq:calF}
\ee    
where $a(D)$ is the inverse of $D(a)$; in passing, we note that ${\cal F}$ is defined by $\fett p= a^2 \partial_t \fett x = {\cal F} \pD \fett x$. Differentiable implementations of \fastpm, such as \textsc{FlowPM}~\cite{2021A&C....3700505M} and \textsc{pmwd} \cite{2022arXiv221109815L}, also exist and can be conveniently interfaced with inference and sampling methods that rely on gradient information, such as Hamiltonian Monte Carlo. We remark that in this paper we only consider the DKD version of \fastpm, and for consistency we perform linear-in-$D$ steps also with \fastpm~(whereas the original implementation uses linear-in-$a$ steps).

In addition to \fastpm, we will also compare our predictions against \textsc{PowerFrog} (PF) recently introduced in ref.~\cite{List:2023jxz}, which is also a $D$-time integrator and comes with the weights
\begin{align}
  \alpha^\PF = 
     \begin{cases}
        - 5 / 7  \,;   \qquad \hspace{3.76cm} n = 0 \,, \\
        \(c_1 D_{n}^{c_2} + c_3 D_{n+1}^{c_2}\) D_{n+ \nicefrac{1}{2} }^{-c_2} + c_4 \,; \qquad n>0 \,,  
     \end{cases}      \,,
    \qquad \beta^\PF = 1- \alpha^\PF \,.  \label{eq:PFcoeff}
\end{align}
Here, the numerical coefficients $c_1 - c_4$ result from solving a transcendental system of equations (see ref.~\cite{List:2023jxz} for details); once the weights are fixed, the \textsc{PowerFrog} integrator is second-order accurate and in general ZA consistent and additionally 2LPT consistent for the first step at $D=0$. We remark that \textsc{PowerFrog} approximates the second-order growth effectively with $E \simeq -(3/7) D^2$ during the first time step, which however is only accurate at sufficiently early times when the impact of the cosmological constant on the matter dynamics is still negligible ($z \gtrsim 10$). We will shortly see that \BF\ rectifies these shortcomings.

All $D$-time integrators have in common that the corresponding simulations can be directly initialised at $D= D_0 = 0$ (the Big Bang), specifically by employing the initial conditions 
\be
  \subcurl{D}{0} \,: \quad\qquad \upcurl{\fett x}{0} = \fett q \,, \qquad \upcurl{\fett v}{0} = -   \nabq \varphi^\ini  \,,  \label{eqs:ICs}
\ee
which correspond to growing-mode initial conditions as employed in LPT (see section~\ref{sec:LPT}). Indeed, the first statement in eqs.\,\eqref{eqs:ICs} implies an initially vanishing density contrast, while the second condition implies ZA motion in the continuous limit. In the literature, these so-called `slaved' initial conditions play a central role in generating initial displacements in standard $N$-body simulations \cite{1983MNRAS.204..891K,1985ApJS...57..241E,2006MNRAS.373..369C,Michaux:2020yis}, and comprise the foundation of demonstrations of LPT convergence \cite{Rampf:2017jan,Saga:2018nud,2021MNRAS.501L..71R}.

\begin{figure}
    \centering
    \resizebox{0.8\textwidth}{!}{    \includegraphics{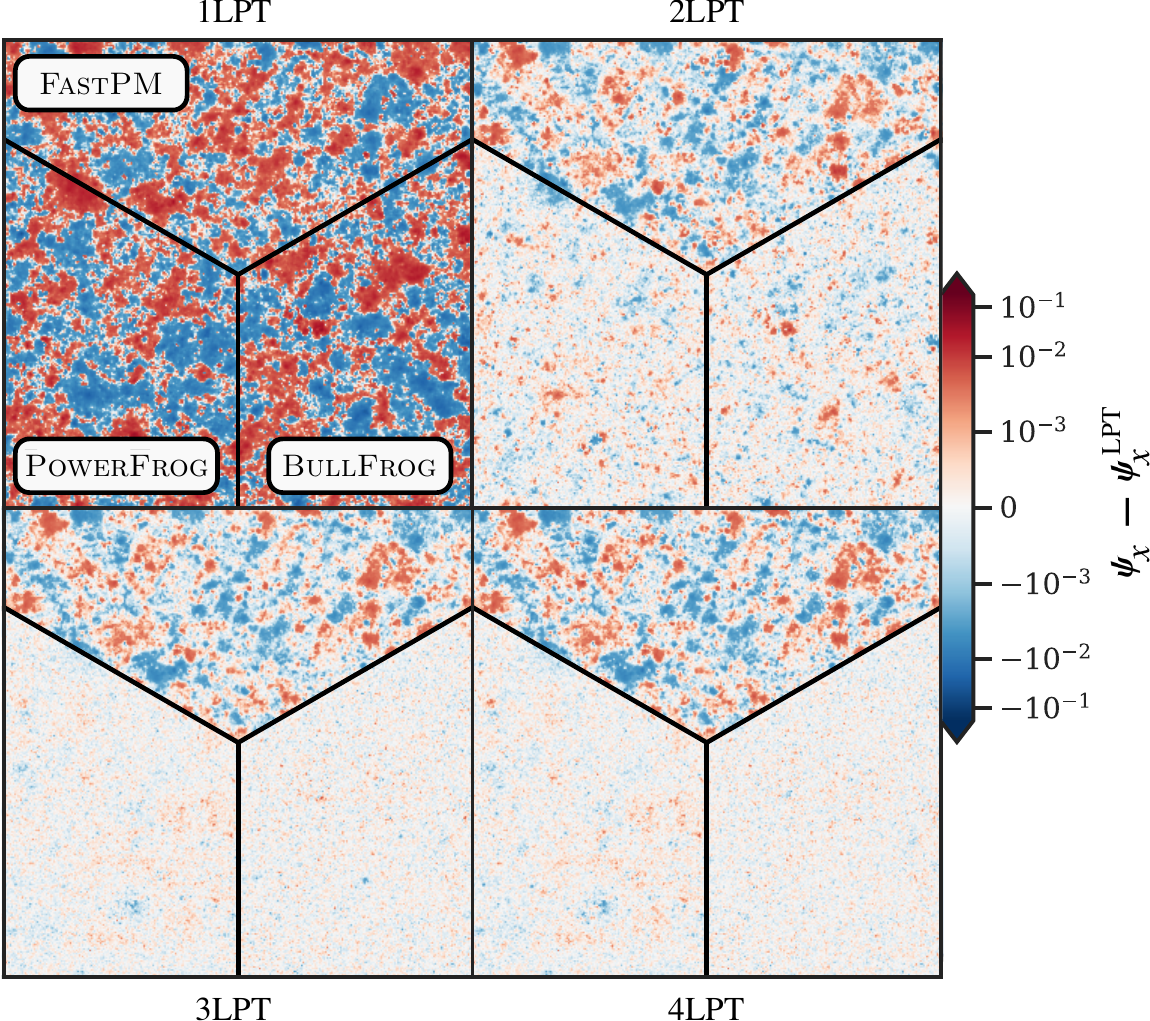}
    }
    \caption{Residuals of the first displacement component $\fett \psi_x$ between the $D$-time integrators for the threefold step $z : \infty \to 50 \to 24 \to 16$ and $n$LPT in a single time step for the orders $n=1-4$. Shown is a razor-thin slice in the Lagrangian $y$-$z$ plane with a side length of $300$\,Mpc$/h$, using $N = 512^3$ particles. Here, all particles are still in the single-stream regime, for which LPT converges towards the exact solution.
    }  \label{fig:field_level}
\end{figure}

Figure~\ref{fig:field_level} shows a performance test of the three $D$-time integrators, conducted at the field level and evaluated at sufficiently high redshifts, where LPT remains very accurate (see section~\ref{sec:res} for details on the numerical implementation and for late-time results). Specifically, we show the difference in the displacement component $\psi_x$ between the $D$-time integrators for the three-step sequence $z : \infty \to 50 \to 24 \to 16$ and $n$LPT in a single time step for the orders $n=1-4$. It is seen that the residuals for both \BF\ and \textsc{PowerFrog} with respect to $3 - 4$LPT are very small, already with as few as three time steps. As expected, the performance is slightly better for \BF\ than for \textsc{PowerFrog} due to the built-in 2LPT matching conditions of the former at each time step, which the latter achieves only for the first time step. Thus, with \BF, only $n$LPT residuals with $n \geq 3$ remain -- which become increasingly smaller as the step size is reduced. By contrast, the residuals of \fastpm~against high-order LPT are clearly populated by a 2LPT contribution, since \fastpm~only incorporates ZA motion for each time step. In summary, figure~\ref{fig:field_level} demonstrates that \BF\ achieves the best agreement with the high-order LPT solution for a given number of time steps in comparison with the other $D$-time integrators.

\section{Details of \BF}\label{sec:main}

Here, we provide details about the new integrator \BF, which comes with the unique choice of weights $(\alpha, \beta)$ within the class of $D$-time integrators that is 2LPT consistent after each completed time step, up to higher-order terms and before shell-crossing. For pedagogical reasons, we first discuss the initial step in section~\ref{sec:BFini}, while we provide derivations valid for any time step in section~\ref{sec:BFlater}. In section~\ref{sec:convergence}, we show that this 2LPT-consistent choice of the weights defines a second-order accurate integrator, implying that \BF\ trajectories converge towards the correct solution at any moment in time, despite the fact that 2LPT becomes invalid after shell-crossing. Then, in section~\ref{sec:loss-conv} we address 
the loss of convergence if approximate displacement growth is used. In the present section we only consider the DKD form; see appendix~\ref{app:KDK-BF} for the KDK version of \BF.

The general strategy is as follows (see also figure \ref{fig:DKD-intro}). Due to temporal discretisation, the gravitational force during the kick step is in general not identical to that evaluated on the 2LPT trajectory (cf.\ eq.\,\ref{eq:A2LPT}), as the particles move with constant $D$-time velocity $\vecb{v}_n$ from $D_n$ to $D_{n+\nicefrac{1}{2}}$ (see eq.~\ref{eq:firsthalfdrift}) -- even if $\vecb{v}_n$ agrees to second order with 2LPT -- because the change of the 2LPT velocity during this time interval is not taken into account in a DKD scheme. An auto-correction for this misplacement can however be applied on the fly, essentially by adjusting the weight functions $(\alpha,\beta)$ appearing in the kick operation (eq.\,\ref{eq:vupdate}), such that the velocity after the kick agrees again with the velocity truncated at order 2LPT, up to higher-order terms.

We will see that this adjustment also implies that the $N$-body particle positions conform to the 2LPT positions at that truncation order, and after each completed time step.

\subsection{Initial time step}\label{sec:BFini}

\noindent \BF\ can be directly initialised at $D=0$ for step $n=0$, which we discuss next. Note that the initialisation step is actually contained as a special case for the calculation for arbitrary $n$ (section~\ref{sec:BFlater}), and thus, nothing `special' needs to be done for the initial step. From the perspective of a perturbative analysis, splitting up the initialisation step from the successive steps is however pedagogically constructive, and this is why we discuss it separately. Nonetheless, we refer fast readers not interested in this subtle difference to the next section, and in particular to eqs.\,\eqref{eqs:weightsn>0} for the results.

The initial step $n: 0 \to 1$ for $D$-time integrators assumes initial conditions of the type~\eqref{eqs:ICs}, which implies that particles are displaced according to the ZA during the first half drift. For the initial set of DKD operations, we have firstly (eqs.\,\ref{eq:pi_integrator} with $n=0$)
\begin{subequations}
\begin{align}
 \upcurl{\fett x}{\nicefrac{1}{2}} &= \fett q + \frac{\Delta D} 2 \upcurl{\fett v}{0} \,, \\
    \upcurl{\fett v}{1} &= \alpha^\BFsmall \,\upcurl{\fett v}{0} + \beta^\BFsmall [\Delta D/2]^{-1} \fett A(\upcurl{\fett x}{\nicefrac{1}{2}}) \,,     \label{eq:kickini}  \\
   \upcurl{\fett x}{1} &= \upcurl{\fett x}{\nicefrac{1}{2}} + \frac{\Delta D}{2} \upcurl{\fett v}{1} \,,   \label{eq:driftsecond}
\end{align}
\end{subequations}
where $\fett x_0 = \fett q$, and  $\upcurl{\fett v}{0} =-   \nabq \varphi^\ini$ at $D_0=0$ (cf.\ eq.\,\ref{eqs:ICs}), while $(\alpha^\BFsmall,\beta^\BFsmall)$ are the weights that we will derive in this section for the initial step. For this, we first need to calculate the gravitational acceleration $\fett A(\upcurl{\fett x}{\nicefrac{1}{2}})$ along the approximate particle trajectories, for which we consider
\be
  \( \nabx \cdot \fett A \)|_{\fett x = \upcurl{\fett x}{\nicefrac 1 2}} = - \delta( \upcurl{\fett x}{\nicefrac{1}{2}} ) \,.
\ee
Actually, since $\upcurl{\fett x}{1/2}$ conforms to the ZA trajectory due to the choice of initial conditions, the given task is completely analogous to the one that we have already performed in section~\ref{sec:LPT}. Specifically, setting $D= \Delta D/2$ in~\eqref{eq:accZA} and evaluating the expression along the discrete particle tracers, we have
\be
 \fett A(\upcurl{\fett x}{\nicefrac{1}{2}}) =  - \frac{\Delta D}{2} \nabq \varphi^\ini 
    - \(\frac{\Delta D}{2} \)^2  \nabqinverse \mu_2  + h.o.t.\,, \label{eq:accZArep}
\ee
where, again, we omit the particle labels for simplicity. Using this result as the input on the r.h.s.\ of the kick update~\eqref{eq:kickini}, as well as equating that expression with the truncated 2LPT velocity~\eqref{eq:vZAand2LPT} {\it evaluated at the completed time step} $D_1 =  \Delta D$, we obtain the following constraint equation which lies at the heart of \BF,
\be \label{eq:matching}
  \fett  v^\twoLPT|_{D= \Delta D}  = \left. \left[ - \nabq \varphi^\ini + \pD E \nabqinverse \mu_2    \right] \right|_{D= \Delta D} \stackrel !=
   - \alpha^\BFsmall \,\nabq \varphi^\ini - \beta^\BFsmall  \left[   \nabq \varphi^\ini + \tfrac{\Delta D} 2 \nabqinverse \mu_2  \right] \,,
\ee
all of which is to be evaluated at the Lagrangian positions of the $N$ tracer particles. Matching the coefficients of the $\nabq \varphi^\ini$ terms leads to $\alpha^\BFsmall+\beta^\BFsmall=1$ implying the aforementioned ZA consistency; executing a similar matching for the 2LPT term then leads to
\begin{subequations} \label{eqs:coeffBF}
\begin{align}
  \alpha^\BFsmall &= 1 + 2\frac{\subcurl{E'}{1}}{\Delta D}  
   = -\frac{5}{7}  -\frac{30 \Delta D^3 \Lambda }{1001} -\frac{15360 \Delta D^6 \Lambda ^2}{3556553}  
   + O(\Delta D^9 \Lambda^3)
   \,,  \\
  \beta^\BFsmall &= 1- \alpha^\BFsmall
   = \frac{12}{7} +\frac{30 \Delta D^3 \Lambda }{1001} +\frac{15360 \Delta D^6 \Lambda ^2}{3556553} + O(\Delta D^9 \Lambda^3)  \,,
\end{align}
\end{subequations}
where $\subcurl{E'}{1} =  [\pD E(D)]|_{D= \Delta D}$. It is intriguing to compare these coefficients and their derivation with those of \fastpm~and \textsc{PowerFrog}. All three integrators have in common that they are all ZA consistent, which is achieved when demanding $\alpha+\beta =1$; this is one of the two conditions required to fix the various $D$-time integrators. However, the three integrators are distinct from each other, as different conditions are imposed to fix the remaining condition. Specifically, for \textsc{PowerFrog}, its coefficients read $\alpha^\PF = -5/7$ and $\beta^\PF = 12/7$ for the initialisation step (cf.\ eq.\,\ref{eq:PFcoeff}), which are obtained by approximating $E\simeq (-3/7)D^2$ in~\eqref{eq:matching}. By contrast, \fastpm~was originally formulated in terms of canonical variables \cite[Section~2.5]{2016MNRAS.463.2273F}, and their coefficient $\alpha$ is kept the same as for the standard leapfrog integrator \cite{1997astro.ph.10043Q} (thus yielding symplecticity), while $\beta$ is adjusted to ensure ZA consistency. In our notation, this leads to $\alpha^\FastPMsmall =  \subcurl{\cal F}{n}/\subcurl{\cal F}{n+1}$ and $\beta^\FastPMsmall = 1 -\alpha^\FastPMsmall$, and one would therefore have $\alpha^\FastPMsmall = 0$ and $\beta^\FastPMsmall = 1$ for the initial step starting from $D_0 = 0$. Actually, as shown in ref.~\cite{List:2023jxz}, \fastpm~is the only $D$-time integrator that is ZA consistent as well as symplectic, while \BF\ enforces 2LPT consistency instead; we will see later that \BF\ generally comes with a better performance than \fastpm~for predicting the LSS.

Finally, we need to verify that \BF\ conforms to the 2LPT trajectories after the completed first time step. For this we plug the ($\alpha^\BFsmall, \beta^\BFsmall$)-coefficients and the acceleration~\eqref{eq:accZArep} into the kick~\eqref{eq:kickini}; using that expression in the second half-drift~\eqref{eq:driftsecond} then leads to
\begin{align}
  \upcurl{\fett x}{1} &\stackrel{\phantom{\text{EdS}}}{=} 
        \fett q - \Delta D  \nabq \varphi^\ini 
        +  \frac{\Delta D}{2}   \subcurl{E'}{1}  \nabqinverse \mu_2 + h.o.t.  \label{eq:traj-after-first-step}  \\ 
        &\stackrel{\text{EdS}}{\asymp} \fett q - \Delta a \nabq \varphi^\ini 
         - \frac{3}{7} \Delta a^2  \nabqinverse \mu_2  + h.o.t.\  \nonumber 
\end{align}
By noting that $E_1 =  (\Delta D / 2) E_1' + O(\Delta D^5)$, it becomes evident that this result agrees to second order with the 2LPT trajectory $\fett x^\twoLPT = \fett q + \fett \psi^\twoLPT$ (eqs.\,\ref{eqs:2LPT}) evaluated at $D=\Delta D$.
\hfill $\blacksquare$

\subsection{Arbitrary time steps}\label{sec:BFlater}

Next we consider all consecutive DKD steps $n \to n+1$ for any $n \geq 0$ (as noted above, the case $n=0$ is actually also included). Assuming that the positions and velocities of the previous completed step~$n$ are 2LPT consistent, we will derive $(\alpha^\BFsmall, \beta^\BFsmall)$ in such a way that the same still holds after the $(n+1)$th step (provided that we are still in the pre-shell-crossing regime).

We begin by considering the first half-drift~\eqref{eq:firsthalfdrift}, which for \BF\ can be written as
\begin{align}
 \upcurl{\fett x}{n+\nicefrac{1}{2}} = \upcurl{\fett x}{n} + \tfrac{\Delta D}{2} \upcurl{\fett v}{n}  &\stackrel{\phantom{\text{EdS}}}{=} 
     \fett q - \left( D_n+ \tfrac{\Delta D}{2} \right) \nabq \varphi^\ini 
     + \left[ \subcurl{E}{n} + \subcurl{E'}{n} \tfrac{\Delta D}{2} \right]  \nabqinverse \mu_2 +h.o.t.\  \label{eq:nexthalfdrift} \\
      &\stackrel{\text{EdS}}{\asymp} 
      \fett q - \left( a_n+ \tfrac{\Delta a}{2} \right)  \nabq \varphi^\ini  - \frac 3 7 a_n (a_n+ \Delta a)  \nabqinverse \mu_2  +h.o.t.\,,
\end{align}      
where, as a reminder, $\subcurl{E}{n} = E(D)|_{D=D_n}$ and $\subcurl{E'}{n} = (\pD E)|_{D=D_n}$, and the second-order growth $E$ is defined in eq.\,\eqref{eqs:2LPT}; furthermore, we note that, for the limiting case of uniform time stepping with $\Delta D = const.$, we have simply $D_n= n\Delta D$ -- here and below. It can be easily seen that the trajectory~\eqref{eq:nexthalfdrift} does not agree with the corresponding trajectory in 2LPT~\eqref{eqs:2LPT} -- not even in the EdS case -- basically because the particles move on straight lines with velocity $\vecb{v}_n$ between time $D_n$ and $D_{n+\nicefrac{1}{2}}$, and the $\upcurl{\fett v}{n}$ term in eq.\,\eqref{eq:nexthalfdrift} is therefore multiplied with just $\Delta D/2$.

The above implies that we need to determine the gravitational acceleration at slightly misplaced positions, for which we consider
\be
    \( \nabx \cdot \fett A  \)|_{\fett x = \upcurl{\fett x}{n+\nicefrac 1 2}} = - \delta(\upcurl{\fett x}{n+\nicefrac{1}{2}})  \,.
\ee
This equation can be solved with analogous perturbative techniques as discussed before, leading to
\begin{align}
  \upcurl{\fett A}{n+\nicefrac{1}{2}} &\stackrel{\phantom{\text{EdS}}}{=}  - \left( D_n+ \tfrac{\Delta D}{2} \right)  \nabq \varphi^\ini
   + \left[ \subcurl{E}{n} +  \subcurl{E'}{n} \tfrac{\Delta D}{2} - \left(  D_n + \tfrac{\Delta D}{2}\right)^2 \right]  \nabqinverse \mu_2 + h.o.t.   \\
  &\stackrel{\text{EdS}}{\asymp}  - \left( a_n+\tfrac{\Delta a}{2} \right)  \nabq \varphi^\ini  - \left[ \tfrac{10}{7}  a_n(a_n+\Delta a) + \tfrac {\Delta a^2} 4 \right] \nabqinverse \mu_2  + h.o.t.
\end{align}
As expected, this result disagrees slightly with the 2LPT acceleration; cf.\ eq.\,\eqref{eq:A2LPT} evaluated at $D=  D_n+ \Delta D/2$. This misplaced result needs to be taken into account in order to autocorrect the particle trajectory of the numerical integrator.

To do so, we plug the above results in the kick update~\eqref{eq:vupdate} and require that the kick coincide with the 2LPT velocity at the completed time $D_{n+1} = D_n+\Delta D$ -- this is the matching condition of \BF\ executed at all time steps (cf.\ previous section where the same is done for the first step). The two individual conditions are easily evaluated, with the first being again $\alpha^\BFsmall+\beta^\BFsmall=1$ (now valid at all times) implying ZA consistency. Enforcing also 2LPT consistency then leads to our central result${}^{\ref{foot:norm}}$
\begin{subequations} \label{eqs:weightsn>0}
\begin{align}
  \alpha^\BFsmall = \frac{\subcurl{E'}{n+1} - \subcurl F {n+\nicefrac 1 2}}{ \subcurl{E'}{n} - \subcurl F {n+\nicefrac 1 2}}  
     \,\stackrel{\text{EdS}}{\asymp} \,\frac{4 \mathfrak{n} ( 4 \mathfrak{n} + 1)-5}{4 \mathfrak{n} ( 4 \mathfrak{n} + 7)+7} \,, 
    \qquad\quad \beta^\BFsmall =  1 -\alpha^\BFsmall
   \,\stackrel{\text{EdS}}{\asymp}\,  \frac{ 24 \mathfrak{n} + 12}{4 \mathfrak{n} (4 \mathfrak{n}+7)+7} 
   \,,  
\end{align}
where $\mathfrak{n} = a_n/\Delta a$ (which is equal to $n$ for uniform time steps), and we have defined
\be
  \subcurl F {n+\nicefrac{1}{2}} =  D_{n+\nicefrac{1}{2}}^{-1}\left( \subcurl{E}{n} + \subcurl{E'}{n}  \Delta D/2 \right)  - D_{n+\nicefrac{1}{2}} \,. \label{eq:F}
\ee 
\end{subequations}
To interpret the meaning of $F_{n+\nicefrac{1}{2}}$, notice that the ratio between the source terms in the ODEs defining the second and first-order growth factors is given by $(E - D^2) / D$ (see eqs.~\ref{eq:ZA} and \ref{eq:ODE_for_E}), and $\subcurl F {n+\nicefrac{1}{2}}$ is the linearisation of exactly this ratio around $D_n$, evaluated at $D_{n+\nicefrac{1}{2}}$. As promised, the calculations leading to~\eqref{eqs:weightsn>0} are actually not only valid for $n>0$, but also for the `initialisation' case with $n=0$, basically since $E_0 = 0 = E_0'$ and thus, the 2LPT growth terms do not contribute in the first drift at that time. Indeed,\ it is easy to verify that for $n=0$, the weights~\eqref{eqs:weightsn>0} agree with those given in eqs.\,\eqref{eqs:coeffBF}. Therefore, nothing `special' has to be done for \BF\ at initialisation time, and the coefficients~\eqref{eqs:weightsn>0} are valid at all times. We emphasise that \BF\ and \textsc{PowerFrog} only agree for the first time step and for sufficiently early times, where both are 2LPT consistent. For later time steps, \textsc{PowerFrog} relies on weights derived from a transcendental equation to enforce second-order accuracy of the integrator. Furthermore, the second-order term of \textsc{PowerFrog} starts deviating from 2LPT for later time steps (section~\ref{sec:res}). By contrast, \BF\ is 2LPT consistent at all time steps while automatically achieving second-order accuracy; see section~\ref{sec:convergence} for details.

Equations~\eqref{eqs:weightsn>0} comprise the main results of the present section. To validate the calculation and to check for consistency, we plug the weights into the last half-drift update~\eqref{eq:secondhalfdriftlatersteps} at time $D_{n+1} = D_n + \Delta D$, which yields after straightforward steps
\begin{align}
     \upcurl{\fett x}{n+1} &\stackrel{\phantom{\text{EdS}}}{=} 
     \fett q - \left( D_n+ \Delta D \right) \nabq \varphi^\ini 
     +\nabqinverse \mu_2  \left[ \subcurl E n + \frac{\Delta D}{2} \left( \subcurl{E'}{n}  +  \subcurl{E'}{n+1} \right) \right]  +h.o.t.\ \label{eq:xn+1LCDM} \\
    & \stackrel{\text{EdS}}{\asymp} 
     \fett q - \left( a_n+ \Delta a \right) \nabq \varphi^\ini 
      - \frac 3 7 (a_n+\Delta a)^2  \nabqinverse \mu_2  +h.o.t.
\end{align}
The square bracketed term in~\eqref{eq:xn+1LCDM} amounts to a two-endpoint approximation for the integral in the formula $\subcurl{E}{n\!+\!1} = \subcurl{E}{n} + \int_{D_n }^{D_{n+1}} E'(\tilde D) \,\dd\tilde D = \subcurl{E}{n} + (\Delta D/ 2) \left[ \subcurl{E'}{n+1}+ \subcurl{E'}{n}\right] + O(\Delta D^3)$. Thus, eq.\,\eqref{eq:xn+1LCDM} agrees to second order with the 2LPT trajectory~\eqref{eqs:2LPT} evaluated at $D_{n+1} = D_n + \Delta D$ as anticipated. \hfill $\blacksquare$

\subsection{Numerical convergence}\label{sec:convergence}

Here, we show that \BF\ is second-order consistent with the standard symplectic DKD integrator as the time step $\Delta D$ approaches zero. Provided that the gravitational force field is sufficiently smooth, this implies that \BF\ is a globally second-order accurate scheme, with its trajectories converging towards the exact solution at any instant in time when reducing $\Delta D$, regardless of the built-in 2LPT matching conditions. However, we note that caustics affect the regularity of the force field \cite{2021MNRAS.505L..90R}, reducing the order of accuracy of {\it any} integrator in the post-shell-crossing regime; see ref.\,\cite{List:2023jxz} for details.

As was shown in ref.\,\cite[prop.\,6]{List:2023jxz}, a $D$-time integrator is second-order accurate if $\alpha+\beta=1$ and, crucially, its $\alpha$-weight has a~$\Delta D$ expansion up to second order that agrees with the one of \fastpm\footnote{\label{foot:alpha}%
 The coefficient $\alpha$ of \fastpm\ is identical to that of the standard leapfrog integrator when formulated in terms of $(\vecb{x}, \vecb{v})$; however, the latter does not satisfy the ZA consistency condition $\beta = 1 - \alpha$ in the kick, nor is its drift consistent with ZA.
}
~(eq.\,\ref{eq:FastPMcoeff}), i.e.,
\begin{align}
  \alpha^\FastPMsmall &=  1-  \alpha_1^\FastPMsmall \Delta D 
    +  \alpha_2^\FastPMsmall \Delta D^2 + O(\Delta D^3)   \,. \label{eq:secondorder} 
\intertext{Here, the $\Delta D$-coefficients read}
    \alpha_1^\FastPMsmall &= \frac{{\cal F}_n'}{{\cal F}_n} \stackrel{\text{EdS}}{\asymp} \frac{3}{2 a_n} \,, \qquad \quad    
    \alpha_2^\FastPMsmall = \frac{2 {{\cal F}'_n}^2 - {\cal F}_n {\cal F}''_n}{2 {\cal F}^2_n}   \stackrel{\text{EdS}}{\asymp}  \frac{15}{8 a_n^2} \,, \label{eq:FPM-DeltaDcoeff}
\end{align}
and we remind the reader that ${\cal F}= a^3 H \partial_a D$. Of course, while all second-order accurate $D$-time integrators as defined in eqs.~\eqref{eq:pi_integrator} must have the same $\Delta D$-coefficients up to second order, their higher-order coefficients are distinct among different integrators such as \BF\ and \fastpm.

By contrast, expanding the $\alpha$ of \BF\ (eq.\,\ref{eqs:weightsn>0}) about $\Delta D=0$, we have
\begin{align}
    \label{eq:alpharep}
  \alpha^\BFsmall &= 
    1 - \alpha_1^\BFsmall \Delta D  + \alpha_2^\BFsmall \Delta D^2 + O(\Delta D^3) \,, 
\intertext{with}
  \label{eqs:alpha12}
    \alpha_1^\BFsmall &= D E'' W^{-1} \,,   \qquad 
    \alpha_2^\BFsmall = - [1/2]\left( D E''' + E'' \right) W^{-1}
       - D^2 E'' W^{-2}\,, 
\end{align}
where $W := E-D (D + E')$ and from here on, the dependence $D=D_n$ is implied and suppressed to ease the notation. Second-order accuracy of \BF\ is achieved if both $\alpha_1^\BFsmall = \alpha_1^\FastPMsmall$ and $\alpha_2^\BFsmall = \alpha_2^\FastPMsmall$ are satisfied simultaneously.

\begin{figure}
    \centering
    \resizebox{0.55\textwidth}{!}{\includegraphics{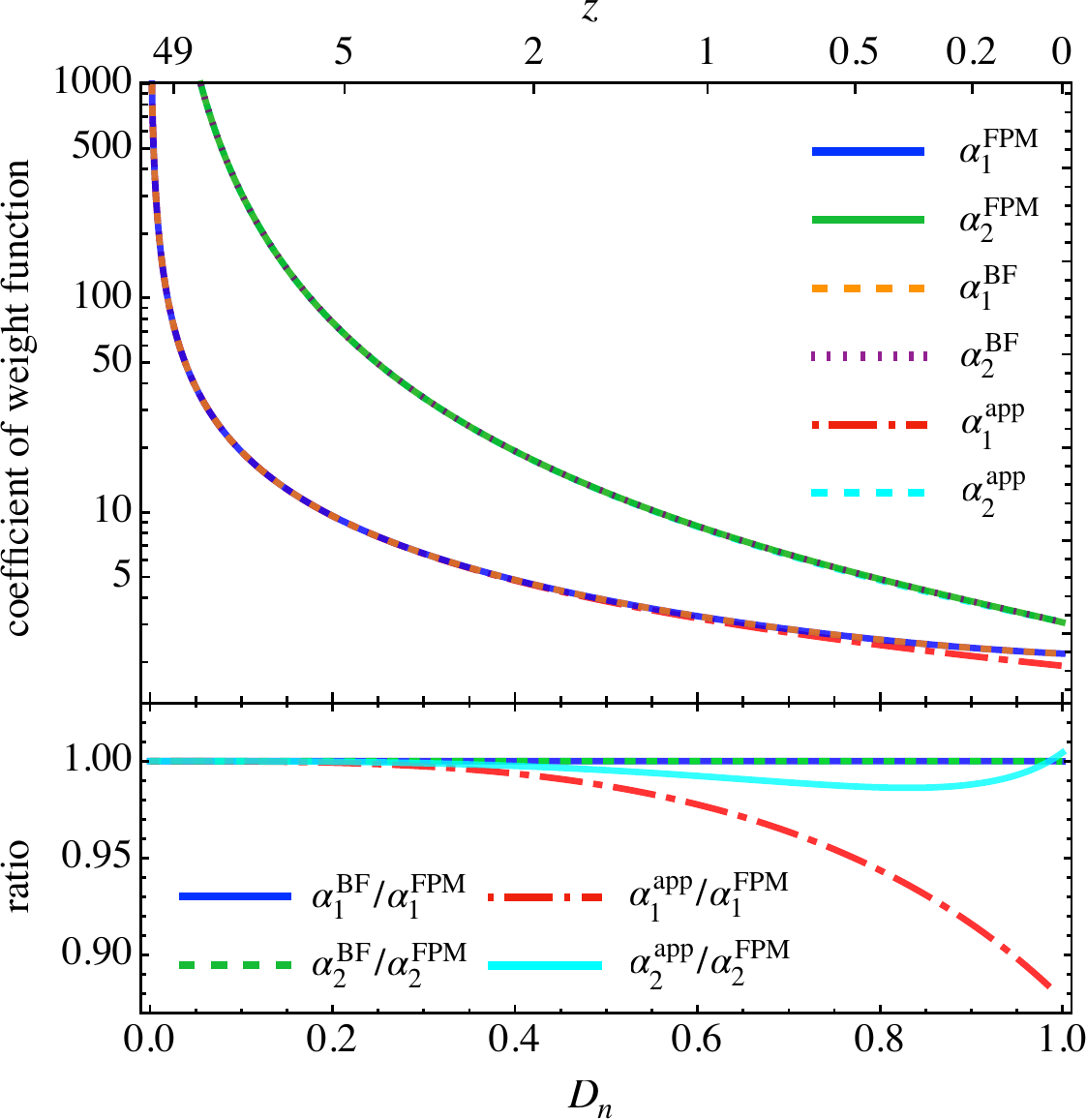}}
    \caption{ 
    {\it Top panel:} Temporal evolution of the two $\Delta D$-coefficients $\alpha_{1,2}^\BFsmall$ of \BF\ (respectively shown in orange dashed and purple dotted lines), compared to the reference solutions for which we take \fastpm\ (solid blue and green lines). Second-order convergence is achieved for the \BF\ coefficients, as they analytically agree with the reference coefficients. We also show approximate coefficients $\alpha_{1,2}^{\rm app}$ that arise if one ignores the explicit $\Lambda$ dependencies in $E$ (red dot-dashed and cyan dashed lines; the `$D^n$ approximation'), which deviate from the reference for a vast range of times and therefore lead to an inconsistent scheme (section~\ref{sec:loss-conv}). We set $\Lambda = 0.698/0.302$, and, for convenience, we employ here the standard normalisation of the linear growth so that it is unity today.
    {\it Bottom panel:} Shown are ratios of the above coefficients versus the reference coefficients.
    }
    \label{fig:weight-coeffs}
\end{figure}

Actually, although the $\alpha_{1,2}^\BFsmall$ coefficients appear to be different from their reference values at first glance, we show analytically in the following that they coincide (see figure~\ref{fig:weight-coeffs} for the numerical confirmation).

We begin with the proof related to $\alpha_1$, for which it is useful to first consider the $D$-time derivative of the definition ${\cal F} = a^2 \dot D$. Specifically, evaluating the temporal operator on the r.h.s.\ of $\pD{\cal F} = \pD(a^2 \dot D)$ and using the ODE for $D$ to get an expression for $\ddot D$ (see eq.\,\ref{eqs:2LPT}), it is straightforward to verify the identity $\pD {\cal F}= 3 H_0^2 \Omega_{\rm m0} a D/(2 {\cal F})$, from which it follows that $\alpha_1^\FastPMsmall$ can be written as
\be
  \alpha_1^\FastPMsmall = \frac{3a D}{2 \mathscr{F}^2} \,,
\ee
where $\mathscr{F}:= {\mathcal F}/(H_0 \Omega_{\rm m0}^{1/2})$. To get the \BF\ weight~$\alpha_1^\BFsmall$ in the same form, it is useful to first rewrite the ODE for $E$ (cf.\ eq.\,\ref{eqs:2LPT}) as a function of $D$:
\be \label{eq:E-Dtime}
  E'' = \frac{3a W}{2\mathscr{F}^2}  \,,
\ee
where we used the definition of $\mathcal F$ as well as the ODE for $D$. Plugging eq.\,\eqref{eq:E-Dtime} into $\alpha_1^\BFsmall$ above, it is then easy to see that  $\alpha_1^\BFsmall = \alpha_1^\FastPMsmall$, which concludes the first part of the proof.

Next we proceed with the proof of $\alpha_2^\BFsmall = \alpha_2^\FastPMsmall$, for which we note, using the before derived identity $\pD \mathscr{F}= 3 a D/(2 \mathscr{F})$, that we can rewrite the reference coefficient as
\be
  \alpha_2^\FastPMsmall  = \frac 3 2 \( \frac{\mathscr{F}'}{\mathscr{F}} \)^2 - \frac{a'}{2a} \frac{\mathscr{F}'}{\mathscr{F}} - \frac 3 4 \frac{a}{\mathscr{F}^2} \,.
\ee
To arrive at this end from the \BF\ side, we need to evaluate $E'''$; taking a $D$-time derivative from eq.\,\eqref{eq:E-Dtime}, this task is simple, leading to $D E''' = W \mathscr{F}' a'/(\mathscr{F} a) - 2 D \mathscr{F}'/\mathscr{F} - 3 W (\mathscr{F}'/\mathscr{F})^2$, where $W = E-D (D + E')$ as above. Using this as well as the above given identities, it is straightforward to confirm that $\alpha_2^\BFsmall = \alpha_2^\FastPMsmall$; thus, we conclude that \BF\ is indeed second-order accurate.  \hfill $\blacksquare$

We remark that the requirement of second-order accuracy is also achieved for \textsc{PowerFrog} (see ref.\,\cite{List:2023jxz} for proofs); however, this requirement has to be explicitly built into the integrator, which involves solving numerically a transcendental equation (eq.\,83 in ref.\,\cite{List:2023jxz}). By contrast, the \BF\ coefficients $(\alpha^\BFsmall, \beta^\BFsmall)$ -- derived as the unique choice that yields 2LPT consistency after each step -- naturally possess the asymptotic behaviour w.r.t.\ $\Delta D$ required for second-order accuracy.

\subsection{Loss of convergence through approximate growth}\label{sec:loss-conv}

It is important to emphasise that \BF\ is only a consistent scheme if the actual $\Lambda$CDM growth functions are used, and not approximations thereof. This requires solving the ODEs for $D$ and $E$ either numerically (as we do) or analytically (i.e., by utilising the exact findings of ref.\,\cite{1995PThPh..94.1151M}). Indeed, if one instead employs the `$D^n$ approximation' $E \simeq (-3/7) D^2$ in the \BF\ weight (dubbed `app' in the following), one finds
\be \label{eqs:alphaEdS}
  \alpha^\EdS = 1 - \alpha_1^\EdS \Delta D + \alpha_2^\EdS \Delta D^2 + O(\Delta D^3) \,,
  \qquad \alpha_1^\EdS = \frac{3}{2 D} \,, \qquad \alpha_2^\EdS = \frac{15}{8 D^2} \,,
\ee
where $D=D_n$ here and below. By contrast, if one formally expands the reference values about $\Lambda =0$, one finds $\alpha_1^\FastPMsmall = 3/(2D) + (3/22) \Lambda D^2 + O(\Lambda^2)$ and  $\alpha_2^\FastPMsmall = 15/(8 D^2) + 3\Lambda D/44 + O(\Lambda^2)$, i.e., {\it an infinite tower of $\mathit \Lambda$ corrections}, which are completely ignored when using the $D^n$ approximation. Since these terms already arise in $\alpha_1$ (i.e.\ at linear order in $\Delta D$), \BF\ with the $D^n$ approximation is not a consistent scheme, but instead only converges to an approximate solution. For similar reasons, convergence is also lost when employing the approximate second-order solution $E\simeq -(3/7) \Omega_{\rm m}^{-1/143}D^2 = -(3/7)D^2 - 3\Lambda D^5/1001 +O(\Lambda^2 D^8)$ of ref.\,\cite{1995A&A...296..575B} (see also~\cite{Rampf:2022tpg}), despite the fact that its~$\alpha_{1,2}$ coefficients would agree with the reference values up to $O(\Lambda^2)$ analytically -- and to within one percent numerically for a standard $\Lambda$CDM cosmology.

In figure~\ref{fig:weight-coeffs} we show the evolution of the two $\Delta D$-coefficients~$\alpha_{1,2}$ as a function of the normalised growth, resulting from the benchmark considerations (solid blue and green lines), compared to the \BF\ coefficients (orange dashed and purple dotted lines), as well as when employing the $D^n$ approximation for \BF\ (dot-dashed red and solid cyan lines). While the \BF\ coefficients agree with the reference values, it is seen that the $D^n$ approximations fare poorly at late times, preventing the integrator from converging to the correct solution as $\Delta D \to 0$.

\begin{figure}
   \centering
   \resizebox{1\textwidth}{!}{
   \includegraphics{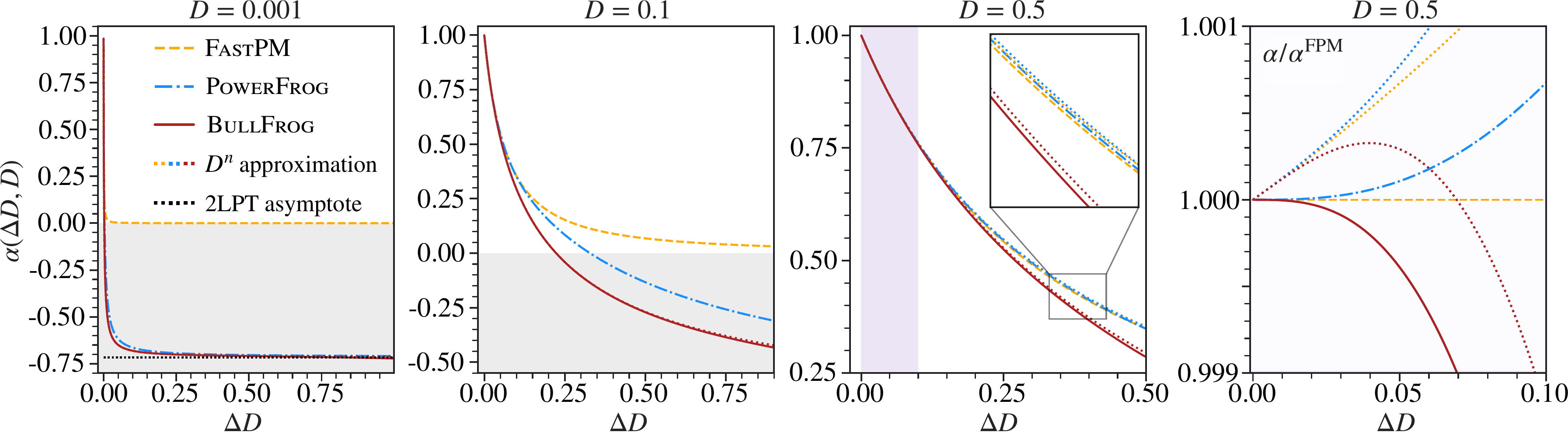}
   }
   \caption{%
   Weight $\alpha$ for various  $D$-time integrators as a function of step size $\Delta D$, respectively shown at the starting times $D=0.001,0.1,0.5$ in the first three panels (we normalise $D$ and $\Delta D$ as in the previous figure). \BF\ (dark-red line) is 2LPT consistent about any starting time, which \fastpm~(yellow dashed line) {\it never} achieves, and \textsc{PowerFrog} (blue dot-dashed line) only at early starting times for $D\to 0$ (black dotted line). The $D^n$ approximation of each $\alpha$ (various dotted lines) does not converge towards the correct solution for $\Delta D\to 0$, as it is seen in the rightmost panel, which shows the ratio $\alpha/\alpha^\FastPMsmall$ as a function of $\Delta D$ for $D=0.5$ (in the purple-shaded time-step range from the third panel): A consistent method that is second-order accurate amounts to $\alpha/\alpha^\FastPMsmall= 1 + O(\Delta D^3)$, which is achieved for the considered $D$-time integrators, but not when approximative growth functions are used, due to `forbidden' contributions of $O(\Delta D)$. The grey shaded area indicates the regime when $\alpha < 0$, implying that the Hubble drag acts so strongly during the time step that the direction of the current velocity is \textit{reversed} before adding the acceleration in the kick (see eq.~\ref{eq:vupdate}).  
   }
   \label{fig:alphas}
\end{figure}

Finally, in figure~\ref{fig:alphas} we show the temporal evolution of $\alpha = \alpha(\Delta D, D_n)$ as a function of $\Delta D$ for $D_n = 0.001, 0.1, 0.5$ (first three panels from left to right). In the first panel for $D = 0.001$, the black dotted line indicates the value $\alpha = -5/7$, required for 2LPT consistency. In contrast to the \fastpm~integrator, both \textsc{PowerFrog} and \textsc{BullFrog} have the correct 2LPT asymptotics at early times. However, \textsc{BullFrog} -- which matches the 2LPT term at \textit{all} times prior to shell-crossing -- prefers smaller values for the $\alpha$-weight than \textsc{PowerFrog} at later times. The dotted lines show the respective $D^n$ approximation for a given weight. At early times, these approximations closely follow their full $\Lambda$CDM counterparts; however, as discussed above, they do \textit{not} lead to consistent methods and hence do not converge towards the correct solution for $\Delta D \to 0$. This becomes clear from the rightmost panel, which shows the ratio of $\alpha$ to $\alpha^\FastPMsmall$ for each stepper, for $D = 0.5$ and the range of time steps $\Delta D$ shaded purple in the previous panel. While the $\Lambda$CDM coefficients all have the same asymptotic behaviour up to second order for $\Delta D \to 0$ (thus yielding second-order accurate methods, see ref.~\cite{List:2023jxz}), already the linear term of the $D^n$ approximations differs.

\section{Implementation and results}\label{sec:res}

\begin{figure}
    \centering
    \resizebox{0.75\textwidth}{!}{\includegraphics{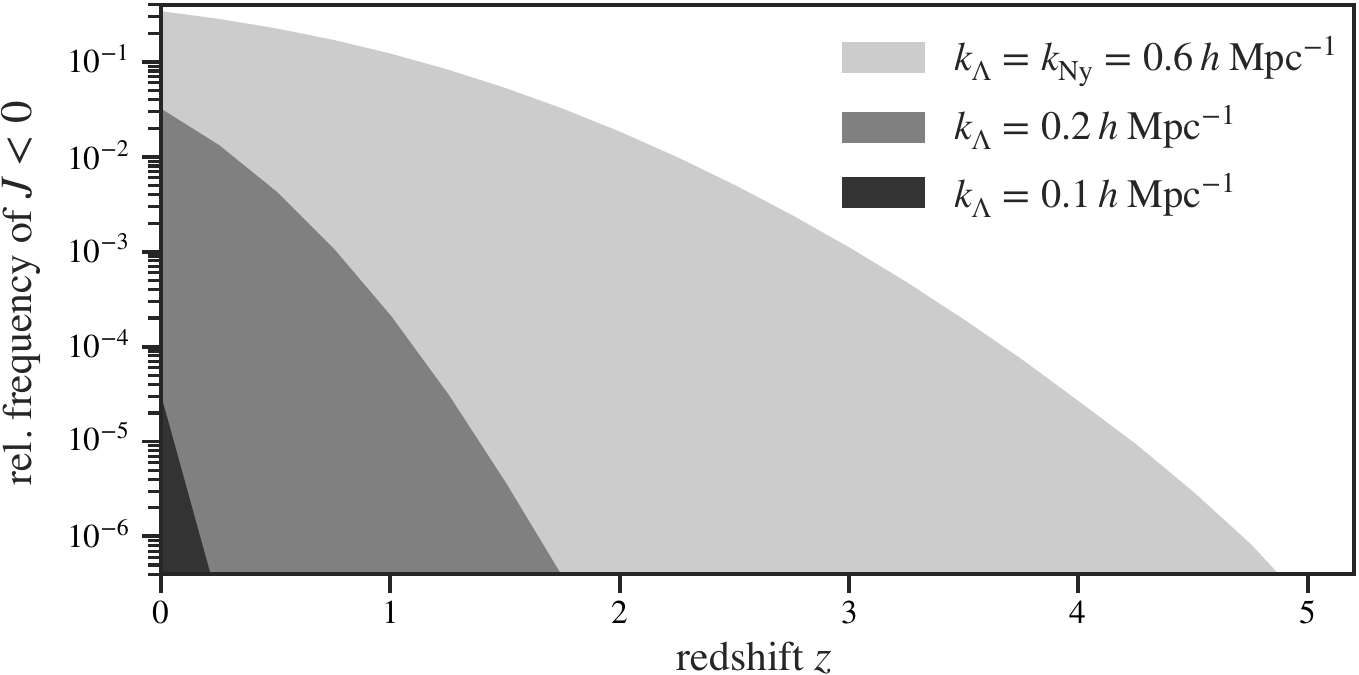}}
    \caption{Fraction of particles with a negative Jacobian determinant $J(\vecb{x}(\vecb{q},z)) < 0$ as a function of redshift~$z$, computed with 3LPT. Recall that $J$ is positive at early times (as $\vecb{x} \to \vecb{q}$ for $z \to \infty$) and crosses zero from above at the instant of shell-crossing. For the present choice of simulation settings and without UV cutoff (`\textit{unfiltered}', light grey; section~\ref{sec:results-unfiltered}), the first shell-crossing occurs around $z\simeq5$, marking the point when LPT starts losing its validity, and more than $30\%$ of the particles have $J < 0$ at $z = 0$. For the filtered cases with a cutoff at $k_\Lambda = 0.2\,h\,\mathrm{Mpc}^{-1}$ (medium grey; section~\ref{sec:results-filter}), the first shell-crossing occurs delayed at about $z\simeq2$, while with $k_\Lambda = 0.1\,h\,\mathrm{Mpc}^{-1}$ (dark grey), almost the whole simulation domain is still single stream at $z=0$. This figure provides a rough rule of thumb for estimating how many particles participate in multi-streaming. 
    }
    \label{fig:det_J_less_than_0}
\end{figure}

We perform two distinct numerical experiments to demonstrate the effectiveness of \BF. For the first experiment, we examine the case of filtered initial conditions with a sharp UV cutoff in Fourier space, which sets all Fourier modes with $|\fett k| > k_\Lambda$ to zero (section~\ref{sec:results-filter}). This filtering removes small-scale information from the initial conditions, which significantly delays the instance of first shell-crossing in the resulting coarse-grained/smoothed fluid: Without smoothing, and under the present simulation settings~(section~\ref{sec:num-imp}), the first shell-crossing occurs around $z\simeq 5$, while with UV cutoff at $k_\Lambda =0.2\,h\,\text{Mpc}^{-1}$, this event is delayed to $z\simeq2$ (see figure~\ref{fig:det_J_less_than_0}). If the filter is instead set at $k_\Lambda =0.1\,h\,\text{Mpc}^{-1}$, the first shell-crossing occurs just shortly before $z=0$. Thus, the applied filtering technique extends the validity of LPT (which becomes unreliable when $J<0$ for the first time) in $\Lambda$CDM, enabling us to perform controlled experiments. From an application point of view, investigating the evolution of coarse-grained fluids is interesting because it opens up the possibility of using \BF\ instead of high-order LPT as a fast, less memory-intense, and easy-to-implement forward model in hybrid approaches that leverage techniques of effective field theory for the small-scale modelling (see e.g.\ \cite{Schmidt:2018bkr,Schmidt:2020ovm,Kokron:2021xgh,Kokron:2021faa} for related avenues).

\begin{table}
\centering
\resizebox{\textwidth}{!}{%
\begin{tabular}{@{}lccccc@{}}
\toprule
                                                                  & LPT         & Standard leapfrog & \textsc{COLA} & \fastpm\ (FPM)  & \BF\ (BF)                   \\ \midrule
UV complete                                                       & \redcross   & \greencheck       & \greencheck   & \greencheck       & \greencheck           \\
Accurate with few orders / steps & \greencheck & \redcross         & \phantom{$^*$} {\tiny \greencheck} $-$ \greencheck$^*$  & {\tiny \greencheck}       & \greencheck           \\
Memory-friendly                                                   & \redcross   & \greencheck       & \orangetilde  & \greencheck       & \greencheck           \\
2LPT consistent & n.a.        & \redcross         & \greencheck   & \redcross         & \greencheck           \\
No additional hyperparameters                                       & \greencheck & \greencheck       & \redcross     & \greencheck       & \greencheck           \\
No need for LPT-based ICs                                         & n.a.        & \redcross         & \redcross     & \phantom{$^\dagger$}{\tiny \greencheck}$^\dagger$ & \greencheck           \\
No spatial discreteness errors       & \greencheck & \redcross         & \redcross     & \redcross         & \phantom{$^{\ddagger}$}\redcross$^{\ddagger}$ \\ \bottomrule
\end{tabular}%
}
\vspace{2mm}
\parbox{\textwidth}{
\begingroup
\footnotesize
$^*$ With \textsc{COLA}, we find large variations in the accuracy depending on the choice of the $n_{\mathrm{LPT}}$ hyperparameter. \\[-0.075cm]
$^\dagger$ \fastpm~can be initialised directly at $a = 0$; however, the mismatch in the 2LPT term hampers its accuracy in this case.  
 \\[-0.075cm]
$^\ddagger$ \BF\ is no different from the other integrators regarding spatial and force-related properties, which is why in this \\[-.125cm] work, we employ discreteness reduction schemes, see the main text. 
\endgroup
}
\caption{An overview of the pros and cons of the employed forward models. LPT is highly accurate and does not suffer from spatial discreteness; however, it is not UV complete, as it does not incorporate shell-crossing. Additionally, calculating the LPT displacement is computationally prohibitive at higher orders. By contrast, the standard leapfrog integrator is oblivious to LPT insights and requires small time steps in order to capture the dynamics on linear scales, thus unnecessarily wasting precious computing resources at high redshifts. This issue is rectified when employing the $D$-time integrators \fastpm\ and \BF, which are respectively ZA and 2LPT consistent. Note that \textsc{PowerFrog} (not listed in table) differs from \BF\ in that it conforms with 2LPT only asymptotically for $a \to 0$, and otherwise is ZA consistent. The modified COLA algorithm of ref.\,\cite{2013JCAP...06..036T} can achieve a high level of accuracy in fast predictions of the~LSS. In contrast to $D$-time integrators, \textsc{COLA} explicitly computes the 2LPT trajectories (which must be kept in memory during the simulation runtime) and corrects them by adding a residual, determined in an $N$-body fashion. Achieving optimal performance with \textsc{COLA} requires fine-tuning of the `$n_{\rm LPT}$ hyperparameter' (see footnote~\ref{footCOLA}) through trial and error. 
}\label{tab:comparison}
\end{table}

For the second experiment, we consider standard/unfiltered initial conditions in $\Lambda$CDM, focusing on comparing \BF\ to other methods; see section~\ref{sec:results-unfiltered} as well as table~\ref{tab:comparison} for an overview of the forward models considered. Apart from \fastpm, we also test against \textsc{COLA} which is not a $D$-time integrator. Instead, it incorporates 2LPT into a standard leapfrog integrator by explicitly shifting the phase-space coordinates to those of a `2LPT observer', and then solving for the residual between the true trajectories and 2LPT \cite{2013JCAP...06..036T,Howlett:2015hfa,Izard:2015dja,Howlett:2015hfa,Koda:2015mca}. \textsc{COLA} requires explicitly computing the 2LPT displacement, which must be stored in memory throughout the simulation. We employ the modified time stepper of the COLA algorithm (see appendix A.3.2 in ref.\,\cite{2013JCAP...06..036T}), since it reportedly performs better than its base stepping scheme. The modified COLA scheme introduces an additional hyperparameter, $n_{\rm LPT}$, which delicately affects the results (see, e.g., refs.~\cite{Howlett:2015hfa,2016MNRAS.463.2273F} and our findings below).\footnote{\label{footCOLA}%
    The $n_{\rm LPT}$ parameter in COLA modifies the temporal weights of the residual terms solved by the $N$-body method, which the authors of ref.\,\cite{2013JCAP...06..036T} suggest is connected to the exponent of any of the six growing and decaying modes that emerge at third order in LPT. However, we respectfully question this justification for the following reasons: (1) Before shell-crossing and using growing-mode initial conditions, the $N$-body residual aligns with the fastest growing mode (modulo discreteness errors), which would make the selection of $n_{\rm LPT}$ unnecessary (in the continuum limit); (2) after shell-crossing, when LPT becomes invalid, the residual is in general not of the 3LPT type, and therefore, none of the growing or decaying modes of 3LPT should be physically relevant in that regime. We conjecture that the observed performance gain from adjusting $n_{\rm LPT}$ may stem from a better control of spatial discreteness errors, though further research is needed to confirm this hypothesis.
}
The authors of \textsc{COLA} recommend values for $n_{\rm LPT}$ ranging between $-4$ and $3.5$, but do not provide explicit instructions for selecting the optimal value, as it can depend non-trivially on factors such as initialisation redshift, box size, force resolution, particle masses, and even subtleties such as the exact procedure for generating the initial conditions. Instead, users should choose an optimal value for $n_{\rm LPT}$ through trial and error.

\subsection{Numerical implementation} \label{sec:num-imp}

Here we provide details of the numerical setup.
Unless stated differently, we consider a simulation box of edge length $L= 2$\,Gpc$/h$, use $384^3$ $N$-body particles with a grid resolution of $768^3$ for the force computation with the PM method, and perform computations in double precision on a single GPU. We do not employ any force softening. We take the cosmology to be spatially flat $\Lambda$CDM with $\Omega_{\rm m0}=0.302$, $H_0 =67.7$\,km$/$s$/$Mpc, $n_{\rm s}=0.968$, and $\sigma_8=0.815$.

Throughout the results section, we will use the shorthand notation $n$BF to denote $n$ time steps executed with \BF, and similarly for \textsc{PowerFrog} (PF), \fastpm~(FPM), and \textsc{COLA}. For $D$-time integrators, we only employ $D$-time spacing uniformly distributed between~$0$ and~$1$ for simplicity. For \textsc{COLA} (which, recall, is not a $D$-time integrator), we employ linear-in-$a$ steps as in the original \textsc{COLA} paper~\cite{2013JCAP...06..036T}, and we use the `modified' \textsc{COLA} scheme in KDK form as described in their appendix~A.3.2 which, according to the authors, performs better than their `standard' scheme (appendix~A.3.1; cf.\ the findings of \cite{Izard:2015dja, Howlett:2015hfa}). It is likely that further performance gains from all fast integrator methods can be achieved by adopting refined time-stepping strategies, which however requires further investigation.

We implemented all numerical integrators, including specifically \textsc{COLA}, within the same code ecosystem. This has the added benefit of ensuring that force computations are identical across all integrators, allowing us to study the intrinsic properties of the time integrators in isolation. Furthermore, to mitigate the discrete nature of $N$-body simulations, we employ several discreteness reduction schemes. Specifically, we use Fourier sheet interpolation \cite{Hahn2016PhaseSpace,2020MNRAS.495.4943S}, the exact Fourier gradient kernel ${\rm i} \fett k$, higher-order mass assignment \cite{2004JCoPh.197..253C}, and de-aliasing via grid interlacing \cite{1988csup.book.....H} to improve the accuracy of the force computation; see ref.\,\cite{List:2023kbb} for details of all these techniques. While we use these techniques at all times for the filtered simulations, we choose to disable them after redshift $z = 25$ for the unfiltered $\Lambda$CDM simulations when the impact of discreteness effects is already significantly lower than at early times. This is mainly for reasons of computational efficiency, but we also note that these techniques should anyhow not be used (without spatial refinement) in scenarios where the phase-space sheet is highly complex, i.e.\ on highly nonlinear scales \cite{Hahn2016PhaseSpace, 2020MNRAS.495.4943S}.

For the LPT implementation, we generate the displacement fields exploiting the recursive relations from ref.~\cite{2021MNRAS.501L..71R} and de-alias them using Orszag's 3/2 rule \cite{1971JAtS...28.1074O}. We have also tested the computationally expensive `leave-no-mode-behind' method explained in ref.\,\cite{Schmidt:2020ovm} (see also \cite{Taruya:2018jtk,Taruya:2021ftd}), which takes into account mode-couplings beyond the original UV cutoff. However, we found negligible impact within our setup and therefore ignore this subtle effect for computational ease. By contrast, we include transverse contributions to the displacement, which were ignored in the results presented in ref.\,\cite{Schmidt:2020ovm}; including these transverse fields is computationally expensive (see below) but introduces a two-loop effect (and higher) to the matter power spectrum, and a one-loop effect to the matter bispectrum \cite{Rampf:2012xb}. Regarding the temporal evolution of the LPT displacement, we perform a single time step from $z=\infty$ to the target redshift, and employ two different schemes: (1) the commonly used `$D^n$ approximation' for modelling the LPT growth at order $n$ (sometimes called `[quasi-]EdS approximation'; see e.g.\ refs.\,\cite{Desjacques:2016bnm,Fasiello:2022lff,Rampf:2022tpg}), and (2) the exact growth based on numerical solutions of the underlying ordinary differential equations up to third order in LPT, with a switch to the $D^n$ approximation for $n$LPT when~$n>3$. We remark that using the actual $\Lambda$CDM displacement growth is ultimately prohibitive at large LPT orders, since the number of spatial terms increases greatly with the order, but must be kept in storage; see ref.\,\cite{Schmidt:2020ovm} for a related implementation of the LPT recursion relation. To determine the Eulerian density given the current particle positions~$\vecb{x}(\vecb{q}, t)$, obtained with a numerical integrator or with LPT, we compute a cloud-in-cell approximation of the exact overdensity
\be
  \delta(\fett x,t) +1  = \int_{(L\mathbb{T})^3} \delta_{\rm D} \left(\fett x - \fett x(\fett q,t) \right) \, \dd^3 q \,,
\ee
where $\delta_{\rm D}$ is the Dirac delta, and $(L\mathbb{T})^3$ denotes the three-dimensional torus of side length $L$. In the following, we will also show results for the matter power spectrum $P$, which we define with
\be
   \left\langle \tilde \delta(\fett k_1) \tilde \delta(\fett k_2)  \right\rangle = (2\pi)^3 \delta_{\rm D} (\fett k_1 + \fett k_2) \,P(k_1) \,,
\ee
where $k = |\fett k|$, and the tilde denotes a Fourier transform with respect to the Eulerian coordinate. The cross-power spectrum $P_{\mathrm{cross}}$ between two fields $\delta_1$ and $\delta_2$ is defined similarly via $\left\langle \tilde \delta_1(\fett k_1) \tilde \delta_2(\fett k_2) \right\rangle$.

\subsection{Simulations with UV cutoff}\label{sec:results-filter}

\begin{figure}
    \centering
    \resizebox{0.98\textwidth}{!}{    \includegraphics{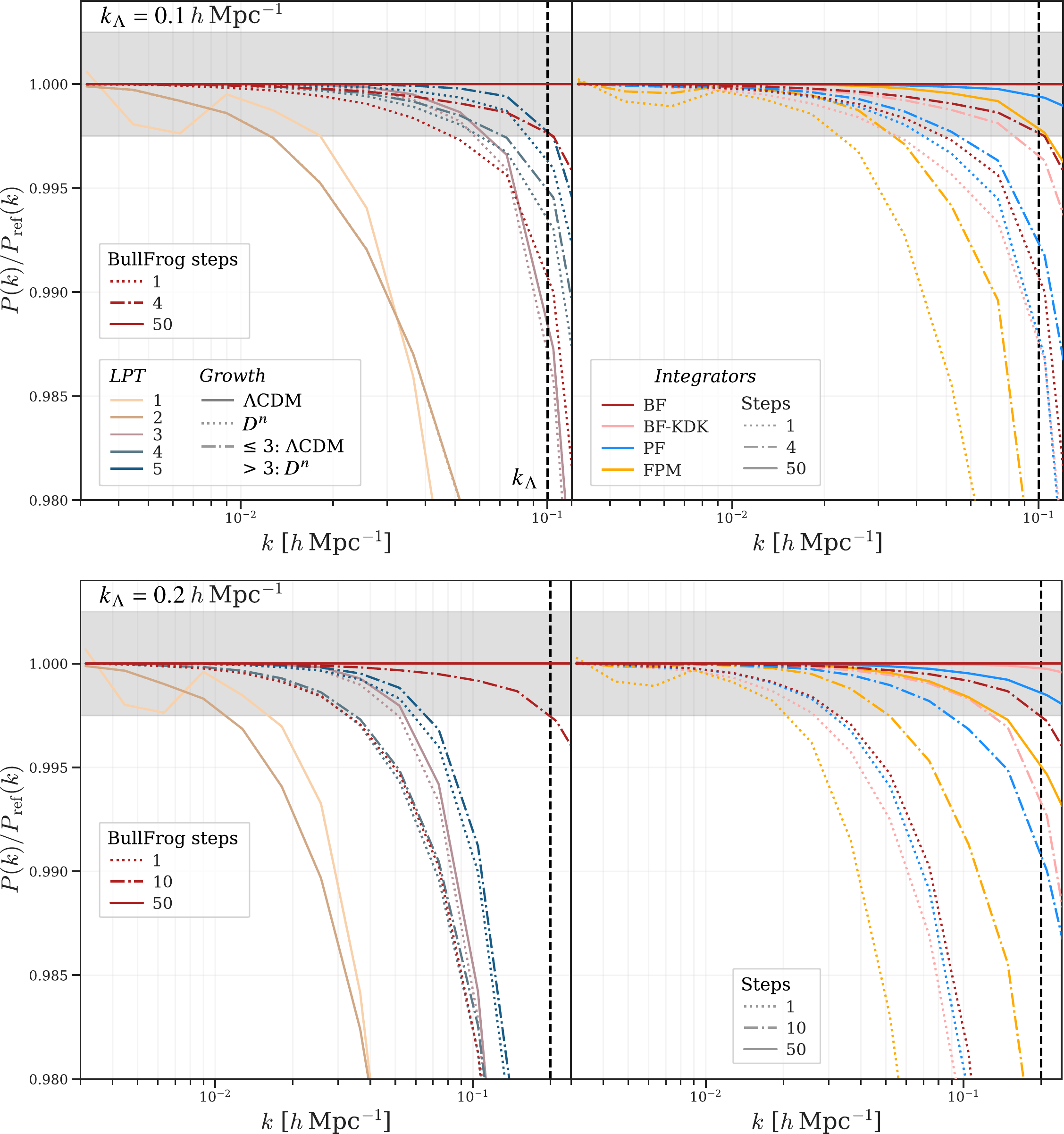}
    }
    \caption{Today's matter power spectrum relative to the \BF\ prediction with 50 steps (50BF). In the upper [lower] panels, the cutoff scale in the ICs is $k_\Lambda\!=\!0.1\  [0.2]\,h\,\mathrm{Mpc}^{-1}$. All integrators were initialised at $z=\infty$, and the discreteness suppression was active for the whole runtime. The left panels show a comparison of BF    with LPT (see main text for the explanation of the growth). For both cutoffs, already a single BF step is better than 2LPT. For $k_\Lambda = 0.1 \,h\,\mathrm{Mpc}^{-1}$, where LPT converges well to the reference solution, the performance of 4BF lies between 4 $-$ 5LPT, with an error of about $0.25\%$ (grey band) at $k_\Lambda$. For the case of $k_\Lambda = 0.2\,h\,\mathrm{Mpc}^{-1}$, the power of high-order LPT drops off already well above $k_\Lambda$, as shell-crossing physics becomes relevant but is not incorporated. By contrast, 10BF is sufficient to keep the error below $0.25\%$ up to $k_\Lambda$. In the right panels, we show results for the KDK variant of \BF\ (BF-KDK; see appendix~\ref{app:KDK-BF}), \textsc{PowerFrog} (PF), and \fastpm~(FPM). The loss of power with 4BF for $k_\Lambda = 0.1\,h\,\mathrm{Mpc}^{-1}$ is comparable to 50FPM, and 10BF outperforms 50FPM for the case with $k_\Lambda = 0.2\,h\,\mathrm{Mpc}^{-1}$. \BF\ also improves on \textsc{PowerFrog} in all cases.
    }
   \label{fig:Pk_comparison_filtered}
\end{figure}

We begin by reporting the results for the case where a sharp UV cutoff in Fourier space is applied to the initial conditions, which is relevant for the study of effective fluids. In figure~\ref{fig:Pk_comparison_filtered} we show ratios of power spectra with respect to a reference simulation done with~50BF, where the latter is well converged for the cutoff choices $k_\Lambda = 0.1\,h\,{\rm Mpc}^{-1}$ (top panels) and $k_\Lambda = 0.2\,h\,{\rm Mpc}^{-1}$ (bottom panels), evaluated at $z=0$. The left panels compare \BF\ (shown in red) with LPT results (other colours). For \BF, the various line styles denote the number of time steps, while for LPT, the line styles designate whether the growth ODEs were solved exactly (solid), or whether the $D^n$-approximation was used (dotted); note that for LPT orders $> 3$, we always use the $D^n$-approximation (dot-dashed).

For the lower cutoff ($k_\Lambda = 0.1\,h\,\mathrm{Mpc}^{-1}$), LPT converges towards the reference result when consecutively increasing the perturbation order. This is expected, as almost the whole spatial domain of the effective fluid is still single stream (see figure~\ref{fig:det_J_less_than_0}), in which case LPT should be fairly accurate. However, the accuracy of LPT comes at a high computational price: achieving 0.25\% accuracy with the lower cutoff choice requires 5LPT, which comes with a whopping 17 spatial kernels in the scalar sector and with 7 transverse contributions in the vector sector. Asymptotically, the number of new terms at each order grows quadratically, making high-order LPT prohibitively expensive in terms of memory and computing time. Also, at such late times, the actual $\Lambda$CDM growth should be taken into account, although we find that the effect is fairly subtle up to third order (see \cite{Schmidt:2020ovm} for incorporating the actual growths at larger orders). 
%
%
%
%
By contrast, 4BF is already converged up to 0.25\% accuracy for the lower cutoff, and even 1BF fares significantly better than 2LPT -- and is almost as good as 3LPT.

A crucial advantage of $n$BF over $n$LPT is its constant memory footprint when increasing the number of steps $n$ (unlike when increasing the order $n$ in the case of LPT). Indeed, when using \BF~as a time integrator, only two arrays of size $N$ (for positions and momenta) are required in principle, along with the resources for force computation, such as a PM grid, just as in any standard $N$-body simulation. The positions and momenta are updated in place by the drift and kick operation, respectively. That said, depending on the desired level of accuracy, the use of discreteness suppression techniques is advisable at early times (see section~\ref{sec:num-imp}), which leads to an increase in runtime and memory (and obviously to more internal parameters that need to be adjusted). However, we remark that the use of discreteness suppression schemes can be circumvented if the simulation is initialised at sufficiently late times using 2 $-$ 3LPT~\cite{Michaux:2020yis, List:2023kbb} -- and for fast approximate simulations, these techniques might be omitted altogether even when starting at $D = 0$.

For the higher cutoff case with $k_\Lambda = 0.2\,h\,\mathrm{Mpc}^{-1}$, we observe emerging non-convergent behaviour of LPT (bottom-left panel in figure~\ref{fig:Pk_comparison_filtered}): 4LPT performs slightly worse than 3LPT with respect to the reference result, while 5LPT comes only with a minor improvement; see also ref.\,\cite{Schmidt:2020ovm} for similar observations. The weaker LPT performance can be explained by the fact that significantly more particles have shell-crossed by $z=0$ as compared to the lower cutoff case (figure~\ref{fig:det_J_less_than_0}). Consequently, it is questionable whether higher-order LPT predictions could significantly alleviate the observed discrepancy, as shell-crossing physics becomes relevant and should be incorporated. By contrast, \BF, being a UV-complete method, converges to the correct solution also after shell-crossing. In fact, we find that already 10BF yields an error of less than $0.25$\% w.r.t.\ the reference result for~$k \leq k_\Lambda$.

In the right panels of figure~\ref{fig:Pk_comparison_filtered}, we compare \BF\ against various $D$-time integrators, all initialised at $a=0$. Regarding the KDK version of \BF\ (BF-KDK; pink lines in figure~\ref{fig:Pk_comparison_filtered}), the late-time performance is fairly comparable to its DKD version, but we note that BF-KDK was initialised at $a=0$ using 1BF(-DKD), for reasons explained in appendix~\ref{app:KDK-BF}. \fastpm\ (i.e., its DKD variant) \cite{List:2023jxz}, by contrast, does not perform so well. This can be explained by the fact that  \fastpm~is only Zel'dovich consistent, which is a disadvantage particularly when initialised at $a=0$. This weakness of \fastpm\ could be slightly mitigated by initialising with, for example, 2LPT or using the DKD variant of BF. Nonetheless, despite such a  `jump start', \fastpm\ introduces a 2LPT error from its first step onwards, causing it to converge significantly more slowly than \BF\ (see figure~\ref{fig:Pk_convergence}).

\begin{figure}
    \centering
    \resizebox{1\textwidth}{!}{    \includegraphics{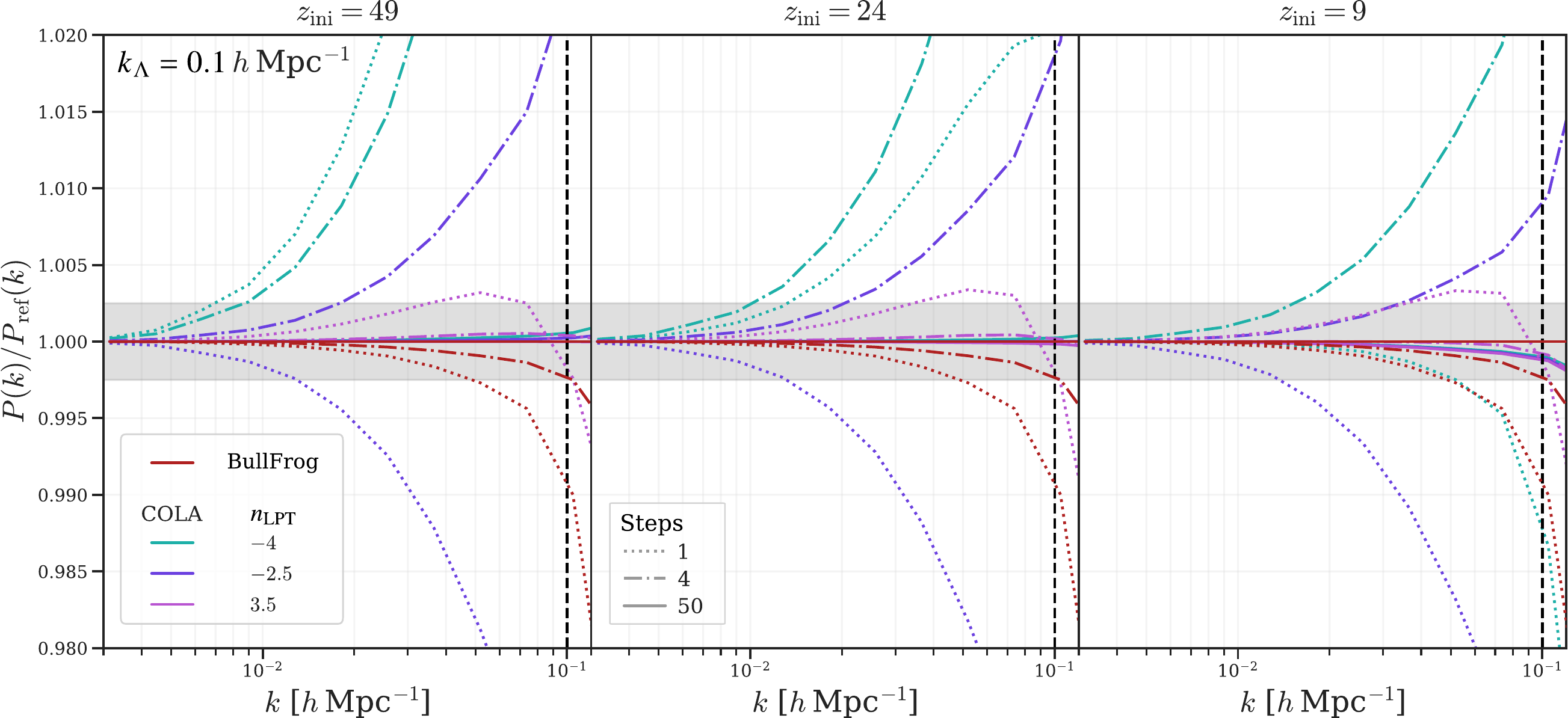}
    }
    \caption{Same \BF\ power spectra as in figure~\ref{fig:Pk_comparison_filtered}, but now in comparison with COLA time stepping, for three different COLA hyperparameters `$n_{\mathrm{LPT}}$' (various colours) and three different redshifts for initialising COLA with 2LPT (from left to right).${}^{\ref{footCOLA}}$ The COLA power spectra are highly sensitive to the choice of $n_{\mathrm{LPT}}$ and can overshoot already with a single step. \BF\ yields results competitive with the best-performing choice of $n_{\textsc{LPT}}$ in COLA (here: $n_{\textsc{LPT}}=3.5$), converges monotonically 
    in terms of power as the number of steps increases, and does not require the computation of spatial 2LPT kernels.
    }
    \label{fig:Pk_comparison_filtered_cola}
\end{figure}

In figure~\ref{fig:Pk_comparison_filtered_cola} we compare \BF\ against various settings within \textsc{COLA} for the cutoff case $k_\Lambda = 0.1\,h\,{\rm Mpc}^{-1}$. The panels from left to right show \textsc{COLA} results initialised at $z_\ini=49,24,$ and $9$. The various line styles indicate different linear-in-$a$  steps (as recommended by the developers), while the three colours denote three choices of the hyperparameter $n_{\rm LPT}$. For all presented combinations of hyperparameters and initialisation redshifts, COLA converges to the benchmark solution. For a low number of time steps, however, \COLA~results depend weakly on the initialisation redshift and strongly on the chosen hyperparameter. Notably, among the three considered choices, we find $n_{\mathrm{LPT}} = 3.5$  performs best in this scenario, which is quite different from the default value of $n_{\mathrm{LPT}} = -2.5$ adopted in refs.\,\cite{2013JCAP...06..036T,Howlett:2015hfa} (the latter value yields favourable results in our experiment without an IC cutoff with 10 steps, however only when starting at $z_\mathrm{ini} = 9$; see figure~\ref{fig:Pk_comparison_unfiltered_cola}).

As a final exercise, and to connect to the full simulations as performed in the following section, we determine the effective speed of sound of matter, $c_s$, which we define through
\be \label{eq:cs2}
  P_{\rm full}(k,t) = \left[ 1- 2c_s^2(k,t; k_\Lambda) k^2 \right] P_\Lambda(k,t) \,\,\quad \Leftrightarrow\quad\,\,  c_s^2(k,t; k_\Lambda)  = \frac{P_\Lambda(k,t) - P_{\rm full}(k,t)}{2 k^2 P_\Lambda(k,t)}  \,.
\ee
Here, $P_\Lambda(k,t)$ is the power spectrum with cutoff at $k_\Lambda$ in the ICs, determined using either $n$LPT or $n$BF as above, while $P_{\rm full}$ denotes a converged simulation result without cutoff. In the following, we compute $P_{\rm full}$ using a 100BF simulation with Nyquist scale $k_{\rm Ny}=0.6\,h\,\text{Mpc}^{-1}$. Physically, $c_s^2$ reflects a contribution to the power spectrum from small-scale physics with modes $k > k_\Lambda$ which, through nonlinear mode coupling, affects also the large-scale modes with $k \leq k_\Lambda$. Within the effective field theory (EFT) framework, $c_s^2$ is the leading-order counterterm if it is scale independent, while residual scale-dependency can be interpreted as higher-order- or EFT counterterm corrections \cite{Baumann:2010tm,Carrasco:2012cv}; see refs.\,\cite{Schmidt:2020ovm,Baldauf:2015tla} for similar results based on eq.\,\eqref{eq:cs2} where, however, $P_\Lambda$ is taken solely from~$n$LPT.

\begin{figure}
    \centering
    \resizebox{1\textwidth}{!}{    \includegraphics{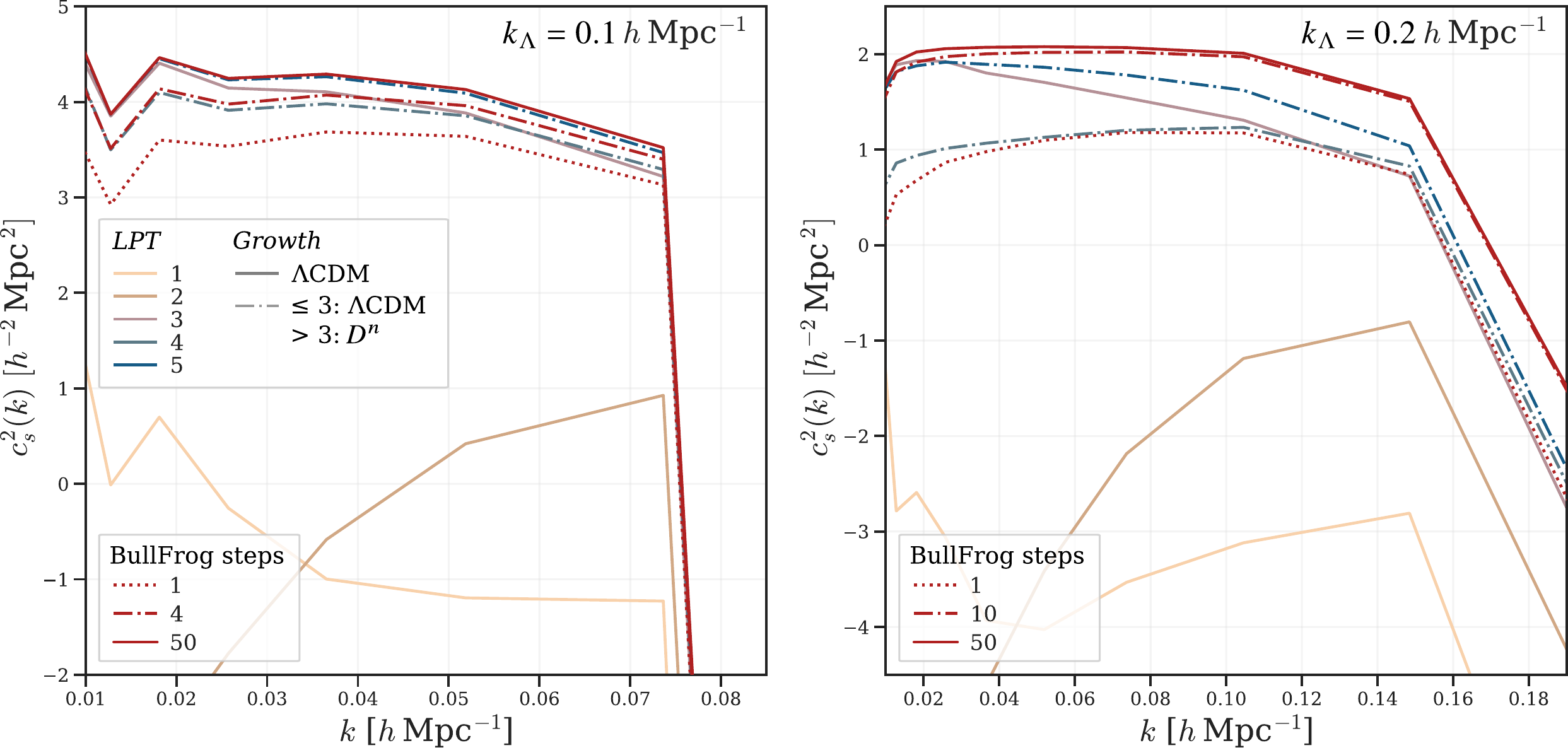}
    }
    \caption{Speed of sound $c_s^2$ as defined through eq.\,\eqref{eq:cs2}, using either $n$LPT or $n$BF as the forward model, adopting the same line styles as in figure~\ref{fig:Pk_comparison_filtered}. 
    The left [right] panel shows results for $k_\Lambda =0.1 [0.2]\,h\,\text{Mpc}^{-1}$. The BF predictions are approximately constant for sufficiently large scales, indicating that $c_s^2$ can be identified with the leading-order EFT counterterm for the present choices of $k_\Lambda$.
    }
    \label{fig:cs_comparison_filtered}
\end{figure}

Figure~\ref{fig:cs_comparison_filtered} shows the predictions for $c_s^2$, with the same colour and line styles as used in figure~\ref{fig:Pk_comparison_filtered}. The left panel displays the results for the lower cutoff ($k_\Lambda = 0.1\,h\,{\rm Mpc}^{-1}$), and the right panel for the higher cutoff ($k_\Lambda = 0.2\,h\,{\rm Mpc}^{-1}$). For both cutoffs, and for \BF\ and LPT, we observe a drop in power of $c_s^2$ for $k$-values approaching $k_\Lambda$ from below. Within an EFT, as argued by refs.\,\cite{Schmidt:2020ovm, Baldauf:2015tla}, this behaviour is expected because, in the current implementation, $c_s^2$ must capture all nonlinear mode couplings, which become non-trivial near the cutoff scale. This scale dependency in $c_s^2$ could be suppressed by including terms of $O([k/k_\Lambda]^4)$ in the relationship between $P_\Lambda$ and $P_{\rm full}$. In contrast, on large scales, sufficiently far from the cutoff, the \BF\ results for $c_s^2$ form a plateau of roughly constant height for any number of time steps, indicating that a $k$-independent counterterm adequately captures the UV feedback on the power spectrum at scales~$k \ll k_\Lambda$, for the present choices of cutoffs.

Regarding the LPT results with the lower cutoff (left panel in figure~\ref{fig:cs_comparison_filtered}), we observe a similar convergence of $n$LPT at increasing perturbation orders as previously noted (see top-left panel of figure~\ref{fig:Pk_comparison_filtered} and \cite{Schmidt:2020ovm}). However, the aforementioned scale independence for $k \ll k_\Lambda$ becomes pronounced only for orders~$n \gtrsim 3$, whereas 1$-$2 LPT leads to (unphysical) negative values of $c_s^2$ (see also \cite{Schmidt:2020ovm}). By contrast, for the larger cutoff (right panel in figure~\ref{fig:cs_comparison_filtered}), the scale independence is not even reached by 5LPT, which in the present case is explained by the impending loss of perturbative control. Indeed, the 4LPT prediction performs worse than 3LPT in comparison to the reference result, indicating that LPT convergence is lost on the considered scales (cf.\ bottom-left panel of figure~\ref{fig:Pk_comparison_filtered}). In this context, we remark that it is well documented that both field-level and power-spectrum predictions of LPT converge monotonically within its validity regime; see e.g.\ refs.\,\cite{Saga:2018nud, Saga:2023vno, Rampf:2017jan, 2021MNRAS.501L..71R, 2023PhRvD.108j3513R, Michaux:2020yis}.

In conclusion, among all tested forward models, \BF\ provides the fastest and most reliable LSS predictions with cutoff. For sufficiently small cutoffs ($k_\Lambda \lesssim 0.1\,h\,\text{Mpc}^{-1}$), where $n$LPT predictions are very accurate for $n \geq 4$ down to $z=0$, we find good agreement with all tested $N$-body integrators, with \BF\ requiring the fewest time steps for reaching a given target accuracy.

\subsection[Full \texorpdfstring{$\Lambda$CDM}{LCDM} simulations]{Full \texorpdfstring{$\fett \Lambda$CDM}{LCDM} simulations}\label{sec:results-unfiltered}

\begin{figure}
    \centering
    \resizebox{1\textwidth}{!}{    \includegraphics{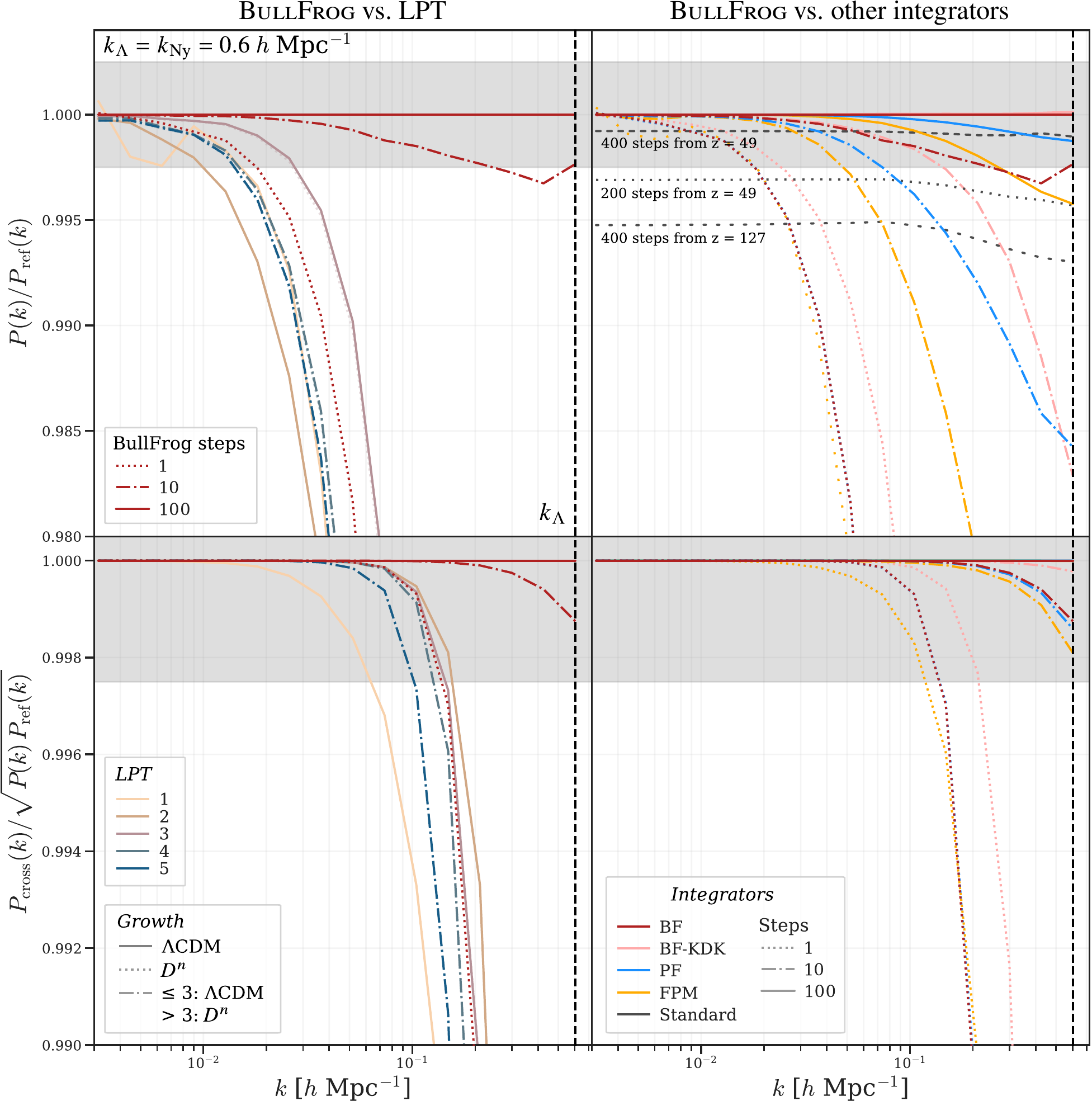}
    }
    \caption{Similar as figure~\ref{fig:Pk_comparison_filtered}, but now showing results for full $\Lambda$CDM simulations at $z=0$, for the power spectrum (upper panels) as well as for the cross-correlation (lower panels). The $D$-time integrators were initialised at $z = \infty$ with discreteness suppression techniques switched off at $z=25$. We also present results using a standard symplectic (DKD) leapfrog integrator (also with discreteness suppression), initialised with 2LPT at $z = 49$ or $z = 127$, whose convergence in terms of the power spectrum is very slow because even the leading-order ZA dynamics are only recovered in the limit $\Delta D \to 0$. The lower panels show the cross-correlation between the density contrast for each case and the reference. For 200 and 400 symplectic integrator steps, the cross-correlation exceeds the accuracy of $99.995\,$\%, and is therefore not shown.
    }
    \label{fig:Pk_comparison_unfiltered}
\end{figure}

We now turn to full $\Lambda$CDM simulations without cutoff, extending down to the particle Nyquist scale $k_{\rm Ny} = \pi N^{\nicefrac 1 3} L^{-1} \simeq 0.6\,h\,\text{Mpc}^{-1}$ (and slightly beyond due to corner modes $|\vecb k| > k_{\rm Ny}$ that we include); see section~\ref{sec:num-imp} for details on the numerical setup. The upper panels of Figure~\ref{fig:Pk_comparison_unfiltered} show ratios of power spectra relative to the reference run, obtained with 100BF. The left panel compares $n$BF with $n$LPT predictions, adopting the same colour and line styles as in figure~\ref{fig:Pk_comparison_filtered}. Unsurprisingly, LPT convergence is lost at such late times (cf.\ figure~\ref{fig:det_J_less_than_0}), which is exemplified by a significant loss in power on a vast range of nonlinear scales. For example, at $k_{\rm Ny}$, the accuracy of the power-spectrum prediction of 3LPT [5LPT] deviates by $66.36\%$  [$69.14\%$] from the reference result. \BF, on the other hand, yields predictions with an accuracy better than $0.33\%$ for $n \gtrsim 10$.

In the upper right panel of figure~\ref{fig:Pk_comparison_unfiltered}, we compare the \BF\ results with the other $D$-time integrators, and against a standard symplectic integrator initialised at the two redshifts $z=127,49$ with 2LPT. Like \BF, the symplectic integrator is UV-complete in the temporal sense, ensuring that it consistently yields the correct solution in the limit of vanishing time steps, unaffected by shell-crossing. However, the symplectic integrator entirely overlooks the accurate description of structure growth provided by LPT at early times and on large scales, necessitating small time steps to capture the dynamics on linear scales, which in turn wastes valuable computing resources at high redshifts. Specifically, even with hundreds of time steps, an (admittedly small) residual in the power spectrum remains, which is remarkably constant as a function of scale and converges very slowly. We have explicitly verified that this residual can be further reduced by a late-time initialisation using 3LPT at $z=24$, where the 200-step result already has an edge over the 400-step result with the 2LPT initialisation at $z=49$ (not shown; see also \cite{Michaux:2020yis}). Regarding the $D$-time integrators, compared to the filtered results (figure~\ref{fig:Pk_comparison_filtered}), the spread between the various predictions is now more pronounced (e.g., 10FPM is only accurate to $4.45\%$ up to $k_{\rm Ny}$). Nonetheless, the qualitative findings remain similar: \BF\ outperforms the other $D$-time integrators for a fixed number of time steps. However, we note that the performance  of \fastpm\ could be slightly improved by initialising at a later time using 1BF or 2 $-$ 3 LPT, rather than directly at $a = 0$ (see below).

The lower panels of figure~\ref{fig:Pk_comparison_unfiltered} show the cross-correlation w.r.t.\ the reference, comparing LPT (left) and other integrators (right). While \BF\ improves upon \textsc{PowerFrog} and \fastpm\ also in terms of the cross-correlation, it is interesting that the BF-KDK variant performs better than BF-DKD in terms of this statistic. 
Furthermore, the cross-correlation for the standard symplectic integrator with $\geq 200$ steps is converged ($\geq 99.995\,\%$, not shown in figure), in contrast to the slower convergence as observed in $P(k)$. This suggests that the lack of knowledge about perturbative structure growth in standard integrators affects the amplitude of density perturbations (as measured by the power spectrum) significantly more than their phases (as measured by the cross-correlation).

\begin{figure}
    \centering
    \resizebox{1\textwidth}{!}{    \includegraphics{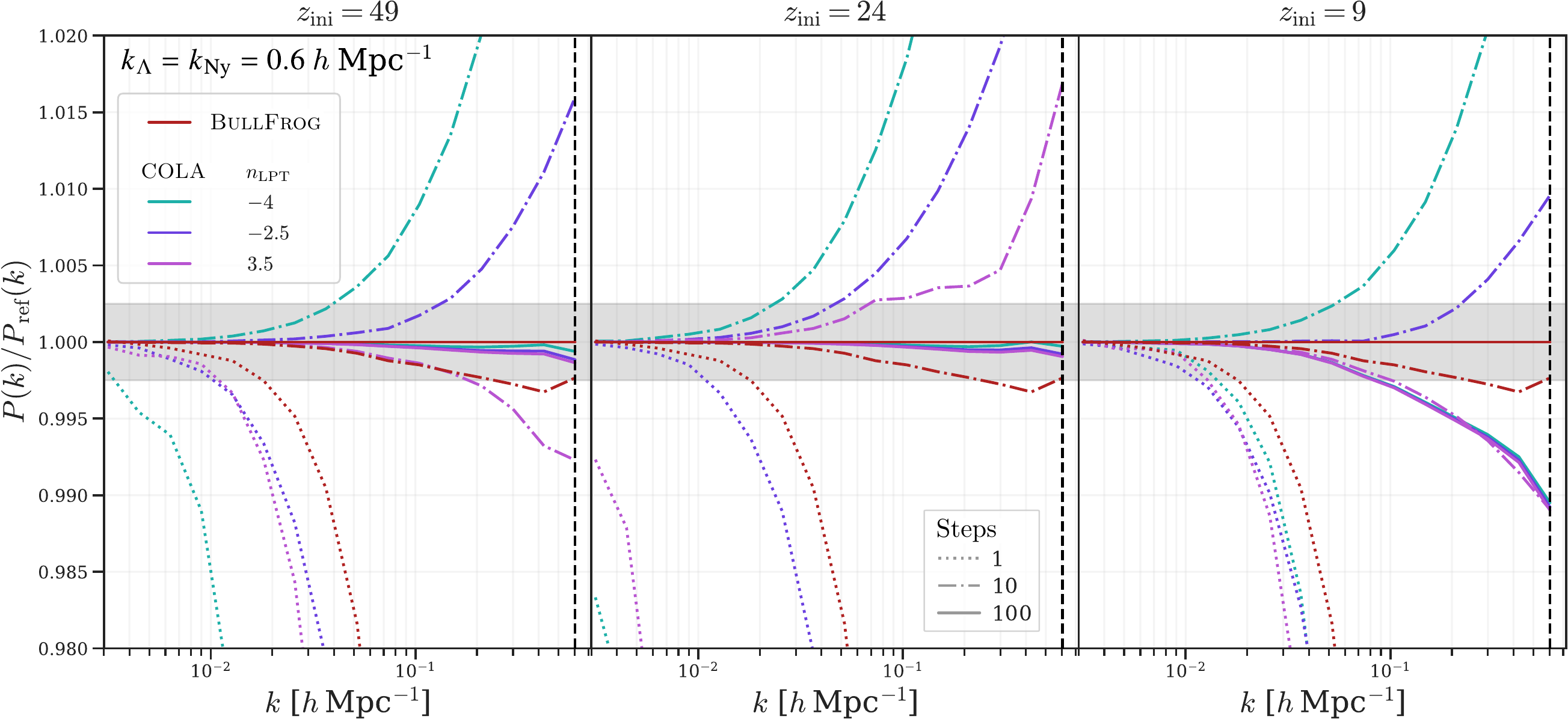}
    }
    \caption{Similar as figure~\ref{fig:Pk_comparison_filtered_cola}, but now for full $\Lambda$CDM simulations. For all three considered hyperparameters, 100COLA agrees with 100BF to better than 0.25\%, however only if COLA is not initialised too late ($z_\ini \gtrsim 24$). For 10COLA, good performance is achieved either with $n_{\rm LPT}=3.5$ at $z_\ini =49$ or with $n_{\rm LPT} = -2.5$ at $z_\ini = 9$, leading to sub-percent agreement with regards to the reference result. 
    }
    \label{fig:Pk_comparison_unfiltered_cola}
\end{figure}

In figure~\ref{fig:Pk_comparison_unfiltered_cola} we compare our findings against COLA in the unfiltered case, using the same line styles and panel structure as in figure~\ref{fig:Pk_comparison_filtered_cola}. We observe that the power spectrum of 100COLA closely approaches the reference result, except in the case when COLA is initialised at $z_\ini =9$~--~regardless of the $n_{\rm LPT}$ choice. This indicates that the 2LPT initialisation was pushed too far (if sub-percent accuracy is required), as the $\geq 3$LPT contributions missing in the initial conditions at $z_\ini = 9$ can never be recovered by the subsequent COLA integration, no matter the number of performed steps. 10COLA, however, performs  well for all considered initialisation redshifts including specifically also $z_\ini=9$, provided that $n_{\rm LPT}$ is suitably adjusted. Overall, for the present experiments and considered scales, we find good performance with 10COLA for either $n_{\rm LPT}=3.5$ and $z_\ini=49$, or for $n_{\rm LPT}=-2.5$ and $z_\ini =9$, resulting in sub-percent agreement with our reference result. The latter observation broadly aligns with the findings of ref.\,\cite{Howlett:2015hfa}, though their reference result was obtained using a different gravity solver (Tree-PM using \textsc{Gadget-2}), whereas we apply an identical force calculation. We remark that it is likely that the COLA performance could be slightly improved for more refined choices of $n_{\rm LPT}$ paired with optimal choices of $z_\ini$ (see also below).\footnote{%
    See also the recent COCA approach that improves the 2LPT reference frame used in COLA by means of a field-level emulator~\cite{Bartlett:2024cra}.
} 
We emphasise that even in this unfiltered case, 10BF consistently achieves smaller $P(k)$ errors than 10COLA across all parameter choices considered.

\begin{figure}
    \centering
    \resizebox{.8\textwidth}{!}{    \includegraphics{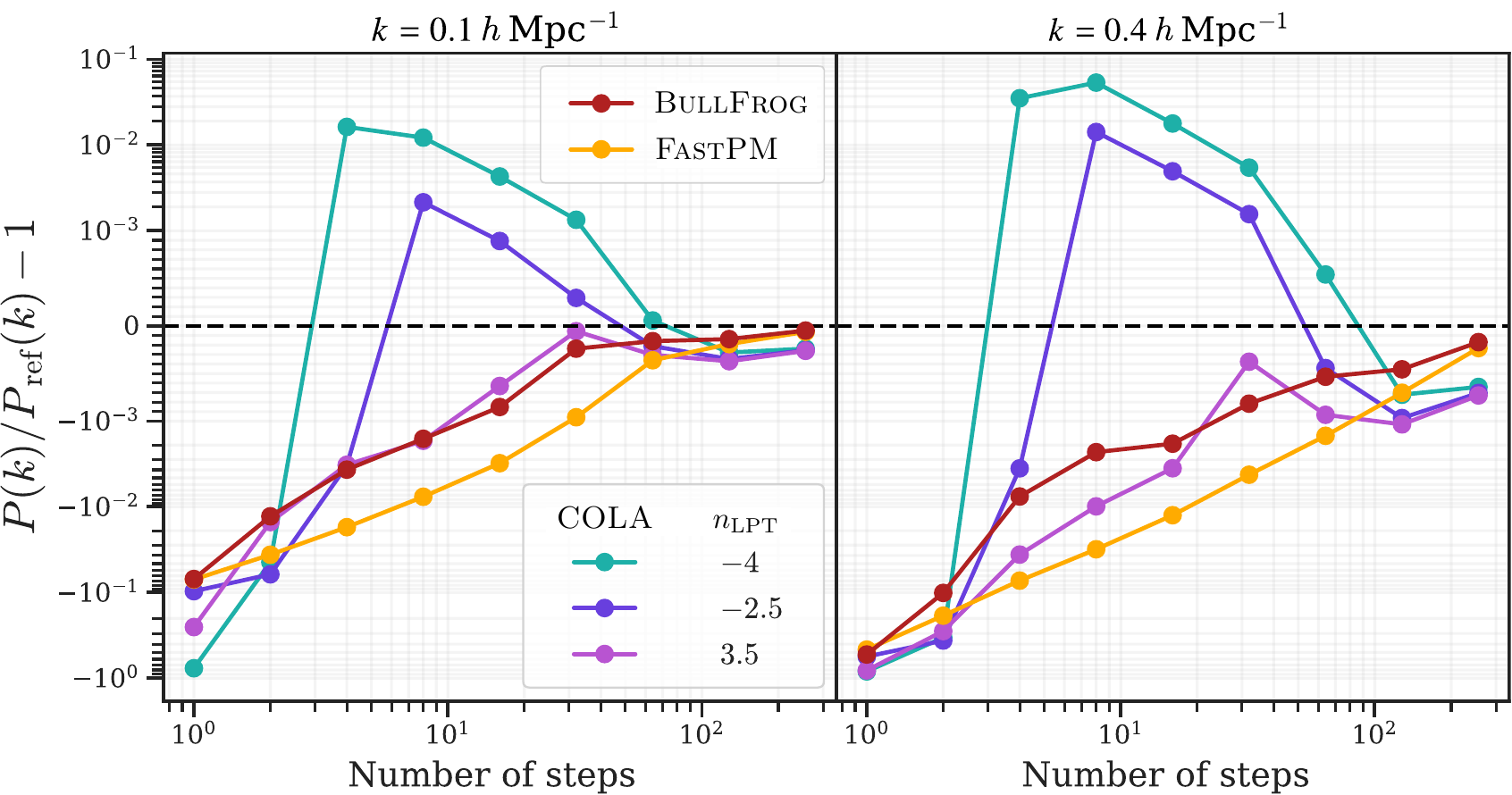}
    }
    \caption{Convergence of the power spectrum for different integrators, plotted in a symmetric logarithmic scale, evaluated for two $k$-bins centred at $k = 0.1\,h\,\mathrm{Mpc}^{-1}$ (left) and $k = 0.4\,h\,\mathrm{Mpc}^{-1}$ (right) as a function of the number of time steps. We take $512$BF as our base, and initialise all simulations at $z_\ini =49$ using 2LPT, to ensure that all results converge towards the same solution for the number of steps $n\to \infty$. }
    \label{fig:Pk_convergence}
\end{figure}

Finally, in figure~\ref{fig:Pk_convergence} we compare the speed of convergence of the power-spectrum predictions for various fast integrators as a function of number of steps~$n$, evaluated for two $k$-bins centred at $k = 0.1\,h\,\mathrm{Mpc}^{-1}$ (left panel) and $k = 0.4\,h\,\mathrm{Mpc}^{-1}$ (right panel). 
Here we take 512BF as our reference. Rather than initialising the $D$-time integrators at $a = 0$ and \textsc{COLA} at some $a > 0$ as above, we now initialise \textit{all} simulations with 2LPT at $z_\ini = 49$ in this experiment. This is in order to ensure that the truncation errors at the time of initialisation are the same for all integrators, and hence the integrators should converge towards the same solution in the limit of infinitely many steps. Depending on the choice of the hyperparameter~$n_{\mathrm{LPT}}$, the \textsc{COLA} power spectrum overshoots already for a few steps, and the convergence behaviour is somewhat erratic. This observation suggests that the optimal value of~$n_{\rm LPT}$ in the modified \textsc{COLA} scheme might benefit from a `lucky coincidence', where the overshooting of the power spectrum with a fairly low number of \textsc{COLA} steps compensates for the lack of power caused by the truncation errors due to the late initialisation with 2LPT. Indeed, for $n_{\rm LPT}=-2.5$, for example, and the $k$-bins as shown in figure~\ref{fig:Pk_convergence}, very accurate COLA results are achieved for $n \simeq 50$, while some accuracy would be lost if $n$ is, at first, just slightly increased (while ultimately converging towards the exact solution for $n \gg 50$). Hence, the modified COLA scheme can end up in an `overdrive mode' for certain tuples of $(n_{\rm LPT}, n)$, where COLA is already very accurate, although the results are clearly not yet converged in $n$.

In contrast, figure~\ref{fig:Pk_convergence} shows that both $n$BF and $n$FPM converge monotonically with $n$, while \BF\ consistently achieves a significantly higher accuracy than \fastpm\ for any fixed $1< n \lesssim 256$. For example, 32FPM [64FPM] reaches an accuracy of $10^{-3}$ for the smaller [larger] $k$-bin relative to 512BF, while \BF\ achieves this already with 16 [32] steps. For very few steps $n \lesssim 10$, the difference at $k = 0.4\,h\,\mathrm{Mpc}^{-1}$ is even more pronounced: sub-percent accuracy is achieved already with 4 \BF\ steps, but only with $>16$ \fastpm\ steps. As a rough rule of thumb, applicable on the large scales considered, we find that \BF\ is 2 $-$ 3 times more efficient with respect to \fastpm\ for reaching a given target accuracy when `jump starting' from $z_\ini=49$ with 2LPT initial conditions. We emphasise that the \BF-performance gain w.r.t.\ \fastpm\ comes virtually for free, as the only difference between these integrators is the different choice of weights (cf.\ eqs.\,\ref{eq:BFcoeff} with eqs.\,\ref{eq:FastPMcoeff}). Nonetheless, the symplecticity of \fastpm\ can be expected to be beneficial when considering smaller scales (see discussion in section~\ref{sec:nbody}), for which reason we do not claim that \BF\ makes \fastpm\ obsolete.

\section{Conclusions}\label{sec:conclusions}

We introduce \BF, the \uline{B}lazingly fast, \uline{U}V-complete, and \uline{L}PT-informed \uline{L}eap\uline{\textsc{Frog}}~integrator for cosmic structures. \BF\ is 2LPT consistent without computing and storing 2LPT trajectories, unlike \textsc{COLA}. Instead, \BF\ uses the first- and second-order $\Lambda$CDM growth functions as sole inputs, which are themselves solutions to ordinary differential equations (eqs.\,\ref{eqs:2LPT}). Within a drift-kick-drift scheme, these growth functions appear in temporal coefficients within the kick update in a way, such that both the particle positions and momenta agree with the truncated 2LPT prediction, after each completed time step, before shell-crossing, and up to higher-order effects. See section~\ref{sec:nut} for a brief explanation how this is achieved, and appendix~\ref{app:KDK-BF} for the kick-drift-kick variant of \BF.

Crucially, \BF\ is UV complete in the temporal sense and converges quadratically against the exact solution regardless of shell-crossing (section~\ref{sec:convergence}), but only if the actual growth functions are used and not approximations thereof (section~\ref{sec:loss-conv}). Also, as with any leapfrog integrator, we emphasise that second-order convergence is in general not achieved whenever the force field is not sufficiently regular~\cite{List:2023jxz}, which in cosmology occurs at caustics~\cite{2021MNRAS.505L..90R}.

In both spirit and implementation, \BF\ can be viewed as the 2LPT extension of \fastpm~\cite{2016MNRAS.463.2273F}, with the latter being consistent with the ZA ($=$1LPT). Furthermore, within the given leapfrog scheme (see below), the difference between \BF\ and \fastpm\ boils down to a choice of temporal coefficients, which are easily calculated in either case. Consequently, integrating \BF\ into code environments that already support \fastpm\ is very straightforward.

\BF\ heavily leverages the framework of $D$-time integrators (or `$\mPi$-integrators') introduced in refs.\,\cite{List:2023jxz, List:2023kbb}, and makes slight improvements to their \textsc{PowerFrog} integrator, which is 2LPT consistent at the initial step, but only ZA consistent afterwards (section~\ref{sec:nbody}). Importantly, while there  is a one-parameter family of ZA-consistent $D$-time integrators (which includes \fastpm\ and \textsc{PowerFrog}), \BF\ is the unique 2LPT-consistent member of this class. This is because in \BF\ the two degrees of freedom in the kick are used to exactly match the 1LPT and 2LPT terms. With no further degrees of freedom left, matching even higher LPT terms would require `sandwiching' additional drifts and kicks within each step; we leave related avenues for future work.

The added benefit of \BF\ in comparison to \fastpm\ is a two- to ten-fold reduction of time steps needed to reach a certain target accuracy in the matter power spectrum on mildly nonlinear scales 
(figs.\,\ref{fig:Pk_comparison_filtered}, \ref{fig:Pk_comparison_unfiltered} and~\ref{fig:Pk_convergence}). Here, the reduction factor depends on the (large) scales considered, but also on the chosen initialisation time. 
Regarding the former, the performance gain of \BF\ is most significant on mildly nonlinear scales ($k \lesssim 0.6\,h\,\text{Mpc}^{-1}$), while the difference to \fastpm~(and other $D$-time integrators) becomes negligible on highly nonlinear scales ($5\,h\,\mathrm{Mpc}^{-1} \lesssim k \lesssim 30\,h\,\mathrm{Mpc}^{-1}$), since 2LPT consistency barely matters in that regime.  Nonetheless, it is worth highlighting that all $D$-time integrators that are at least ZA consistent
significantly outperform the standard symplectic leapfrog integrator, even on these highly nonlinear scales, as demonstrated in figure~10 of ref.\,\cite{List:2023jxz}.

All considered $D$-time integrators (in DKD form) support the start of simulations directly at time zero, but there, \fastpm\ performs the weakest due to the missing 2LPT consistency.  In all our LSS experiments, the performance of \BF\ is best among the $D$-time integrators, followed closely by \textsc{PowerFrog}, and not so closely by \fastpm. Nonetheless, \fastpm\ is the only $D$-time integrator that is symplectic~\cite{List:2023jxz}, which is a desired property for modelling virialised systems such as halos.

We also carried out tests against \textsc{COLA}, which is not a $D$-time integrator, but incorporates 2LPT by performing a coordinate boost onto the explicitly computed 2LPT trajectories at the level of a standard leapfrog stepper~\cite{2013JCAP...06..036T}. We confirm previous findings in the literature that \textsc{COLA} can achieve fast and accurate predictions for the LSS, which however sensitively depend on the choice of the $n_{\rm LPT}$ hyperparameter${}^{\ref{footCOLA}}$ that appears within their modified scheme. \BF, by contrast, has no free parameter -- other than the number of time steps and their spacing -- and has a slightly smaller computational footprint than COLA, as 2LPT consistency comes virtually at no extra cost.

From the perspective of perturbation theory, and before shell-crossing, \BF\ amounts to implementing a multi-step LPT scheme within an $N$-body simulation. The `tasks' of the $N$-body integrator can be consistently analysed throughout in Lagrangian space, i.e., without the need for swapping temporarily back to Eulerian coordinates for determining the force field on a regular grid (section~\ref{sec:main}). The swapping between Lagrangian (i.e., comoving with  particles) and Eulerian (fixed) coordinates leads to advection terms, but through a careful analysis we find that these effects are actually accommodated for, at least up to second-order in perturbation theory; see appendix~\ref{sec:multi-time-steppingLPT} for complementary perturbative considerations in semi-Lagrangian form that mimic these swapping effects. After shell-crossing, when some or many particles are not single stream anymore, \BF\ trajectories start deviating from those predicted by LPT, since \BF\ accounts for multi-streaming in the force computation, while LPT does not. In fact, this departure from perturbative trajectories is essential for \BF\ to converge toward the correct solution, regardless of shell-crossing.

We also compared \BF\ against LPT, which is known to provide accurate results as long as most particles remain in the single-stream regime \cite{Rampf:2017jan,Saga:2018nud,2021MNRAS.501L..71R,2023PhRvD.108j3513R}. For $\Lambda$CDM simulations with a low UV cutoff ($k_\Lambda = 0.1\,h\,\text{Mpc}^{-1}$), where essentially all particles are still single stream down to $z=0$ (figure~\ref{fig:det_J_less_than_0}), we find very good agreement between \BF\ and LPT. However, achieving comparable accuracy with LPT requires high-order solutions, which become prohibitive in terms of memory and computational time due to the need for de-aliasing and storing the rapidly increasing number of perturbation kernels with increasing order. Even worse, for less strong filtering ($k_\Lambda = 0.2\,h\,\text{Mpc}^{-1}$), we observe already the first signatures of loss of LPT convergence at $z=0$, owing to the fact that many particles have left the single-stream regime. \BF\ and the other $D$-time integrators, by contrast, have a constant memory footprint when increasing the number of time steps.

From a practical standpoint, \BF\ is an-easy-to-implement, fast, and accurate forward model, and we imagine it to be useful in a host of cosmological applications. Indeed, we have seen that \BF\ converges {\it monotonically} towards the exact solution when increasing the number of time steps, with the residue being pushed to smaller / nonlinear scales. This means that with \BF, one simply needs to adjust the (low) number of time steps required to achieve a desired accuracy at a given scale. Moreover, \BF\ could replace LPT as a superior forward model in many applications, such as within effective-fluid descriptions, data inference, machine-learning tasks, or hybrid approaches thereof (e.g.\ refs.\,\cite{Ramanah:2018eed,He:2018ggn,Elsner:2019rql,Schmidt:2020ovm,Kokron:2021xgh,Kokron:2021faa,Boruah:2023fph}).

Let us comment on possible future extensions of \BF. The current implementation assumes a spatially flat $\Lambda$CDM Universe with a cosmological constant. Relaxing the flatness assumption or allowing for a time-evolving dark-energy background component is feasible, likely by simply updating the linear growth rate within the $D$-time integrator as done by the \textsc{FastPM}-based approach of Ref.\,\cite{2021A&C....3700505M}, supplemented with revisiting the second-order matching conditions against analytical LPT solutions. Similar techniques could also be applied to allow for non-zero neutrino masses in \BF. Also here, an integrator with a refined time variable is expected to be highly beneficial (see ref.\,\cite{2021JCAP...01..016B} for a \fastpm-based approach). However, owing to the scale-dependent nature of the multi-fluid problem at hand, it is likely that further alterations would be necessary, such as modifications to the source term in the Poisson equation; see e.g.\ refs.\,\cite{Elbers:2022tvb,Chen:2020kxi,Partmann:2020qzb,Heuschling:2022rae} for potentially fruitful starting points. Finally, redshift-space distortions and lightcone effects could be incorporated directly at the level of the integrator, which would further assist in bridging the gap between the theoretical modelling and observational data; we leave related implementations to future work.

\acknowledgments
We thank Giovanni Cabass, St\'ephane Colombi, Fabian Schmidt, Jens St\"{u}cker, Zvonimir Vlah, and Yvonne Y.Y.\ Wong for related discussions and/or comments on the manuscript. A differentiable implementation of \textsc{BullFrog} using \textsc{Jax}, on which the results in this work are based, will be made publicly available in the near future as part of the \textsc{Disco-DJ} framework (see ref.~\cite{hahn2023disco} for the differentiable Einstein--Boltzmann solver). Furthermore, a particle-mesh code implementing \BF\ into the MPI-parallel $n$LPT IC generator \href{https://bitbucket.org/ohahn/monofonic/}{\textsc{MonofonIC}} is available upon request.


\appendix

\section[KDK version of \BF]{KDK version of \BF\ (BF-KDK)}\label{app:KDK-BF}

In section~\ref{sec:main} we have determined the \BF-weights based on a DKD scheme. Here we consider \BF\ within a KDK scheme, with updates from step $n \to n+1$ according to
\begin{subequations} \label{eqs:KDK}
\begin{align}
   \fett v_{n+\nicefrac{1}{2}} &= \ma \,\fett v_n + \mb \,D_n^{-1} \fett A(\fett x_n) \,, \\
   \fett x_{n+1}   &= \fett x_n + \Delta D\, \fett v_{n+\nicefrac{1}{2}} \,, \\
   \fett v_{n+1}   &= \mabar \,\fett v_{n+\nicefrac{1}{2}} + \mbbar \, D_{n+1}^{-1} \fett A(\fett x_{n+1}) \,. \label{eq:vn+1}
\end{align}
\end{subequations}
Here, $\nabx \cdot \fett A = - \delta$ as before, while $\ma,\mb,\mabar$ and $\mbbar$ are weights that we determine in the following appendix, such that the integrator is 2LPT consistent. In appendix~\ref{sec:convKDK}, we then verify that the resulting integrator is second-order accurate.

We will see that the BF-KDK has some disadvantage for initialising simulations at $D=0$ as opposed to its DKD variant (which in the main text we often abbreviate just with BF), for which reason we do not recommend using its KDK version for that purpose. However, the KDK variant has a conceptional advantage at later time steps: For BF-KDK, the acceleration $\fett A$ is always evaluated at integer time steps -- as opposed to the DKD variant where $\fett A$ is evaluated at the half step. Thus, provided that the 2LPT matching condition is incorporated at integer time steps, $\fett A$ is by construction exactly evaluated at the truncated 2LPT positions in the KDK case. By contrast, in the DKD scheme and for finite $\Delta D$,  $\fett A$ is {\it never} exactly evaluated at the truncated 2LPT position due to temporal discretisation (see figure~\ref{fig:DKD-intro}), which then needs to be rectified by choosing appropriate weights to autocorrect for this slight misplacement. Moreover, KDK integrators are favourable when using individual time steps \cite{2005MNRAS.364.1105S}, which are routinely implemented in industry-standard codes. Thus, extending these codes with \BF\ might be more straightforward and give rise to interesting extensions of the present work, such as the development of hybrid schemes that follow \BF\ at early times / on large scales and employ a symplectic scheme at late times / on small scales.

We remark that, in all considered experiments, we observed that the power spectrum obtained with BF-DKD consistently converged from below, while BF-KDK convergence can occur from above, particularly when initialised at early times ($z_{\rm ini} \simeq 49$).\footnote{We thank Mayeul Aubin for pointing this out.}

\subsection{Determining the weights}

Similarly as discussed in section~\ref{sec:BFini}, \BF\ could be initialised at $D=0$ when formulated in KDK form, however we choose not to, for the following reason: if we start at $n=0$ and employ the initial conditions~\eqref{eqs:ICs}, the acceleration field $\fett A(\fett x_0)$ within the first kick step would be exactly zero; as a direct consequence, $\fett v_{\nicefrac{1}{2}}$  and the following drift update~$\fett x_1$ could only be set to agree with ZA motion at fixed truncation order. Thus, initialising BF-KDK at $D = 0$ is tantamount to starting on the Zel'dovich trajectory at the end of the first time step $\Delta D$. While the weights appearing in the second kick update could then be adjusted to agree with the 2LPT velocity at fixed truncation order, we would be left with a somewhat imbalanced situation since the positions would be lagging one order behind in terms of LPT. This imbalance could be rectified in subsequent time steps, such that the KDK integrator initialised at $D=0$ matches both the truncated 2LPT positions and velocities at later times steps (namely from $n = 2$ onwards), however this would require individually designed weights for the first few time steps, which we choose to avoid for the sake of simplicity.

Therefore, in what follows we consider exclusively the case $n \geq 1$ and thus assume that the particle positions and velocities are 2LPT consistent after the first completed time step, i.e., the matching conditions are ($n \geq 1$)
\begin{align} \label{eq:matchingKDK}
    \fett x_{n} \stackrel{!}{=} \fett x^\twoLPT|_{D = D_{n}} + h.o.t. \,, \qquad \quad
    \fett v_{n} \stackrel{!}{=} \fett v^\twoLPT|_{D = D_{n}} + h.o.t. \,,
\end{align}
where the 2LPT solutions are listed in eqs.\,\eqref{eqs:2LPT}. Of course, the required initialisation process for this could be done in various ways, specifically by using a single \BF\ step in DKD form (eqs.\,\ref{eqs:DKDintro} for $n=0$), or simply by initialising the system using a standard 2LPT initial condition generator (see e.g.\ refs.\,\cite{2006MNRAS.373..369C,Michaux:2020yis}).

To determine the weights, we first observe that within the KDK scheme, the acceleration field is always evaluated at the temporal boundaries of the step. Imposing our matching 2LPT conditions~\eqref{eq:matchingKDK} thus allows us to estimate $\fett A$ perturbatively by directly applying the perturbative considerations from section~\ref{sec:LPT}. Specifically, evaluating the truncated 2LPT acceleration~\eqref{eq:A2LPT} at $D=D_n$ and using the matching conditions~\eqref{eq:matchingKDK} in the first kick and drift updates, it is straightforward to derive
\be \label{eq:vn+1/2}
   \fett v_{n+\nicefrac{1}{2}} = - \nabq \varphi + \frac{E_{n+1}- E_n}{\Delta D}  \nabqinverse \mu_2 + h.o.t.,
\ee
which implies the following form for the first pair of weights,
\begin{subequations}\label{eqs:weightsKDK}
\begin{align}  \label{eq:alpha_full_lcdm_kdk}
  &\boxed{ \ma = \frac{D_n (E_{n+1}- E_n) + \Delta D (D_n^2 - E_n) }{\Delta D \left( D_n^2 + E_{n}' D_n - E_n\right)} 
  \,\stackrel{\text{EdS}}{\asymp} \, 1-\frac{3}{4\mathfrak{n}}
  \,,
     \qquad
     \mb = 1- \ma  
     \,\stackrel{\text{EdS}}{\asymp} \, \frac{3}{4\mathfrak{n}}
     \,, }  
\intertext{where we remind the reader that $\mathfrak{n}=a_n/\Delta a$ (which is equal to $n$ for uniform time steps). Similarly, using eq.\,\eqref{eq:vn+1/2} as well as the 2LPT matching conditions and plugging them in the last kick update~\eqref{eq:vn+1}, we obtain for the second pair of weights}
   &\boxed{
     \mabar = \Delta D \frac{D_{n+1}^2 + D_{n+1} E_{n+1}' - E_{n+1}}{D_{n+1} (E_{n+1}-E_n) + \Delta D (D_{n+1}^2 - E_{n+1} )} 
    \,\stackrel{\text{EdS}}{\asymp} \,
     \frac{4 + 4\mathfrak{n}}{7+4\mathfrak{n}} \,,
    \qquad
     \mbbar = 1- \mabar 
     \,\stackrel{\text{EdS}}{\asymp} \,
     \frac{3}{7+4\mathfrak{n}}
     \,,
   }
\end{align}
\end{subequations}
which concludes the derivation. Equations~\eqref{eqs:weightsKDK} are the weights for the KDK version of \BF, to be used as sole inputs in equations~\eqref{eqs:KDK}.

For completeness, we mention that in the EdS case, the BF-KDK is remarkably similar to the $D$-time integrator `\textsc{LPTFrog}' introduced by ref.~\cite{List:2023jxz}, with the only difference that with \BF\ the velocity in the friction term (see their eqs.~I.6) is evaluated at the interval \textit{boundaries} $D_n$ and $D_{n+1}$ in the first and second kick half, respectively, instead of the interval midpoint as for \textsc{LPTFrog}. As discussed in that reference, \textsc{LPTFrog} can be viewed as a slight generalisation of the leapfrog method discussed in \cite[Example~2]{Vermeeren2019} to the cosmological case. That method, in turn, is derived from the same discrete Euler--Lagrange equation that also forms the basis of `contact integrators', which respect the contact structure of the underlying system. Consequently, the same connection to contact geometry also exists for \BF, and while a rigorous analysis in this direction is beyond the scope of this work, it is remarkable that building 2LPT into an integrator leads to a method that could also have be obtained by approaching the Vlassov--Poisson system from a contact-geometrical perspective.

\vskip1cm

\subsection{Numerical convergence} \label{sec:convKDK}

To validate that BF-KDK is globally second-order accurate as $\Delta D \to 0$, we need to establish two distinct findings, namely that 
\begin{enumerate}
    \item[\circled{1}] $\ma$ and $\mabar$ are separately first-order accurate, and that
    \item[\circled{2}] the combination $\ma \mabar =: \alpha_{\rm eff}$ is second-order accurate.
\end{enumerate}
Since the remaining two coefficients $\mb$ and $\mbbar$ in the kick are defined via the Zel'dovich consistency conditions $\mb = 1 - \ma$ and $\mbbar = 1 - \mabar$, these conditions on $\ma$ and $\mabar$ are sufficient for obtaining a second-order scheme, similar to \cite[Proposition~6]{List:2023jxz}. The first condition is strictly necessary to ensure global second-order accuracy; indeed it is easy to construct integrator weights that violate $\circled{1}$ but satisfy~$\circled{2}$, which however would imply a drift with an inconsistent velocity, spoiling second-order accuracy. 

To demonstrate that both conditions are separately satisfied, we expand the \BF\ weights in powers of $\Delta D$ and compare the resulting coefficients with the
reference values from a second-order accurate KDK integrator. For the latter, we take the weights for the KDK version of \fastpm,${}^{\ref{foot:alpha}}$ which read \cite{List:2023jxz}
$\ma^{\FastPMsmall} = {\cal F}_n/{\cal F}_{n+\nicefrac{1}{2}}$ and $\mabar^{\FastPMsmall} = {\cal F}_{n+\nicefrac{1}{2}}/{\cal F}_{n+1}$ such that $\alpha^\FastPMsmall = \ma^{\FastPMsmall} \mabar^{\FastPMsmall}$ as before (see section~\ref{sec:convergence}).

We begin by verifying the condition $\circled{1}$ for which we expand the reference values as
\be
  \ma^{\FastPMsmall} = 1 - \ma_1^{\FastPMsmall} \Delta D + O(\Delta D^2) \,,
  \qquad
  \mabar^{\FastPMsmall} = 1 - \mabar_1^{\FastPMsmall} \Delta D + O(\Delta D^2) \,, \label{eqs:ma&mabar}
\ee
where 
\be
   \ma_1^{\FastPMsmall} = \mabar_1^{\FastPMsmall} =  \frac{{\cal F}'}{2{\cal F}}
    = \frac{3a D}{4 \mathscr{F}^2} \,,
\ee
which is precisely equal to $\alpha_1^{\FastPMsmall}/2$ (cf.\ eq.\,\ref{eq:FPM-DeltaDcoeff}); here and in the following, functional dependencies are w.r.t.\ $D=D_n$ which we omit for notational simplicity. We compare the coefficients~\eqref{eqs:ma&mabar} against those of \BF, which are
\be
  \ma = 1 - \ma_1 \Delta D + O(\Delta D^2) \,,
  \qquad
  \mabar = 1 - \mabar_1 \Delta D + O(\Delta D^2) \,,
\ee
with
\be \label{eq:ma1}
  \ma_1 = \mabar_1 = [1/2] D E'' W^{-1} \,,
\ee
where $W= E-D (D + E')$. Equation~\eqref{eq:ma1} is equal to $\alpha_1^\BFsmall/2$ (cf.\ eq.\,\ref{eqs:alpha12}) and thus, by virtue of identical derivations as carried out in section~\ref{sec:convergence}, we conclude that the \BF\ weights are locally first-order accurate, as required by condition $\circled{1}$.

Secondly and finally, to verify that condition $\circled{2}$ is met, we consider the second-order expansion of the reference value $\alpha^\FastPMsmall = \ma^{\FastPMsmall} \mabar^{\FastPMsmall}$, which we repeat here for convenience,
\begin{align}
  \alpha^\FastPMsmall &=  1-  \alpha_1^\FastPMsmall \Delta D 
    +  \alpha_2^\FastPMsmall   \Delta D^2 + O (\Delta D^3 )   \,,\label{eq:secondorderRep} 
\intertext{where}
    \alpha_1^\FastPMsmall &= \frac{{\cal F}'}{{\cal F}}  \,, \qquad \quad    
    \alpha_2^\FastPMsmall = \frac{2 {{\cal F}'}^2 - {\cal F} {\cal F}''}{2 {\cal F}^2}  \,.
\end{align}
By contrast, expanding the effective \BF\ weight $\alpha_{\rm eff} = \ma \mabar$ to second order in the $\Delta D$ expansion, we have
\begin{align}
  \alpha_{\rm eff} &= 1 - \alpha_{1,\rm eff} \Delta D  + \alpha_{2,\rm eff} \Delta D^2 + O(\Delta D^3)\,,
\intertext{where}
 \alpha_{1,\rm eff} &= D E'' W^{-1} \,,   \qquad 
    \alpha_{2,\rm eff} = - [1/2]\left( D E''' + E'' \right) W^{-1}
       - D^2 E'' W^{-2} \,.
\end{align}
Actually, these two $\Delta D$-coefficients coincide exactly with those as derived in the DKD case of \BF\ (cf.\ eq.\,\ref{eqs:alpha12}); thus, we conclude that also BF-KDK is second-order accurate.  \hfill $\blacksquare$

\vskip1cm

\section{Multi-time LPT and \texorpdfstring{$\fett N$}{N}-body simulations}\label{sec:multi-time-steppingLPT}

In the main text, we have determined the acceleration $\fett A$ appearing in the kick step of \BF\ by a perturbative calculation that was performed strictly in Lagrangian coordinates (section~\ref{sec:main}). The reader might be surprised how this is possible or even justified, especially considering that cosmological $N$-body simulations employ a `semi-Lagrangian scheme' (elucidated below), and thus, such simulations are not strictly Lagrangian.

While there are many different implementations of semi-Lagrangian schemes in the literature, let us explain in words how such a scheme can be viewed in light of DKD integrators:
  
$N$-body particles are displaced in Lagrangian coordinates. To determine the resulting gravitational acceleration from this displacement, one temporarily swaps to a Eulerian reference frame to solve the Poisson equation $\nabx \cdot \fett A = - \delta$.  This swapping from Lagrangian to Eulerian space is achieved using interpolation methods; most common is the use of the cloud-in-cell interpolation, but there are of course many other methods (see e.g.\ \cite{2022LRCA....8....1A} for a recent review). Once the Poisson equation is solved, one again uses an interpolation scheme, thereby swapping back to Lagrangian coordinates where the resulting acceleration $\fett A$ is assigned to the $N$-body particles. One then updates their velocity during the kick and advances them in the second half drift, all in Lagrangian space, completing the DKD operation. (All this is repeated for consecutive time steps.)

In the subsequent sections, we analyse the aforementioned semi-Lagrangian method by analytical means. For this we introduce a time-shifted variant of LPT, which comes with the crucial addition that temporal expansions are performed not around $a=0$ but any $a_s>0$. Here and in the following, we work with an EdS cosmology for simplicity, but this could be straightforwardly rectified, for which we give explicit instructions when appropriate.

\begin{figure}
    \centering
    \resizebox{0.92\textwidth}{!}{
    \includegraphics{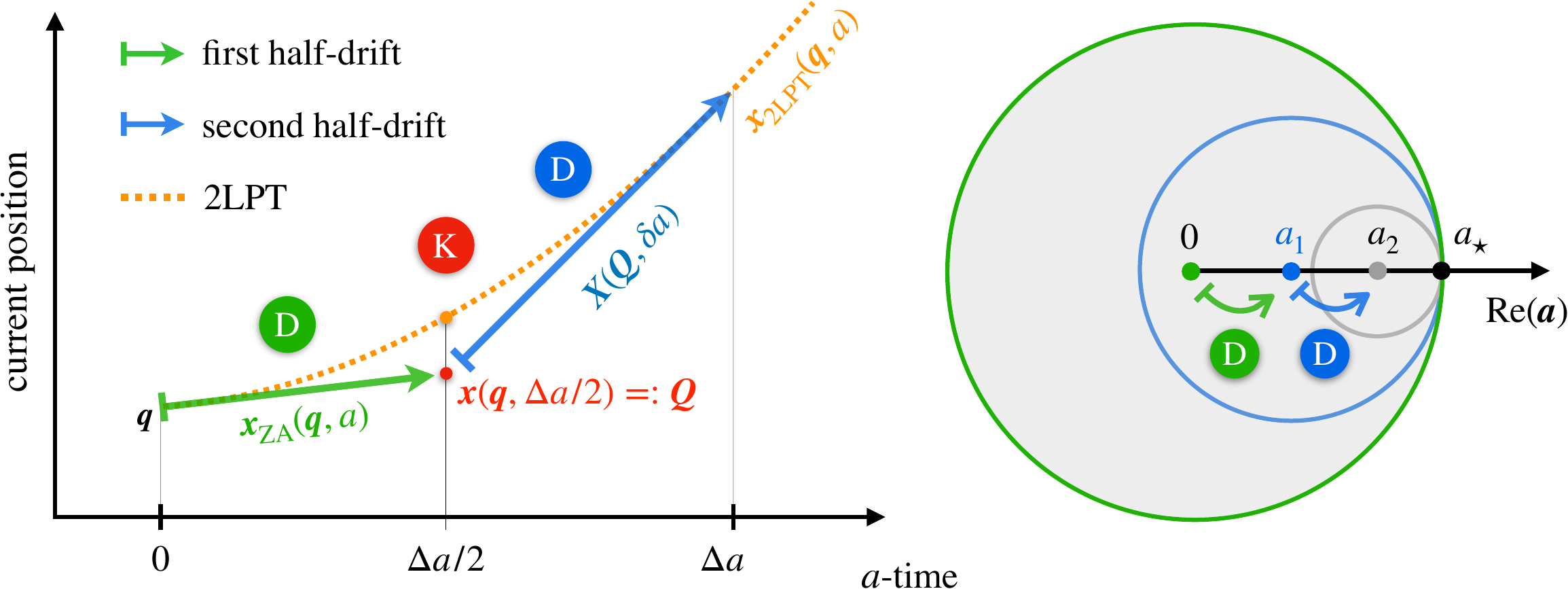}
    }
    \caption{{\it Left sketch:} First DKD step of \BF\ (green and blue arrows), as compared against the truncated 2LPT solution (orange dashed) in an EdS universe. In \BF, the first drift is executed within the ZA, therefore, particles will be at a misplaced position at the half-time $a=\Delta a/2$ w.r.t.\ 2LPT. To determine the resulting acceleration at half time, we employ time-shifted LPT (TSLPT, appendix~\ref{app:TSLPT}), which allows us to initialise the second drift (blue arrow) with the density and velocity inputted from the previous drift (green arrow), expressed in `fresh' Lagrangian coordinates denoted $\fett Q := \fett x(\fett q, \Delta a/2)$, and current position $\fett X(\fett Q,\delta a)$, where here $\delta a = a- \Delta a/2$.    Evaluating then $\fett X(\fett Q,\delta a)$ at $\delta a \to 0^+$ allows us to determine the acceleration at half time (appendix~\ref{app:force}). This strategy closely resembles the semi-Lagrangian nature inherent in the first DKD step of \BF. 
    {\it Right sketch:} Shown are various discs of convergence of displacement power series in the formally complexified $a$-time variable, denoted with $\fett a$ in the following. Specifically, the grey disc indicates the disc of convergence of standard LPT which is limited by a mathematical singularity located at complex time~$\fett a = a_\star$ (which generically arises after shell-crossing \cite{2021MNRAS.501L..71R,2023PhRvD.108j3513R}). The blue circle centred at $\fett a = a_1$ denotes the convergence radius of TSLPT at that time step; and so on. The outlined strategy invokes a generalisation of the analytic continuation theorem of Weierstrass, which in the present case amounts to displacing particles within a multi-time stepping approach.
    }
    \label{fig:DKD-LPT}
\end{figure}

In the following, we introduce the time-shifted equations of motion together with a time-shifted LPT expansion, which we dub TSLPT. In appendix~\ref{app:force}, we apply the TSLPT to the calculation of the force during the initial step of \BF. See figure~\ref{fig:DKD-LPT} where we show an illustration of our strategy to compute the force at the first half step (left panel), as well as the mathematical idea of TSLPT to drift particles within a multi-step setup (right panel).

\subsection{Time-Shifted Lagrangian Perturbation Theory}\label{app:TSLPT}

Let us begin with the basic equations that are at the core of standard (i.e., non-shifted) LPT, given here for an EdS universe for simplicity,
\begin{align} \label{eqs:standardEoMs}
  \frac 2 3 a^2 {\fett{x}}''(\fett q, a)  + a {\fett{x}}'(\fett q, a)  &= -  \nabx \phi , \qquad  \delta(\fett q, a) = \frac{1}{J} - 1,  \qquad
  \nabx^2 \phi(\fett q, a) = \delta(\fett q, a)  \,,
\end{align}
where from here on, temporal derivatives are w.r.t.\ the scale-factor time $a$ and denoted with a prime. In standard LPT, these equations are solved with $\fett x - \fett q = \fett \psi = \sum_{n=1}^\infty \fett \psi^{(n)}\,a^n$, i.e., a power series expansion about $a=0$; see eqs.\,\eqref{eqs:2LPT} for the (well-known) EdS results up to second order in the limit $\Lambda \to 0$. We remark that, beyond second order, one needs to solve for the zero-vorticity constraint (the Cauchy invariants) in order to incorporate transverse displacements; see, e.g., \cite{Rampf:2021rqu}. 

By contrast, with TSLPT we aim to enable perturbative expansions about a shifted expansion point $a_s \geq 0$, which we take to be a constant and $s$ is an integer counter. Here, $a_s$ can be arbitrarily large and thus is not a perturbative quantity. To accommodate the necessary framework for TSLPT, we proceed twofold. First, we define a `fresh' Lagrangian map $\fett Q \mapsto \fett X(\fett Q,\delta a)$ from a `fresh initial time' $\delta a=0$ to evolved time $\delta a$ (our expansion parameter), where $a= a_s + \delta a$ (see left panel in figure~\ref{fig:DKD-LPT} for $a_s=\Delta a/2$). Second, we shift eqs.\,\eqref{eqs:standardEoMs} according to $a \to a_s + \delta a$. The resulting `time-shifted' equations of motion are then
\begin{subequations} \label{eqs:time-shiftedEoMs}
    \begin{align}
       \frac{2}{3}\left( a_s + \delta a \right)^2 {\fett X}''(\fett Q, \delta a) + \left(a_s + \delta a  \right)  {\fett X}'(\fett Q, \delta a)  &=  - \nabX \phi(\fett Q,\delta a ) \,, \label{eq:Eulershifted} \\ 
       \delta(\fett Q, \delta a) &= \frac{1+ \delta_s(\fett Q)}{J(\fett Q, \delta a)} -1 \,,  \label{eq:massshifted}\\
       \fett \nabla_\text{\fontsize{6}{6} \!\!$\fett X$\!}^2 \phi(\fett Q, \delta a) &= \delta(\fett Q, \delta a)\,,
\end{align}
\end{subequations}
where, importantly, we have added in eq.\,\eqref{eq:massshifted} the `initial' density $\delta_s := \delta(\delta a=0)$ provided at $a=a_s$, which is in general nonzero (as opposed to eq.\,\ref{eqs:standardEoMs} where initial quasi-homogeneity is assumed). We remark that, as in the standard LPT case, one also needs to solve for the zero-vorticity constraint $\nabx \times \fett v = \fett 0$, which in Lagrangian coordinates becomes $\varepsilon_{ijk} X_{l,j} X_{l,k}'=0$ for the shifted map. For our initial conditions, however, the resulting transverse displacement contributions only become nonzero beyond the second order, therefore we do not consider this constraint in the following.

To solve the shifted equations~\eqref{eqs:time-shiftedEoMs}, we define the `shifted' displacement $\fett S = \fett X- \fett Q$, for which we impose the Ansatz
\be
 \fett S(\fett Q,\delta a) = \sum_{n=1}^\infty \fett S^{(n)}(\fett{Q})\, {\delta a}^{n} \,.
\ee
The first-order solution follows directly from the definition of the shifted map or, put differently, from momentum conservation evaluated at $\delta a=0$, i.e., 
${\fett S}' (\fett Q, \delta a )|_{\delta a =0} = \fett S^{(1)} (\fett Q) = \fett{v}_s(\fett Q)$, where $\fett v_s := \fett v(\delta a=0)$ is the velocity at time $a=a_s$. The first-order solution is then used as an input to solve the evolution equations at second order. Note that in TSLPT, a second-order treatment is the absolute minimum, otherwise the second input field $\delta_s(\fett Q)$ is not passed over, thereby violating the second-order structure of the underlying differential equation. In summary, we find for the first and second-order solutions respectively
\begin{subequations} \label{eqs:TSLPT1-2}
\begin{align}
   S_{l,l}^{(1)}(\fett Q) &= \nabQ \cdot \fett{v}_s 
   = - \varphi_{,ll}^\ini(\fett Q)  - \frac{\Delta a}{2} \left( \varphi_{,lm}^\ini\varphi_{,lm}^\ini + \varphi_{,l}^\ini\varphi_{,lmm}^\ini\right)
   \,, \\
  S_{l,l}^{(2)}(\fett Q) &= - \frac{3}{4 a_s} \left[ \frac{\delta_s(\fett Q)}{a_s} + S^{(1)}_{l,l}(\fett Q) \right]
  = - \frac 3 4 \mu_2(\fett Q) \,.
\end{align}
\end{subequations}
In summary, the time-shifted displacement truncated at second order is then
\be
  \fett S(\fett Q,\delta a) =  - \delta a\, \fett {\nabQ} \varphi^\ini(\fett Q)
   - \delta a \,a_s \nabQinverse \left( \varphi_{,lm}^\ini\varphi_{,lm}^\ini + \varphi_{,l}^\ini\varphi_{,lmm}^\ini\right)
   - \frac 3 4 \delta a^2   \nabQinverse \mu_2 + O(\delta a^3)\,, \label{eq:Psishifted2nd}
\ee
where $\nabQinverse := \fett \nabla_{\fett Q}^{-2}\fett{\nabQ}$. Next, as promised, we briefly outline the steps needed to generalise TSLPT to a $\Lambda$CDM cosmology. A potentially good starting point for this is the equations of motion formulated in $D$-time, which are given in ref.~\cite[eqs.\,5-7]{Rampf:2022tpg}. These equations can be shifted according to $D \to D_1 + \Delta D$ and solved by an Ansatz for $\fett S$ in powers of $\Delta D$, which is a straightforward exercise. However, the shifted displacement results should be resummed, so as to include all $\Lambda$ terms proportional to the given spatial terms to fixed order in perturbation theory; see ref.~\cite[Secs.\ 4.1-4.2]{Rampf:2022tpg} where this has been done for the nonshifted LPT case. We leave related avenues for TSLPT for future work.

Before concluding this appendix, let us outline a (delayed) motivation and underlying strategy for TSLPT; see the right panel of figure~\ref{fig:DKD-LPT} for a related illustration. Mathematically speaking, one does not need TSLPT to begin with, since LPT converges until the time of first shell-crossing and slightly beyond \cite{Rampf:2017jan,Saga:2018nud,2021MNRAS.501L..71R,2023PhRvD.108j3513R}: Lagrangian perturbation theory, which employs effectively a time-Taylor series, is not limited by complex-in-$a$-time singularities, but by a mathematical singularity located on the real-time axis. For generic initial conditions, this `real singularity' occurs after shell-crossing, implying that LPT is time-analytic and thus converges for the whole real-time branch from initial time, and at least until shell-crossing. However, evaluating the LPT displacement at times close to its final time of convergence requires large orders in order to get sufficiently accurate representations for the displacement (except in 1D where ZA is the only nonzero term in the Taylor series). In practice, generating LPT solutions at large orders is computationally prohibitive (see main text).

Time-shifted LPT avoids this computational drawback, by replacing a mega-time step in LPT with consecutive TSLPT steps: since by definition the TSLPT steps will be smaller, a lower-order truncation of the shifted displacement should remain accurate, basically since the fresh input fields are assigned at each new time step. The underlying mathematical idea is rooted in a generalisation of the Weierstrass analytic continuation theorem for analytic functions; see ref.\,\cite{2016JCoPh.306..320P} for numerical implementations of highly related schemes, however applied to incompressible Euler flow.

\subsection[\texorpdfstring{$N$}{N}-body force computation in  light of time-shifted LPT]{\texorpdfstring{$\fett N$}{N}-body force computation in light of time-shifted LPT}\label{app:force}

Finally, let us proceed with the main task of this appendix, which is illustrated in the left panel of figure~\ref{fig:DKD-LPT}: Given the LPT density and velocity at the evolved position $a=\Delta a/2$, we seek to employ TSLPT to determine the acceleration field at that half step. The result is identical to the one outlined in the main text, but the involved technical procedure is certainly very different. The present complementary derivation serves two main purposes, namely (1) to validate TSLPT, (2) and as a consistency check for the acceleration computation in the main text.

To achieve this, and to closely resemble the initialisation process of \BF\ (and for any $D$-time integrator as well), we assume here that the first drift operation has been executed assuming Zel'dovich dynamics (see section~\ref{sec:nbody} and in particular eq.\,\ref{eqs:ICs}); this restriction can easily be alleviated if needed. For the first half drift at $a=0$, we can employ just standard LPT, which implies
\begin{align}
    \textcolor{darkgreen}{\fett x(\fett q, \Delta a/2)} = \fett q - \tfrac{\Delta a}{2} \nabq \varphi^\ini \,, \qquad \qquad  \circledgreen{D} 
\end{align}
which allows us to determine the density and velocity at the half-time, 
\begin{align}
       \textcolor{darkgreen}{ \delta(\fett x(\fett q,\Delta a/2))}
          &= \frac{1}{\det [\delta_{ij} - (\Delta a/2) \varphi^\ini_{,ij}]} - 1  \nonumber \\ 
          &= \frac{\Delta a}{2} \varphi_{,ll}^\ini(\fett q)- \left( \frac{\Delta a}{2} \right)^2 \mu_2 + \left( \frac{\Delta a}{2} \right)^2 \varphi_{,ll} ^\ini\varphi_{,mm}^\ini + h.o.t. =: \textcolor{mblue}{\delta_{\nicefrac{1}{2}}(\fett q) } \,, \label{eq:delta1/2} \\
         \textcolor{darkgreen}{ \fett v(\fett x(\fett q,\Delta a/2))} 
           &= - \nabq \varphi^\ini(\fett q)
           =: \textcolor{mblue}{\fett v_{\nicefrac{1}{2}}(\fett q)} \,, \label{eq:v1/2}
\end{align}
where quantities related to the first [second] drift half are coloured in green [blue]. Before employing TSLPT, we need to make sure that the fields for the second drift half are passed over by taking the fresh coordinates into account. Since we only consider results to second-order accuracy, that task is achieved by synchronising the coordinates at $a=\Delta a/2$ with
\be
  \textcolor{darkgreen}{\fett q - \frac{\Delta a}{2}\nabq \varphi^\ini(\fett q) }=: \textcolor{mblue}{{\fett Q}} 
  \qquad \Rightarrow\qquad    \fett q = {\fett Q} + \frac{\Delta a}{2}\nabQ \varphi^\ini({\fett Q}) + h.o.t. \,,
\ee
Plugging the second expression into the evolved fields~\eqref{eq:delta1/2}--\eqref{eq:v1/2}, one then gets the input fields as a function of the fresh coordinates,
\begin{align}
  \begin{aligned}
    \textcolor{mblue}{\delta_{\nicefrac{1}{2}}(\fett Q)} 
     &= \frac{\Delta a}{2} \varphi_{,ll}^\ini(\fett Q)
      + \left( \frac{\Delta a}{2} \right)^2 \varphi_{,l}^\ini \varphi_{,lmm}^\ini
     - \left( \frac{\Delta a}{2} \right)^2 \mu_2 + \left( \frac{\Delta a}{2} \right)^2 \varphi_{,ll}^\ini \varphi_{,mm}^\ini + h.o.t. 
     \,, \\
    \textcolor{mblue}{\fett v_{\nicefrac{1}{2}}(\fett Q)} &= - \nabQ\varphi^\ini(\fett Q) -\frac{\Delta a}{2} \varphi_{,l}^\ini \nabQ \varphi_{,l}^\ini + h.o.t.
 \end{aligned}  \label{eqs:ICsshifted}
\end{align}
Plugging the input fields~\eqref{eqs:ICsshifted} in the TSLPT results~\eqref{eqs:TSLPT1-2} while setting $a_s = \Delta a/2$ leads directly to the time-shifted displacement
\be
  \textcolor{mblue}{\fett S(\fett Q,\delta a)} =  - \delta a\, \fett {\nabQ} \varphi^\ini(\fett Q)
   - \delta a \,\frac{\Delta a}{2} \nabQinverse \left( \varphi_{,lm}^\ini\varphi_{,lm}^\ini + \varphi_{,l}^\ini\varphi_{,lmm}^\ini\right)
   - \frac 3 4 \delta a^2   \nabQinverse \mu_{2} \,, \quad \circledblue{D}\label{eq:Psishifted}
\ee
truncated up to second order. Equipped with the shifted displacement, we can determine the acceleration at the half step in three simple steps. First, similar as in the main text, we exploit the equivalence principle and express the gravitational field, here defined with $\fett A = - \nabX \phi$, in terms of the acceleration:
\be
  \frac{2}{3} \left( \Delta a/2 + \delta a \right)^2 {\fett X}''
  + \left( \Delta a/2 + \delta a \right) {\fett X}' = \fett A  \label{eq:A}
\ee
(eq.\,\eqref{eq:Eulershifted} for $a_s = \Delta a/2$).
Second, we replace $\fett X = \fett Q + \fett S$ and use eq.\,\eqref{eq:Psishifted}; evaluating eq.\,\eqref{eq:A} then at $\delta a  \to 0^+$ leads to the acceleration at the half step,
\be
   \textcolor{red}{\fett A(\fett Q,\delta a=0)} = - \frac{\Delta a}{2} \fett {\nabQ}\varphi^\ini(\fett Q) - \left( \frac{\Delta a}{2} \right)^2 \nabQinverse \left[ \mu_2 + \varphi_{,lm}^\ini  \varphi_{,lm}^\ini  + \varphi_{,l}^\ini  \varphi_{,lmm}^\ini \right]  \,.
\ee
Third and finally, we wish to express the acceleration in terms of the original Lagrangian coordinates; swapping back can be achieved by evaluating the last equation at the position $\fett Q = \fett q + (\Delta a/2) \nabq \varphi^\ini+h.o.t.$, which after straightforward calculations leads to our final result of this appendix,
\be  \label{eq:Afin}
  \boxed{ \textcolor{red}{\fett A(\fett q, \delta a=0)}  =  - \frac{\Delta a}{2} \nabq\varphi^\ini({\fett q})  - \left( \frac{\Delta a}{2} \right)^2 \nabqinverse \mu_2 \,. 
  }
\ee
In deriving this result, the following identity is useful: $\varphi_{,l}^\ini \nabq \varphi_{,l}^\ini = [1/2]  \nabqinverse ( \varphi_{,l}^\ini \varphi_{,l}^\ini )_{,mm}$.
Evidently, eq.\,\eqref{eq:Afin} agrees with~\eqref{eq:accZArep} in the limiting case of an EdS universe, although the respective derivations are quite distinct: In the main text, we performed the acceleration calculation in a completely Lagrangian fashion, while here we have determined it using a semi-Lagrangian approach that relies on swapping between Lagrangian and Eulerian steps in the derivation.

\bibliographystyle{JHEP}
\bibliography{biblio}

\end{document}